\documentclass{aa}  
\usepackage{graphicx}
\usepackage{txfonts}
\usepackage{color}
\usepackage{wasysym}
\usepackage{natbib}
\begin{document}
    \title{    Edging towards an understanding of CH/CH$_2$ on nano-diamonds:  }
    \subtitle{Regular and semi-regular polyhedra and diamond network models} 

    \author{A.P. Jones 
 %          \inst{1} 
%           \and
}

    \institute{Universit\'e Paris-Saclay, CNRS,  Institut d'Astrophysique Spatiale, 91405, Orsay, France.\\
               \email{anthony.jones@universite-paris-saclay.fr}
%         \and
%           University \\
%             \email{ }
%             \thanks{}
              }

    \date{Received ? : accepted  ?}

   \abstract
% context
{Nano-diamonds have been observed in only a handful of circumstellar regions $10-100$\,\tiny{ A.U.} from moderately bright stars ($T_{\rm eff} \sim 8,000-10,000$\,K). They have also been extracted from primitive meteorites; some of these are clearly pre-solar, that is to say that they formed far from the solar system and therefore traversed the interstellar medium, where they must exist but, because we see no evidence of them, must be extremely well hidden.}
%aims
{Our goal is to understand if it is possible to constrain the sizes and shapes of nano-diamonds in circumstellar media using the observed ratio, [CH]/[CH$_2$], of their surface CH$_2$  and CH  infrared bands at $\simeq 3.43\,\mu$m and $\simeq 3.53\,\mu$m, respectively.}
% methods
{We calculated the CH and CH$_2$ abundances on nano-diamonds using two approaches. The first  assumes regular and semi-regular polyhedra (tetrahedra, octahedra, and cubes and their truncated forms). The second uses a diamond bonding network to derive the structures of tetrahedral and octahedral particles, and their truncated variants, and also of spherical nano-diamonds.}
% results
{As a function of the particle size and shape, and for the two different calculation methods, we derived the relative abundance ratio [CH]/[CH$_2$], which can then be weighted by their laboratory-measured infrared band intensities. The two methods give good agreement and indicate that the spread in values, over the different particle forms, is more that an order of magnitude for any size.}
% conclusions
{We conclude that the ratio [CH]/[CH$_2$], and their infrared band ratio, strongly depend upon particle size and shape. For a given shape or size, the ratio can vary by more than an order of magnitude. It may therefore be difficult to constrain nano-diamond sizes using the observed $3-4\,\mu$m spectra alone. James Webb Space Telescope (JWST) mid-infrared spectra may help, but only if bands are size-specific.}
   \keywords{ISM:dust,extinction  -- ISM:abundances }

    \maketitle
%
%________________________________________________________________

%||||||||||||||||||||||||||||||||||||||||||||||||||||||||||||||||||||||||||||||||||||||||||||||||||||||||||||||||||||||||||||||||||||||||||||||||||||||||||||||||||
\section{Introduction}
%||||||||||||||||||||||||||||||||||||||||||||||||||||||||||||||||||||||||||||||||||||||||||||||||||||||||||||||||||||||||||||||||||||||||||||||||||||||||||||||||||

Pre-solar nano-diamonds with median radii of $1.3-1.5$\,nm \citep{1996GeCoA..60.4853D} have been extracted from primitive meteorites in abundances of up to $\simeq 1400$\,ppm \citep{1995GeCoA..59..115H}. Their isotopically anomalous Xe content (Xe-HL) is considered characteristic of the nucleosynthetic processes in supernovae \citep{Lewis_etal_1987},  and their  $^{15}$N depletion and low C/N ratios are consistent with carbon-rich stellar environments \citep{1997AIPC..402..567A}. The observation of a tertiary carbon CH stretching mode at $3.47\,\mu$m towards dense regions was originally thought to be an indicator of the presence of nano-diamonds in the interstellar medium (ISM) \citep{1992ApJ...399..134A}, but this band most likely has another origin \citep{1996ApJ...459..209B}. However, we do today have conclusive observational evidence for the presence of nano-diamonds in proto-planetary discs where they are identified by their characteristic CH$_n$ ($n = 1,2$) stretching modes at $3.43$ and $3.53\,\mu$m 
\cite[e.g.][]{1999ApJ...521L.133G,2002A&A...384..568V,2004ApJ...614L.129H}. 

In the laboratory, diamond-like materials and nano-diamonds have long been synthesised and studied and there exists a vast literature on this subject.\footnote{Here we desist from giving any obviously limited and highly-selective citation of this literature.} In most of these laboratory studies, we have the luxury of being able to directly analyse the textures,  particle sizes, and shapes through a myriad of sophisticated techniques (e.g. scanning and tunnelling electron microscopy and atomic force microscopy). In astronomical observations this option is not available to us and so we must progress using solid-state material models. Our understanding is therefore critically constrained by the limitations of our knowledge and our models. 

In order to model, analyse, and interpret the spectra of the nano-diamonds observed in circumstellar proto-planetary discs, we required well-determined wavelength- and size-dependent optical constants: the complex indices of refraction. These were then used to calculate the optical properties of nano-diamonds and thence their extinction, absorption, and scattering cross-sections, which in turn were used to derive their temperatures in a given stellar radiation field.  In the following paper, we derive the nano-diamond optical constants and, in this paper, as a key input to this modelling, we concern ourselves with approximating nano-diamond structures using regular and semi-regular polyhedral shapes, principally tetrahedra (T) and octahedra (O), and their truncated forms (tT and tO), but also consider cuboctahedra (cO), cubes (C), and truncated cubes (tC). Additionally we developed a diamond bonding network model to derive the structures of tetrahedral and octahedral particles and their truncated forms. 

A key question in unravelling the essential characteristics of the $3-4\,\mu$m infrared bands that have been attributed to nano-diamonds and, in particular the $3.53\,\mu$m/$3.43\,\mu$m ratio, is how their size and the shape (e.g. rounded, angular, crystal-faceted [euhedral], \ldots) can affect this ratio. Here we consider this question from a purely geometrical aspect, that is by considering the surface and edge structures of euhedral forms modelled as regular polyhedra and diamond-bonded networks. The principal product of both of our nano-diamond modelling approaches is the relative abundance ratio [CH]/[CH$_2$] of nano-diamonds as a function of the particle size and shape. 

This paper is structured as follows: 
Section \ref{sect_properties} describes the nano-diamond physical properties, 
Section \ref{sect_shape} considers particle shapes and surfaces, 
Section \ref{sect_regulars} details the polyhedral methodology for a range of regular forms,
Section \ref{sect_irregulars} describes semi-regular truncated polyhedra,
Section \ref{sect_networks} presents the diamond-bonded network methodology,
Section \ref{sect_CH-CH2} discusses the predicted CH/CH$_2$ ratios in nano-diamonds and their use as a ruler, 
Section \ref{sect_dehydrogenation} considers nano-diamond stability and surface dehydrogenation effects, 
Section \ref{sect_discussion} discusses the implications of the results and speculates upon their utility, and
Section \ref{sect_conclusions} summarises the work and presents the conclusions.

%||||||||||||||||||||||||||||||||||||||||||||||||||||||||||||||||||||||||||||||||||||||||||||||||||||||||||||||||||||||||||||||||||||||||||||||||||||||||||||||||||
\section{Nano-diamond physical properties}
\label{sect_properties}
%||||||||||||||||||||||||||||||||||||||||||||||||||||||||||||||||||||||||||||||||||||||||||||||||||||||||||||||||||||||||||||||||||||||||||||||||||||||||||||||||||

The diagnostic potential of interstellar dust species is principally driven by their characteristic infrared spectra. In order to fully utilise this potential for hydro-carbonaceous dust, be it aliphatic-rich, aromatic-rich or (nano)diamond, we need to understand how the particle size, composition, and morphology determine its spectroscopic signatures. For nano-diamonds there is a wealth of laboratory data to aid us \citep[e.g.][]{1989Natur.339..117L,1994A&A...284..583C,1995MNRAS.277..986K,1995ApJ...454L.157M,Reich:2011gv,1998A&A...330.1080A,2000M&PS...35...75B,1998A&A...336L..41H,2002JChPh.116.1211C,2002ApJ...581L..55S,2004A&A...423..983M,Jones:2004fu,2007ApJ...661..919P,2011ApJ...729...91S,Usoltseva:2018er,1997PhRvB..55.1838Z}. The hope is that we can, at the very least, use these data to enable us to determine nano-diamond sizes in interstellar media through the ratio of the $3.43\,\mu$m and $3.53\,\mu$m band strengths \citep[e.g.][]{2002JChPh.116.1211C,2002ApJ...581L..55S,2007ApJ...661..919P} because, as these works show, this ratio is size dependent. However, the band ratio is also morphology-dependent \citep[e.g.][]{2007ApJ...661..919P} and  also depends on the nature and degree of the surface hydrogen coverage. Further, and given that nano-diamonds are observed in emission close to hot stars, this ratio will be temperature-dependent because of the underlying thermal continuum and  the possibility of differential excitation and/or the de-hydrogenation of surface CH and CH$_2$ groups.

%||||||||||||||||||||||||||||||||||||||||||||||||||||||||||||||||||||||||||||||||||||||||||||||||||||||||||||||||||||||||||||||||||||||||||||||||||||||||||||||||||
\subsection{Possible degeneracies and critical assumptions} 
%\label{sect_degeneracies}
%||||||||||||||||||||||||||||||||||||||||||||||||||||||||||||||||||||||||||||||||||||||||||||||||||||||||||||||||||||||||||||||||||||||||||||||||||||||||||||||||||

It is essential to understand and, where possible, to try and break the inherent degeneracies that must exist between the effects of particle size, shape, structure, composition, and degree of surface (de-)hydrogenation. In the latter case, we clearly need a detailed understanding of how (de-)hydrogenation may (dis)proportionately effect the CH/CH$_2$ surface concentration  ratio, that is its effect on the relative contributions of CH$_n$ groups at particles vertices, edges, and faces \citep[e.g.][]{2007ApJ...661..919P}. 

The $3.43\,\mu$m and $3.53\,\mu$m band strengths and ratios have not yet been completely calibrated because the particle shapes in most infrared (IR) studies were not determined \citep[e.g.][]{2002JChPh.116.1211C,2002ApJ...581L..55S}. This was not the case for  the small diamondoids ($N_{\rm C} < 100$) studied by \cite{2007ApJ...661..919P} where the particle shapes were principally tetrahedral and the $3.53/3.43\,\mu$m band ratio appears to be approximately linearly-dependent upon the CH/CH$_2$ surface concentration ratio. However, it should be noted that a tetrahedral form for larger diamondoids or nano-diamonds may not be the most likely or even the most stable form in either a fully hydrogenated or a fully dehydrogenated state \citep[e.g.][]{2003JChPh.118.5094B,2004JChPh.121.4276B}.

In this work we, preliminarily at least, assume that the intrinsic $3.53\,\mu$m CH band peak intensity per CH group is $\approx 1.2$ times that of the $3.43\,\mu$m CH$_2$ group, as indicated in the \cite{2007ApJ...661..919P} simulations.\footnote{The experimental results for the $3.53/3.43\,\mu$m band ratio in the poly-amantanes in the \cite{2007ApJ...661..919P} study are, however,  less clear-cut. } This calibration, based on the smaller, tetrahedral diamondoids, may no longer hold true for different shapes and/or larger nano-diamonds. However, we have no choice but to ignore this possibility until such time as suitable laboratory data become available to refute this supposition.

% FIGURE 
\begin{figure*}[h]
\centering
\includegraphics[width=18cm]{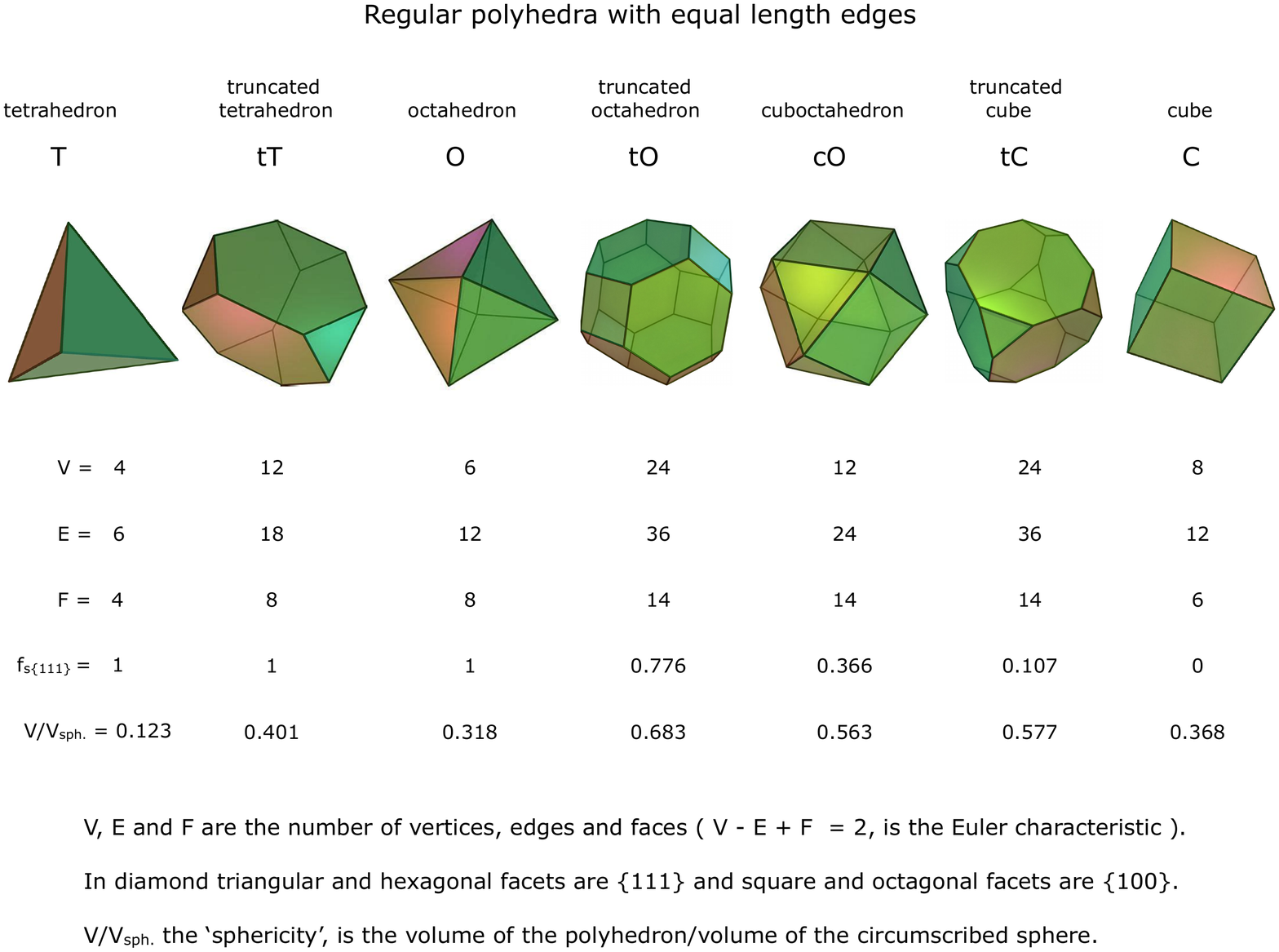}
\caption{Regular polyhedra and their truncated forms along with some of their properties. $f_{\rm s}$\{111\} is the fraction of the particle surface in triangular and hexagonal facets. We note that all square and octagonal facets are \{100\}.}
\label{fig_shapes}
\centering
\end{figure*}

%||||||||||||||||||||||||||||||||||||||||||||||||||||||||||||||||||||||||||||||||||||||||||||||||||||||||||||||||||||||||||||||||||||||||||||||||||||||||||||||||||
\subsection{Bulk density, number of C atoms, and surface coverage}
\label{sect_density}
%||||||||||||||||||||||||||||||||||||||||||||||||||||||||||||||||||||||||||||||||||||||||||||||||||||||||||||||||||||||||||||||||||||||||||||||||||||||||||||||||||

Using the regular diamond lattice structure we can construct particles with a given number of C atoms as a model for interstellar nano-diamonds. The number of constituent C atoms in a particle can be determined from the particle volume, for example $(4/3) \, \pi \, abc$, for ellipsoids ($a \ne b \ne c$), spheroids ($a \ne b = c$), and spheres ($a = b = c$), and the bulk diamond specific density (here taken to be $\rho_{\rm bulk} = 3.51$\,g\,cm$^{-3}$).  Nevertheless, the density of a particle does exhibit some dependence on its size, as shown experimentally for $30 - 200$\,nm radius silica (SiO$_2$) nano-particles, which  have densities of $\sim 1.9$\,g\,cm$^{-3}$, which is $14-30$\% lower than that of the parental solid \citep[$2.2-2.7$\,g\,cm$^{-3}$,][]{2014IAC_silica_nps}. Thus, and by analogy with silica, a structure not too dissimilar to that of diamond, and as in the THEMIS modelling \citep{2017A&A...602A..46J}, we should perhaps assume a  diamond density reduction of the order of $20$\% for sub-$\mu$m particles. However, for the present purposes, given that the density must be depth-dependent and that we consider polyhedral shapes, we assume the bulk diamond density because our network modelling approach assumes a sp$^3$ diamond C$-$C bond length of 0.154\,nm and therefore a  density of $3.51$\,g\,cm$^{-3}$.

In order to determine the number of C atoms in a particle, $N_{\rm C}$, and because we are dealing with H atoms only at the surfaces, we assume that the mean atomic mass of the `bulk' interior of a nano-diamond is that of a C atom, $A_{\rm C} = 12$ a.m.u. The number of C atoms per particle is then given by 
\begin{equation}
N_{\rm C} =  \ \frac{  V_{\rm nd}{\rm (S)} \ \rho_{\rm bulk} }{ A_{\rm C} \ m_{\rm H} }  
\simeq 734 \left( \frac{a}{[1\,{\rm nm}]} \right)^3, 
\end{equation}
where $V_{\rm nd}{\rm (S)}$ is the nano-diamond shape-dependent volume and the right hand expression gives $N_{\rm C}$ for a spherical nano-diamond. Given that nitrogen is the most common diamond hetero-atom, the inclusion of a number of N atoms, $N_{\rm N}$, within the bulk can be expressed as a fraction, $f_{\rm N}$, of the total number of C atoms, that is $N_{\rm N} = f_{\rm N} \times N_{\rm C}$. In this case the mean atomic mass of the `bulk' would then need to be adjusted accordingly, that is $A_{\rm C(N)} = (1-f_{\rm N}) \times A_{\rm C} + f_{\rm N} \times A_{\rm N}$ where $A_{\rm N} = 14$. 

The fraction of C atoms that are at the particle surface depends upon the shape and will therefore be determined on a shape-by-shape basis. For (nano)diamond particles this depends upon the numbers of CH and CH$_2$ groups (per unit edge length or per unit surface area) on particle vertices, edges and faces.  

For carbon atoms within the diamond bulk the C$-$C$-$C bond angle is 109.5$^\circ$, the tetrahedral angle, the projected distance of the 0.154\,nm C$-$C bond length (d$_{\rm C-C}$) onto a \{100\} facet\footnote{A facet is a particular crystallographic plane, face or facet indicated by \{$h,k,l$\} where $h$, $k$, and $l$ are the Miller indices, which are the minimised integer distances along the respective $x$, $y$, and $z$ crystalline axes.} is 0.126\,nm and  the distance between  CH$_2$ groups on this surface or along \{111\}-\{111\} and \{111\}-\{100\} edges is $(2 \times 0.126) = 0.252$\,nm ($= D_{\rm CH_2}$). The area per CH$_2$ group on a \{100\} facet is then $(2 \times 0.126)^2$\,nm$^2 = 0.0635$\,nm$^2$ because, viewed perpendicularly, the \{100\} surface appears as a 0.126\,nm\,$\times$\,0.126\,nm square grid with one CH$_2$ group per four squares. Therefore, \{100\} facets present one CH$_2$ group per 0.0635\,nm$^2$  ($= A_{\rm CH_2}$), or one H atom per  0.0318\,nm$^2$. In contrast, \{111\} facets present a hexagonal grid with projected C$-$C bond length sides of 0.126\,nm, and area $(3/2) \surd 3 \times 0.126^2$\,nm$^2 = 0.0412$\,nm$^2$, with one CH bond per hexagon, and  one CH group or one H atom per 0.0412\,nm$^2$  ($= A_{\rm CH}$). The H atom density on \{100\} facets is $\approx 30$\% greater than on \{111\} facets but CH groups are $\approx 35$\% more abundant on \{111\} facets. 

In  our calculations we derived the shape-dependent group abundance ratio, \{[CH]/[CH$_2$]\}$_{\rm S}$,\footnote{For simplicity this ratio is hereafter written as [CH]/[CH$_2$].} and the model-predicted IR band ratio is then 
\begin{equation}
I(3.53\,\mu{\rm m})/I(3.43\,\mu{\rm m})  = R_{\rm IR} \times \{ {\rm [CH]/[CH}_2] \}_{\rm S},  
\end{equation}
where $R_{\rm IR}$ is the measured ratio of the  $3.53\,\mu$m and $3.43\,\mu$m band intensities, that is $I_{3.53}$(CH)/$I_{3.43}$(CH$_2$) $\simeq 1.2$ \cite[e.g.][]{2007ApJ...661..919P}.\footnote{N.B., This is equivalent to a ratio of 0.6 when normalised to the number of hydrogen atoms involved.} In the following we calculated and present the abundance ratio, [CH]/[CH$_2$], leaving the user to adjust the required numbers by their preferred value for $R_{\rm IR}$.

%||||||||||||||||||||||||||||||||||||||||||||||||||||||||||||||||||||||||||||||||||||||||||||||||||||||||||||||||||||||||||||||||||||||||||||||||||||||||||||||||||
\section{Shape and surfaces}
\label{sect_shape}
%||||||||||||||||||||||||||||||||||||||||||||||||||||||||||||||||||||||||||||||||||||||||||||||||||||||||||||||||||||||||||||||||||||||||||||||||||||||||||||||||||

The infrared bands observed to date in proto-planetary discs that are attributed to nano-diamond are, primarily, the $3.43$ and $3.53\,\mu$m CH$_n$ stretching bands, which are characteristic of diamond \{100\} and \{111\} surfaces or facets passivated by H atoms in CH$_2$ and CH groups, respectively. Given this facet-specificity we clearly must consider all of the possible and likely euhedral forms that nano-diamonds are known to exhibit in the laboratory along with their respective surface CH$_n$ functionalisation. 

From images of synthetic nano-diamonds and CVD diamond coatings it appears that a wide range of particle shapes is possible, including: truncated octahedral, cuboctahedral, tetrahedral, and cubic particles. For completeness we therefore consider all possible euhedral forms, from tetrahedral to cubic and all intervening polyhedra, including: regular tetrahedra, octahedra, and cubes and their regular and semi-regular truncated forms. For brevity in the following, we define the following designators for the various forms: T, tT, O, tO, cO, tC, and C, for tetrahedra, truncated tetrahedra, octahedra, truncated octahedra, cuboctahedra, truncated cubes, and cubes, respectively (see Fig.~\ref{fig_shapes}). The edges of regular particles are all of equal length, while semi-regular truncated polyhedra have truncated vertex facets of variable edge length, that is all of the edges bordering the triangular and square truncation faces are of the same length and can differ in length from the other edges. In truncated polyhedra the hexagonal and octagonal faces are transposed into six- and eight-sided faces with alternating edges of differing lengths. The regular polyhedral forms, with all edges of equal length, are shown in Fig.~\ref{fig_shapes} along with their fundamental characteristics. In addition to these forms, we also considered `spherical' nano-diamonds.

% TABLE
\begin{table*}
\caption{Polyhedral nano-diamond particle vertex (V), edge (E) and face (F) properties.}
\begin{center}
\begin{tabular}{cclclclcc}
     &    &    &    &    &    &    &    &    \\[-0.35cm]
\hline
\hline
     &    &    &    &    &    &    &    &   \\[-0.3cm] 
      shape   &  V & vertices  &  E & edges  &  F & faces  &   $f_{\rm s}$\{111\}  & $r_{\rm eff} / a_{\rm nd}$  \\[0.05cm]
\hline
     &    &    &    &    &    &    &    &    \\[-0.25cm]
   T  & 4 &  4 isolated CH  & 6 &  \ \ $6 \, \times$\{111\}/\{111\}, CH$_2$  &  4  &  $4 \, \times$\{111\} triangular, \ CH &   1  & 2.013 \\[0.2cm]
   tT  & 12 &  $\equiv$ edge CH$_2$  & 18  & \ \ $6 \, \times$\{111\}/\{111\}, CH$_2$  &  8  &  $4 \, \times$\{111\} hexagonal, CH   &  1  & 1.356 \\
        &   &   &  &    $12 \, \times$\{111\}/\{111\}, CH \ \ on faces &    &  $4 \, \times$\{111\} triangular, \ CH &    \\[0.2cm]
   O  & 6 &  6 isolated CH$_2$  & 12 &  $12 \, \times$\{111\}/\{111\}, CH \ \ on faces & 8 &    $8 \, \times$\{111\} triangular, \ CH  &  1  & 1.465  \\[0.2cm]
   tO  & 24 &  $\equiv$ face CH  & 36 & $24 \, \times$\{111\}/\{100\}, CH$_2$ on faces  &  14  &  $6 \, \times$\{100\} square, \ \ \ \ \ \ CH$_2$  &    &  1.135  \\
        &   &   &  &    $12 \, \times$\{111\}/\{111\}, CH \, on faces  &    &  $8 \, \times$\{111\} hexagonal, CH  & 0.776 &   \\[0.2cm]       
   cO  & 12 &  $\equiv$ face CH$_2$  & 24 & $24 \, \times$\{111\}/\{100\}, CH$_2$ on faces  &  14  &    $6 \, \times$\{100\} square, \ \ \ \ \ \ CH$_2$   &    &  1.211  \\
        &   & &      &  \ \ \ \ \ \ (all cO edges are equivalent)  &   &  $8 \, \times$\{111\} triangular, \ CH
          &  0.366 &  \\[0.2cm]              
   tC  & 24 &  $\equiv$ face CH$_2$  & 36 &    $24 \, \times$\{111\}/\{100\}, CH$_2$ on faces  & 14 &  $6 \, \times$\{100\} octagonal, \ CH$_2$  &      &  1.396   \\
      &      &  &    &   $12 \, \times$\{100\}/\{100\}, CH$_2$ on faces &  &   
      $8 \, \times$\{111\} triangular, \ CH  &  0.107        
      \\[0.2cm]    
      
   C  & 8 &  4 CH$_2$ + 4 CH$_3$  &  12  & $12 \, \times$\{100\}/\{100\}, CH$_2$  on faces &  6  &  $6 \, \times$\{100\} square, \ \ \ \ \ \ CH$_2$ &  0  &  1.201   \\ 
     &    &    &    &    &    &    &     &    \\[-0.35cm]
\hline
\end{tabular}
\tablefoot{Only \{111\} and \{100\} facets are considered in the modelling where $f_{\rm s\{111\}}$ is the fraction of the particle surface in  \{111\} facets and therefore $f_{\rm s\{100\}} = (1 - f_{\rm s\{111\}})$. The considered regular polyhedral shapes are tetrahedra (T), truncated tetrahedra (tT), octahedra (O), truncated octahedra (tO), cuboctahedra (cO), truncated cubes (tC), and cubes (C). For  vertices `$\equiv$ edge(face) CH$_n$' indicates that their CH$_n$ groups can be considered as forming an integral part of the adjacent edges (faces). The last column gives the ratio of the  radius of the circumscribed sphere, $r_{\rm eff}$, with the same volume as a sphere, $a_{\rm nd}$ (see text for details).}
\end{center}
\label{table_struct_summary}
\end{table*}

% TABLE
\begin{table}
\caption{Nano-diamond polyhedra vertex (V), edge (E), and face (F) heirarchies as a function of CH$_n$ group composition and facet geometry.}
\begin{center}
\begin{tabular}{cllllc}
     &    &    &    &    &    \\[-0.35cm]
\hline
\hline
     &    &    &    &    &    \\[-0.3cm] 
      shape   &  CH$_n$  &  V  &  E  &  F  & [CH]/[CH$_2$]  \\[0.05cm]
\hline
     &    &    &    &    &    \\[-0.25cm] 
     
T   &  CH$_2$  &  --   &  {\bf 6}  &  --  &  $\propto l$  \\
     &  CH          &  {\bf 4}  &  --  &  {\bf 4}$\triangle$  &  \\[0.2cm]

     &    &    &    &    &    \\[-0.3cm] 
tT  &  CH$_2$  &  12$\rightarrow$   &  {\bf 6}  &  --  &  $\propto l$  \\
     &  CH          &  --  &  12$\rightarrow$  &  {\bf 4}$\triangle${\bf 4}$\varhexagon$  &  \\[0.2cm]

     &    &    &    &    &    \\[-0.3cm] 
O  &  CH$_2$  &  {\bf 6}   &  --  &  --  &  $\propto l^2$  \\
     &  CH          &  --  &  12$\rightarrow$  &  {\bf 8}$\triangle$  &  \\[0.2cm]

     &    &    &    &    &    \\[-0.3cm] 
tO  &  CH$_2$  &   --  &  24$\rightarrow$  & {\bf 6}$\square$  &  $2 \surd 3$  \\
     &  CH          &  24$\rightarrow$   &  12$\rightarrow$  &  {\bf 8}$\varhexagon$  &  \\[0.2cm]

     &    &    &    &    &    \\[-0.3cm] 
cO  &  CH$_2$  &   12$\rightarrow$  &  24$\rightarrow$  & {\bf 6}$\square$  &  $\frac{1}{3} \surd3$  \\
     &  CH          &  --   &  --  &  {\bf 8}$\triangle$  &  \\[0.2cm]

     &    &    &    &    &    \\[-0.3cm] 
tC  &  CH$_2$  &   24$\rightarrow$  &  36$\rightarrow$  & {\bf 6}$\octagon$  &  $\frac{\surd 3}{6(\surd 2 + 1)}$  \\
     &  CH          &  --   &  --  &  {\bf 8}$\triangle$  &  \\[0.2cm]

     &    &    &    &    &    \\[-0.3cm] 
C  &  CH$_{2,3}$  &   8$\rightarrow$  &  12$\rightarrow$  & {\bf 6}$\square$  &  0  \\
     &  CH          &  --   &  --  &  --  &  \\[0.2cm]
\hline
\end{tabular}
\tablefoot{Rightward arrows indicate that a given quantity is subsumed into the quantity to its right, i.e. V$\rightarrow$E, E$\rightarrow$F  or V$\rightarrow$E$\rightarrow$F and therefore that only the boldface quantities need to be included in order to avoid double counting. The [CH]/[CH$_2$] ratio behaviours are indicated in the right hand column.}
\end{center}
\label{table_struct_heirarchy}
\end{table}

Based on the studies of \cite{Barnard:2005dt} and \cite{2007ApJ...661..919P}, but primarily following the modelling of regular and semi-regular polyhedra and their truncated forms presented here, we make the following observations: 
%\begin{itemize}
%\item 
vertices are terminated with CH (T) or CH$_2$ (O) groups, 
%\item 
the edges between \{111\} facets are CH$_2$ (T) or CH (O), 
%\item 
all three- and six-sided faces are \{111\} facets, 
%\item 
all four- and eight-sided faces are \{100\} facets,    
%\item 
\{111\} facets exhibit only coherently-directed CH bonds and 
%\item 
\{100\} facets are CH$_2$-covered. 
%\end{itemize}
In the non-specified cases vertex CH$_n$ structures form part of the adjacent edges and edge CH$_n$ structures form part of the adjacent faces.\footnote{This is strictly only valid for infinitely small vertices and infinitely thin edges but is a good approximation for particle sizes $\ggg$ d$_{\rm C-C}$.} These results are summarised in Table \ref{table_struct_summary}.

For a given polyhedron the fractional surface area in triangular and/or hexagonal \{111\} facets is denoted as $f_{\rm s \{111\}}$ and that in square and/or octagonal \{100\} facets as $f_{\rm s \{100\}}$. The number of surface C and H atoms per particle, $N_{\rm Cs \{111\}}$, $N_{\rm Hs \{111\}}$, $N_{\rm Cs \{100\}}$ and $N_{\rm Hs \{100\}}$ in \{111\} and \{100\} facets respectively, are then:
\begin{equation}
  N_{\rm Cs \{111\}}  =  f_{s \{111\}} \  N_{\rm Cs},   \ \ \ \ \ \ \ \ \ \
  N_{\rm Hs \{111\}}  =  N_{\rm Cs \{111\}} \ f_{\rm H},  
\end{equation}
and 
\begin{equation}
  N_{\rm Cs \{100\}}  =  f_{s \{100\}} \  N_{\rm Cs},   \ \ \ \ \ \ \ \ \ \
  N_{\rm Hs \{100\}}  =  2 \ N_{\rm Cs \{100\}} \ f_{\rm H}, 
\end{equation}
where $f_{\rm H}$ is the degree of surface hydrogenation, with $f_{\rm H} = 1$ indicating a maximally-hydrogenated surface and $f_{\rm H} = 0$ a completely dehydrogenated surface.

%||||||||||||||||||||||||||||||||||||||||||||||||||||||||||||||||||||||||||||||||||||||||||||||||||||||||||||||||||||||||||||||||||||||||||||||||||||||||||||||||||
\section{Regular polyhedral particles}
\label{sect_regulars}
%||||||||||||||||||||||||||||||||||||||||||||||||||||||||||||||||||||||||||||||||||||||||||||||||||||||||||||||||||||||||||||||||||||||||||||||||||||||||||||||||||

Here we consider the case of regular polyhedral particles, and their truncated variants, where all edges are of equal length, $l$. The full expressions for the properties and characteristics of regular polyhedral particles can be found in Appendix~\ref{app_regulars}. We recall that regular triangular and hexagonal \{111\}  facets  have areas of $\surd 3 /4 \, l^2$ and $3 \surd 3 /2 \,l^2$, respectively, and that regular square and octagonal \{100\}facets have areas of $l^2$ and $2 (\surd2 +1)\, l^2$, respectively.\footnote{In this and the following section we implicitly assumed that the polyhedral edge lengths, $l$, are in units of the C--C bond length d$_{\rm C-C}$, even though they are not actually discretised in this way.}

For all of the polyhedral nano-diamond particle shapes considered here the total number of CH groups is obtained by summing over their number on vertices  (V$_{\rm CH}$), edges (E$_{\rm CH}$), and triangular and hexagonal \{111\} facets (F$_{\{111\}} (\triangle \varhexagon)$) and the number of CH$_2$ groups by summing over those on vertices  (V$_{{\rm CH}_2}$), edges (E$_{{\rm CH}_2}$), and square and octagonal \{100\} facets (F$_{\{100\}} (\square {\tiny \octagon})$). The [CH]/[CH$_2$] ratio is then obtained via, 
\begin{equation}
\frac{\rm [CH]}{\rm [CH_2]} 
= \frac{\rm V_{\rm CH} \ + \ E_{\rm CH} \ + \ F_{\{111\}} (\triangle \varhexagon) }
          {\rm V_{{\rm CH}_2} \ + \ E_{{\rm CH}_2} \ + \ F_{\{100\}} (\square {\tiny \octagon}) }.   
\end{equation}
From the observations drawn at the end of the previous section, it follows that where vertices (edges) are of the same CH$_n$ type as an adjacent edge (facet) then they are subsumed into that edge (facet). 

To determine the various polyhedral behaviours the reader should make extensive reference to Fig. \ref{fig_shapes} and Tables \ref{table_struct_summary} and \ref{table_struct_heirarchy} to derive the quantities V$_{\rm CH_n}$, E$_{\rm CH_n}$, F$_{\{111\}}$ and F$_{\{100\}}$, and this should be taken as implicit in the following subsections and is therefore not be repeated at every instance. In Table \ref{table_struct_summary} polyhedral vertices indicated as `$\equiv$ edge CH$_n$' or `$\equiv$ face CH$_n$' do not count as independent CH$_n$ groups because they are of the same type as an adjacent edge or face and therefore already counted as a part of that edge or face, which is to their right in the table. Similarly, edges labelled as `on faces' are already counted as part of the adjacent face. This is generally manifest where horizontally adjacent V and E or E and F entries in the table have identical CH$_n$ group types. These dependencies are also shown more explicitly and concisely as arrows in Table~\ref{table_struct_heirarchy}. In the calculation of the polyhedral [CH]/[CH$_2$] ratios only the boldfaced quantities in Table~\ref{table_struct_heirarchy} were included in order to avoid double counting CH$_n$ groups. 

The number of vertices for a given polyhedron is fixed, the total edge length scales with $l$ and the facet area scales with $l^2$. Thus, for large polyhedra, large with respect to the C--C bond length (i.e. $l \gg$\,d$_{\rm C-C}$), it is the nature of the edges and facets that determine the [CH]/[CH$_2$] ratio and that in very large polyhedra ($l \ggg$\,d$_{\rm C-C}$) the size dependence of this ratio is principally driven by the $l^2$ dependence of the relevant facet areas.

%||||||||||||||||||||||||||||||||||||||||||||||||||||||||||||||||||||||||||||||||||||||||||||||||||||||||||||||||||||||||||||||||||||||||||||||||||||||||||||||||||
\subsection{ Regular tetrahedral (T) particles}
%||||||||||||||||||||||||||||||||||||||||||||||||||||||||||||||||||||||||||||||||||||||||||||||||||||||||||||||||||||||||||||||||||||||||||||||||||||||||||||||||||

We first consider regular tetrahedral particles, the simplest regular polyhedra, with four CH vertices, V$_{\rm CH} =4$, six  CH$_2$ edges, E$_{\rm CH_2} = 6\,l$, and a total surface area of F$_{\{111\}}(\triangle) = 4 \times \surd 3 /4 \, l^2$.  In this case V$_{\rm CH_2}$, E$_{\rm CH}$ and F$_{\{100\}}$ are all zero and so we have  
\begin{equation}
\frac{\rm [CH]}{\rm [CH_2]} 
= \frac{\rm V_{\rm CH} \ + \ F_{\{111\}} (\triangle) }{\rm E_{\rm CH_2}}
= \frac{4 \ + \ \surd 3 \, l^2}{6\,l},   
\end{equation}
which reduces to $\approx (\surd 3 / 6) \, l$ for large $l$. Thus, for regular tetrahedral nano-diamonds [CH]/[CH$_2$] increases with particle size as shown in Fig.~\ref{fig_3D_geo_CHs}.

%||||||||||||||||||||||||||||||||||||||||||||||||||||||||||||||||||||||||||||||||||||||||||||||||||||||||||||||||||||||||||||||||||||||||||||||||||||||||||||||||||
\subsection{Regular truncated tetrahedral (tT) particles}
%||||||||||||||||||||||||||||||||||||||||||||||||||||||||||||||||||||||||||||||||||||||||||||||||||||||||||||||||||||||||||||||||||||||||||||||||||||||||||||||||||

Regular truncated tetrahedral particles are tetrahedra with the four vertices truncated into equilateral triangular faces. They have twelve CH$_2$ vertices counted within the CH$_2$ edges,  and therefore V$_{\rm CH_2} =0$, eighteen edges of which the 12 CH edges are counted within the adjacent \{111\} facets, yielding E$_{\rm CH} = 0$ and E$_{\rm CH_2} = 6\,l$, and a surface comprised of four triangular \{111\} facets and four hexagonal \{111\} facets. Thus, F$_{\{111\}}(\triangle \varhexagon) = 4 \times \surd 3 /4 \, l^2 + 4 \times 3 \surd 3 /2 \,l^2
= 7 \surd 3 \, l^2$, and with V$_{\rm CH}$ and F$_{\{100\}}$ equal to zero, we have
\begin{equation}
\frac{\rm [CH]}{\rm [CH_2]} 
= \frac{\rm F_{\{111\}} (\triangle \varhexagon) }
          {\rm E_{CH_2} }
= \frac{7 \surd 3}{6} \ l. 
\end{equation}
Thus,  as indicated in Fig.~\ref{fig_3D_geo_CHs}, the [CH]/[CH$_2$] ratio for regular truncated tetrahedral nano-diamonds increases with $l$ (increasing size) and with the same slope as for T particles.

%||||||||||||||||||||||||||||||||||||||||||||||||||||||||||||||||||||||||||||||||||||||||||||||||||||||||||||||||||||||||||||||||||||||||||||||||||||||||||||||||||
\subsection{Regular octahedral (O) particles}
%||||||||||||||||||||||||||||||||||||||||||||||||||||||||||||||||||||||||||||||||||||||||||||||||||||||||||||||||||||||||||||||||||||||||||||||||||||||||||||||||||

Turning to regular octahedral particles with six CH$_2$ vertices, V$_{\rm CH_2} =6$, 12 CH edges all of which are counted within the adjacent \{111\} facets ($\therefore$ E$_{\rm CH} = 0$) and eight triangular \{111\} facets, F$_{\{111\}}(\triangle) = 8 \times \surd 3 /4 \, l^2 =  2 \surd 3 \, l^2$. With V$_{\rm CH}$, E$_{\rm CH_2}$, and F$_{\{100\}}$ all zero, we have
\begin{equation}
\frac{\rm [CH]}{\rm [CH_2]} 
= \frac{\rm F_{\{111\}} (\triangle) }
          {\rm V_{CH_2}}
= \frac{\surd 3}{3}  \ l^2. 
\end{equation}
As for tetrahedral nano-diamond polyhedra, the [CH]/[CH$_2$] ratio increases with size but in this case with a steeper $l^2$ dependence (Fig.~\ref{fig_3D_geo_CHs}).

%||||||||||||||||||||||||||||||||||||||||||||||||||||||||||||||||||||||||||||||||||||||||||||||||||||||||||||||||||||||||||||||||||||||||||||||||||||||||||||||||||
\subsection{Regular truncated octahedral (tO) particles}
%||||||||||||||||||||||||||||||||||||||||||||||||||||||||||||||||||||||||||||||||||||||||||||||||||||||||||||||||||||||||||||||||||||||||||||||||||||||||||||||||||

A common nano-diamond particle shape is the regular truncated octahedron \cite[e.g.][]{Barnard:2005dt}, an octahedron with its six vertices truncated into square \{100\} facets. This polyhedron has 24 CH vertices and 36 edges (24 CH$_2$ and 12 CH) all of which can be counted within the adjacent edges and facets, respectively ($\therefore$ V$_{\rm CH}$, V$_{\rm CH_2}$, E$_{\rm CH}$, E$_{\rm CH_2}$ are all zero). The surface is comprised of eight hexagonal \{111\} facets and six square \{100\} facets, that is F$_{\{111\}}(\varhexagon) = 8 \times 3 \surd 3 /2 \,l^2 = 12 \surd 3 \, l^2$ and F$_{\{100\}}(\square) = 6 \, l^2$, and 
\begin{equation}
\frac{\rm [CH]}{\rm [CH_2]} 
= \frac{\rm F_{\{111\}} (\varhexagon) }
          {\rm F_{\{100\}} (\square) }
= 2 \surd 3,  
\end{equation}
for regular truncated octahedral nano-diamonds the [CH]/[CH$_2$] ratio is therefore independent of size (Fig.~\ref{fig_3D_geo_CHs}), in contrast to all of the previously considered polyhedral forms.

%||||||||||||||||||||||||||||||||||||||||||||||||||||||||||||||||||||||||||||||||||||||||||||||||||||||||||||||||||||||||||||||||||||||||||||||||||||||||||||||||||
\subsection{Regular cuboctahedral (cO) particles}
%||||||||||||||||||||||||||||||||||||||||||||||||||||||||||||||||||||||||||||||||||||||||||||||||||||||||||||||||||||||||||||||||||||||||||||||||||||||||||||||||||

Nano-diamonds also commonly form cuboctahedra, which are a limiting form of tO particles where  the hexagonal faces are reduced to triangular faces. These interesting polyhedra have 12 CH$_2$ vertices and 24 CH$_2$ edges, all of which can be counted within the adjacent square \{100\} facets (i.e. V$_{\rm CH_2}$ and E$_{\rm CH_2} = 0$). Their surfaces exhibit eight triangular \{111\} facets and six square \{100\} facets, F$_{\{111\}}(\triangle) = 8 \times \surd 3 /4 \, l^2 = 2 \surd 3 \, l^2$, F$_{\{100\}}(\square) = 6 \, l^2$, and with V$_{\rm CH}$ and E$_{\rm CH}$ each zero, we have
\begin{equation}
\frac{\rm [CH]}{\rm [CH_2]} 
= \frac{\rm F_{\{111\}} (\triangle) }
          {\rm F_{\{100\}} (\square) }
= \frac{\surd 3}{3},  
\end{equation}
and for this common nano-diamond polyhedral shape the [CH]/[CH$_2$] ratio is also independent of size, as per the closely related truncated octahedral polyhedron (Fig.~\ref{fig_3D_geo_CHs}).

%||||||||||||||||||||||||||||||||||||||||||||||||||||||||||||||||||||||||||||||||||||||||||||||||||||||||||||||||||||||||||||||||||||||||||||||||||||||||||||||||||
\subsection{Regular truncated cube (tC) particles}
%||||||||||||||||||||||||||||||||||||||||||||||||||||||||||||||||||||||||||||||||||||||||||||||||||||||||||||||||||||||||||||||||||||||||||||||||||||||||||||||||||

\begin{figure}[h]
\centering
\includegraphics[width=9.5cm]{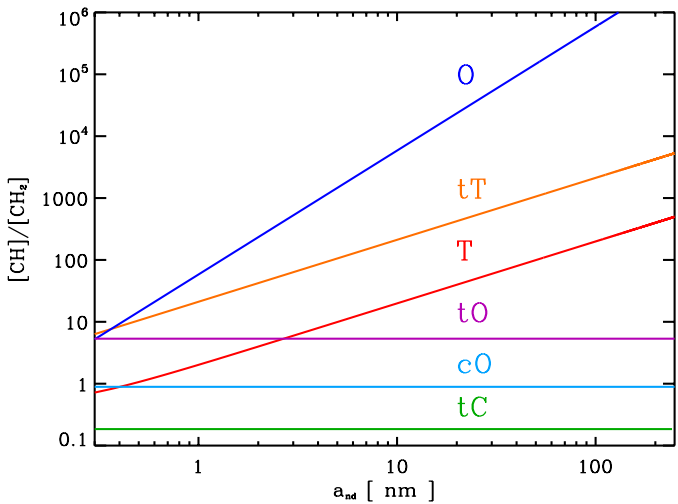}
\caption{[CH]/[CH$_2$] ratios for regular polyhedra as a function of the effective nano-diamond radius, $a_{\rm nd}$ (nm), for T (red), tT (orange), O (blue), tO (purple), cO (cobalt), and tC (green) forms. For the cubic form (C) this ratio is zero.}
\label{fig_3D_geo_CHs}
\centering
\end{figure}

For completeness we finally consider the regular truncated cube, a cube with its eight vertices truncated into triangular facets. This polyhedron has 24 CH$_2 $ vertices and 36 CH$_2 $ edges,  all of which can be counted within the adjacent octagonal \{100\} facets (i.e. V$_{\rm CH_2}$ and E$_{\rm CH_2} = 0$). Their surfaces consist of eight triangular \{111\} facets and six octagonal \{100\} facets), F$_{\{111\}}(\triangle) = 8 \times \surd 3 /4 \, l^2 = 2 \surd 3 \, l^2$, F$_{\{100\}}({\tiny \octagon}) = 6 \times 2 (\surd2 +1)\, l^2 = 12 ( \surd 2 + 1) \, l^2$, and with V$_{\rm CH}$ and E$_{\rm CH}$ both zero, we have
\begin{equation}
\frac{\rm [CH]}{\rm [CH_2]} 
= \frac{\rm F_{\{111\}} (\triangle) }
          {\rm F_{\{100\}} ({\tiny \octagon}) }
= \frac{\surd 3}{6 ( \surd 2 + 1)},  
\end{equation}
and for truncated cubic nano-diamonda the [CH]/[CH$_2$] ratio is also independent of size, as per tO and cO polyhedral particles (Fig.~\ref{fig_3D_geo_CHs}).

%||||||||||||||||||||||||||||||||||||||||||||||||||||||||||||||||||||||||||||||||||||||||||||||||||||||||||||||||||||||||||||||||||||||||||||||||||||||||||||||||||
\subsection{Regular cube (C) particles}
%||||||||||||||||||||||||||||||||||||||||||||||||||||||||||||||||||||||||||||||||||||||||||||||||||||||||||||||||||||||||||||||||||||||||||||||||||||||||||||||||||

Another typical diamond shape is the regular cube, with 8 vertices, four of which are CH$_2$ and four CH$_3$, 12 CH$_2$ edges and 6 CH$_2$ facets, that is 
\begin{equation}
\frac{\rm [CH]}{\rm [CH_2]} 
= \frac{\rm 0}
          {\rm V_{\rm CH_3} \ + \ V_{\rm CH_2} \ + \ E_{\rm CH_2} \ + \ F_{\{100\}} (\square) }. 
\end{equation}
Thus, given that there are no CH groups on perfectly cubic nano-diamonds their [CH]/[CH$_2$] ratios are zero.

% FIGURE 

%||||||||||||||||||||||||||||||||||||||||||||||||||||||||||||||||||||||||||||||||||||||||||||||||||||||||||||||||||||||||||||||||||||||||||||||||||||||||||||||||||
\subsection{The spatial properties of regular polyhedra}
\label{sect_poly_props}
%||||||||||||||||||||||||||||||||||||||||||||||||||||||||||||||||||||||||||||||||||||||||||||||||||||||||||||||||||||||||||||||||||||||||||||||||||||||||||||||||||

In the preceding sub-sections we briefly derived the expected [CH]/[CH$_2$] ratios for regular polyhedral nano-diamonds. The results are graphically summarised in  Fig. \ref{fig_3D_geo_CHs} where the ratio is plotted as a function of the effective radius $a_{\rm nd}$ for T (red), tT (orange), O (blue), tO (purple), cO (cobalt), and tC (green) particle forms (see Appendix~\ref{app_regulars} for the derivation of the effective nano-diamond radii, $a_{\rm nd}$). In Fig. \ref{fig_3D_geo_CHs} we take into account the discrete nature of nano-diamonds and calculate the [CH] and [CH$_2$] surface abundances using the parameters $D_{\rm CH_2}$, $A_{\rm CH_2}$, and $A_{\rm CH}$ defined in Section \ref{sect_density}. This figure indicates that [CH]/[CH$_2$] for the closely related tO, cO, and tC polyhedra\footnote{N.B., tO, cO, and tC polyhedra are transformable tO\,$\leftrightarrow$\,cO\,$\leftrightarrow$\,tC through variations in the relative area of the square faces, which become octagonal in the cO\,$\leftrightarrow$\,tC transition.} are independent of particle size. However, for the T, tT, and O polyhedra the [CH]/[CH$_2$] ratio, and hence the $3.53\,\mu$m(CH)/$3.43\,\mu$m(CH$_2$) IR band ratio, does depend on size, and more strongly so for O than the T and tT polyhedra, and can span orders of magnitude. 

The primary motivation for this study was to better understand the $3.53\,\mu$m(CH)/$3.43\,\mu$m(CH$_2$) IR band intensity ratio, through a study of the surface [CH]/[CH$_2$] abundance ratio on nano-diamond \{111\} and \{100\} facets. It is therefore interesting to note the wide variation in the CH-covered, \{111\} facet surface fraction, $f_{s\{111\}}$, which while not strongly dependent on polyhedral form, is independent of size and spans about an order of magnitude ($0.107-1.000$, see Fig.~\ref{fig_shapes}).  

We conclude that, if the observed nano-diamond shapes are polyhedral, but the distribution of polyhedral forms is not well determined, deriving the intrinsic [CH]/[CH$_2$] ratio can be uncertain by orders of magnitude. However, if the [CH]/[CH$_2$] ratio and size are known then it would be possible to usefully constrain the form or, conversely, to constrain the size if the [CH]/[CH$_2$] ratio and form are known. Ideally, and in order to maximise the amount of information from nano-diamond IR spectra, we therefore need to know the particle size and shape distributions. Unfortunately, astronomical observations cannot access either of these critical quantities and it is therefore going to be difficult to lift this degeneracy other than through detailed nano-particle modelling. Additionally, current presolar nano-diamond studies, while constraining particle sizes, do not provide sufficient information on the particle shapes.\footnote{For even this scant information to be of use we would have to assume that there is an as yet unsubstantiated direct relationship between the presolar nano-diamonds and those responsible for the observed circumstellar $3.53\,\mu$m(CH) and $3.43\,\mu$m(CH$_2$) IR bands.}

In this modelling scheme the forms are contiguous, meaning that the polyhedral dimensions can be scaled arbitrarily, that is the equations of these structures are not discretised. However, the numerical results presented in Fig.~\ref{fig_3D_geo_CHs} do assume the appropriate C$-$C bond length, d$_{\rm C-C}$, and CH$_n$ group areas, A$_{\rm CH}$, and A$_{\rm CH_2}$. A more detailed treatment of the polyhedral particle geometries can be found in Appendices \ref{app_regulars} and \ref{app_irregulars}. In our later diamond-bonded network modelling (Section~\ref{sect_networks}) all of the calculations are necessarily discretised.

% FIGURE 
\begin{figure}[h]
\centering
\includegraphics[width=9cm]{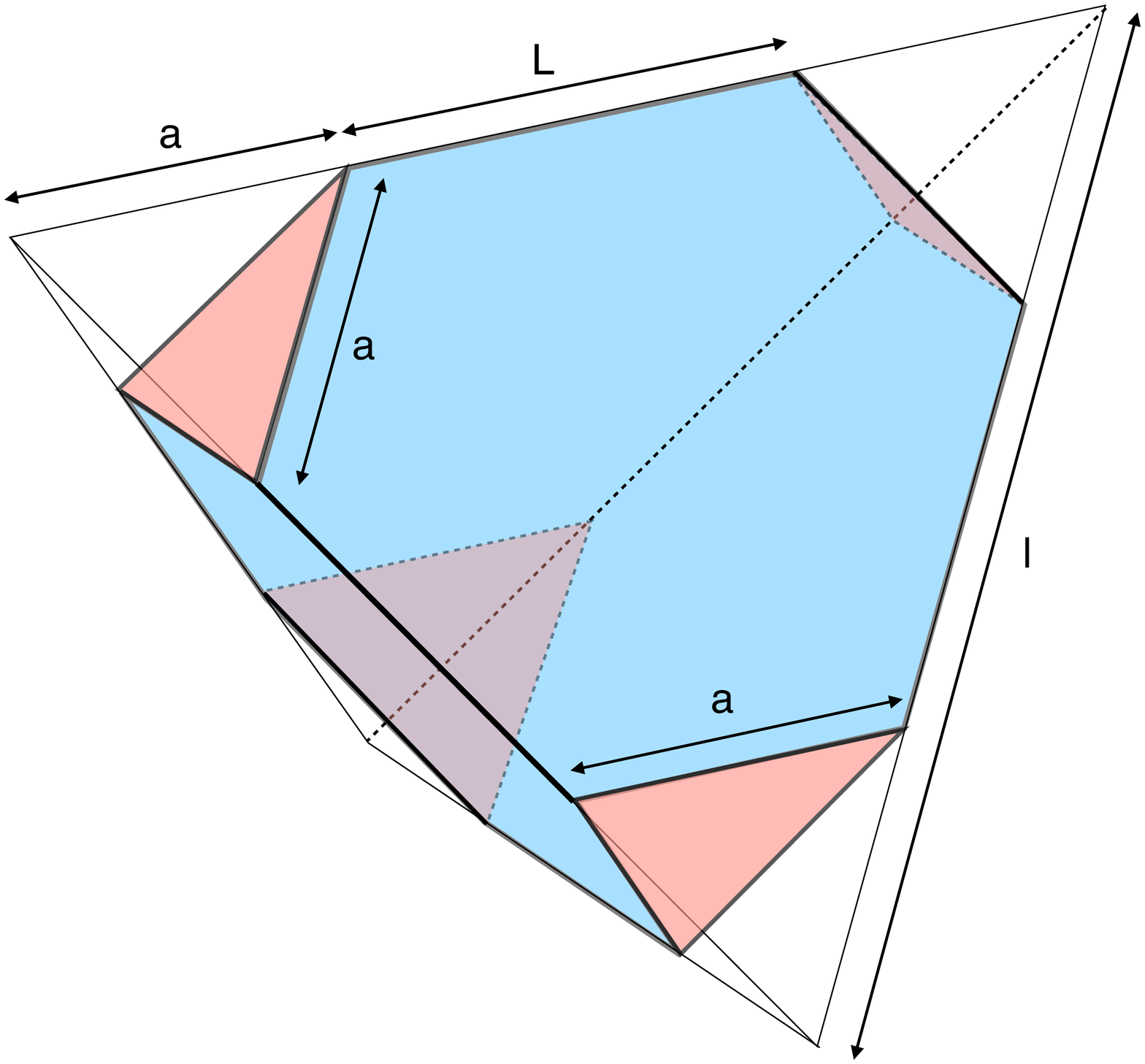}
\caption{Semi-regular truncated tetrahedron.}
\label{fig_trunc_tetr}
\centering
\end{figure}

% FIGURE 
\begin{figure}[h]
\centering
\includegraphics[width=9cm]{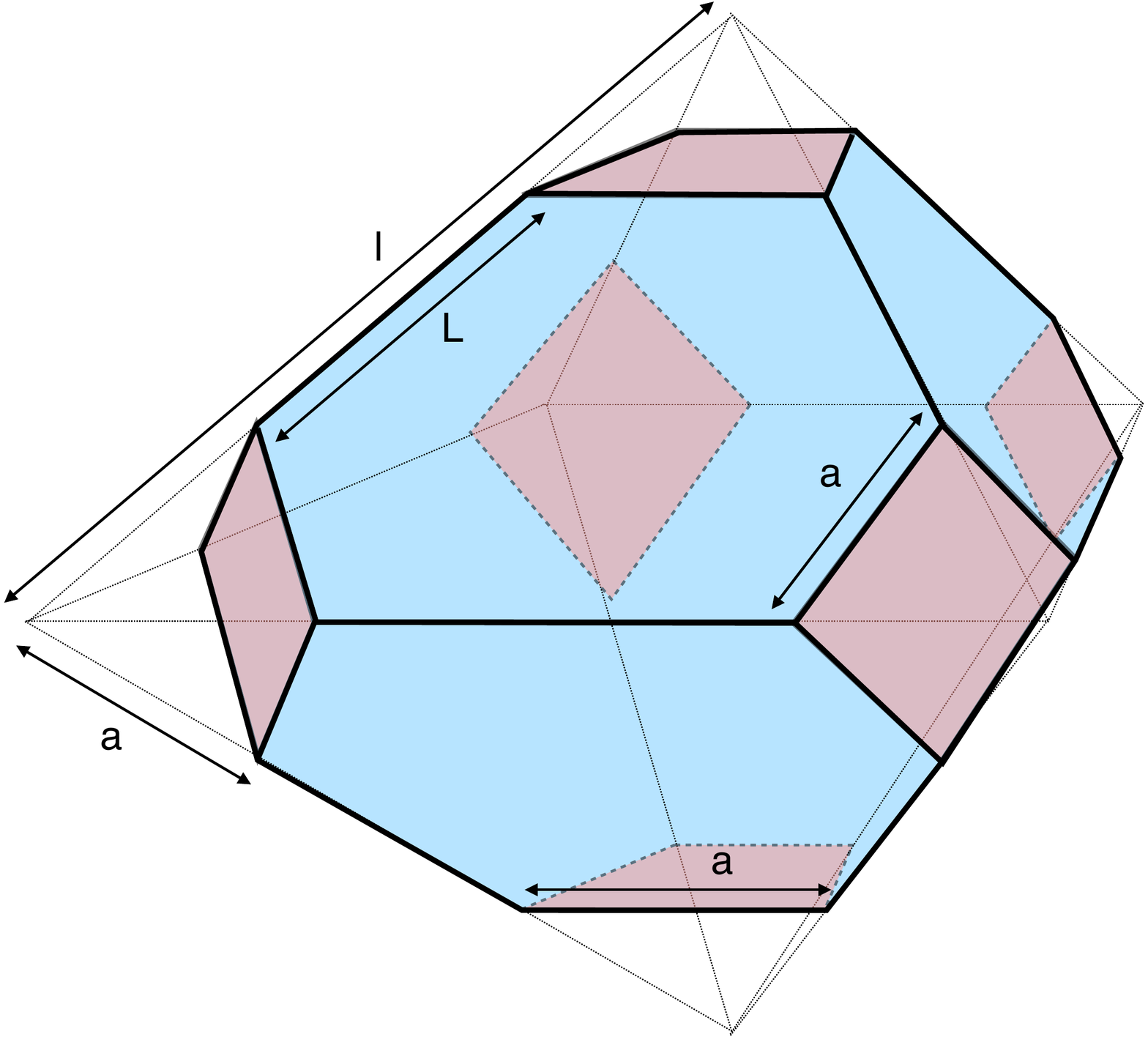}
\caption{Semi-regular truncated octahedron.}
\label{fig_trunc_octa}
\centering
\end{figure}

% FIGURE 
\begin{figure}[h]
\centering
\includegraphics[width=9cm]{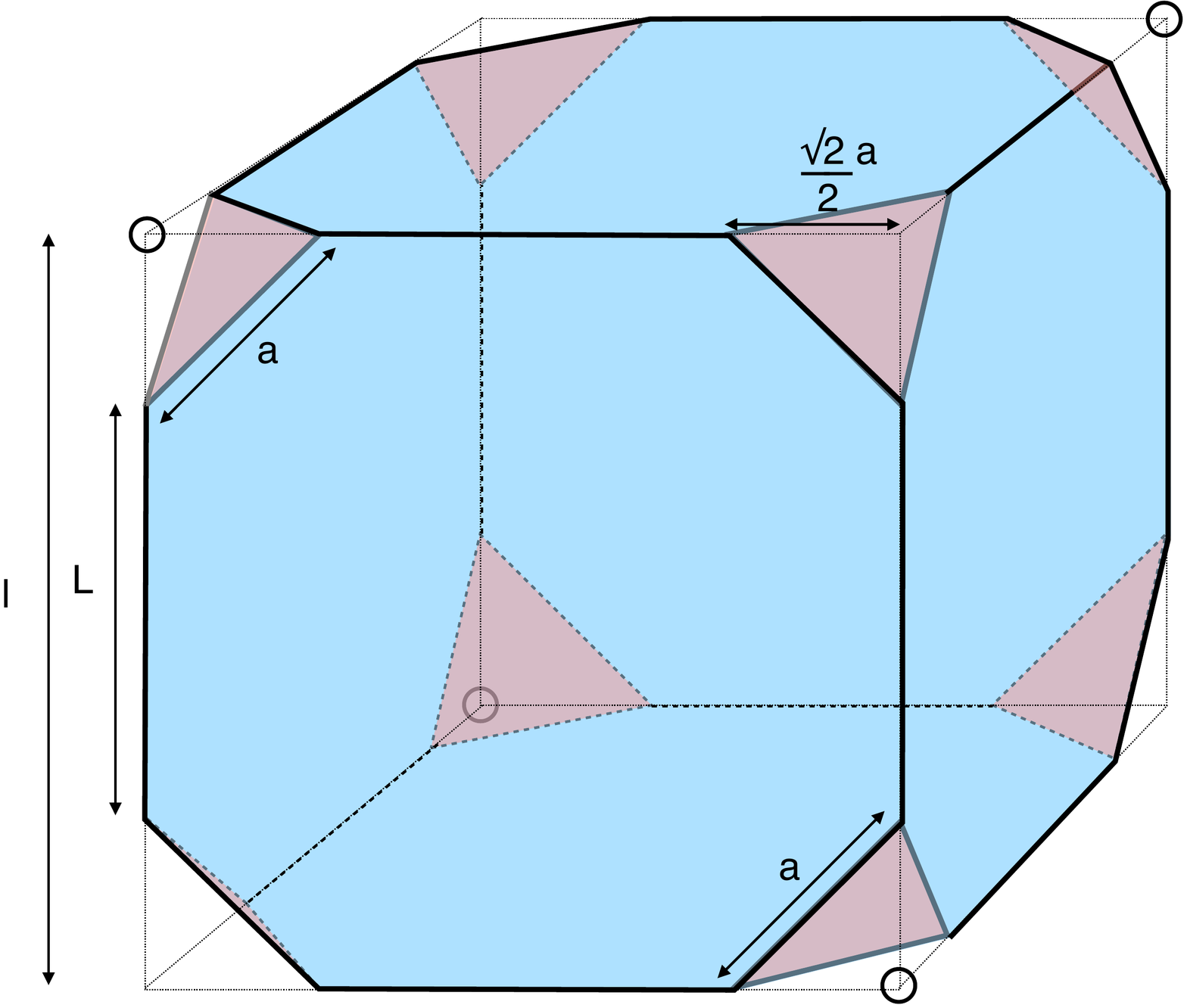}
\caption{Semi-regular truncated cube.}
\label{fig_trunc_cube}
\centering
\end{figure}

%||||||||||||||||||||||||||||||||||||||||||||||||||||||||||||||||||||||||||||||||||||||||||||||||||||||||||||||||||||||||||||||||||||||||||||||||||||||||||||||||||
\section{Semi-regular polyhedral particles}
\label{sect_irregulars}
%||||||||||||||||||||||||||||||||||||||||||||||||||||||||||||||||||||||||||||||||||||||||||||||||||||||||||||||||||||||||||||||||||||||||||||||||||||||||||||||||||

Wishing to generalise the method, but probably also complicate matters a little further, we now turn our attention to semi-regularly truncated polyhedra, which are semi-regular in the sense that, with respect to the regular parent polyhedron (i.e. tetrahedron, octahedron, and cube), the truncated facets are of arbitrary size but remain parallel to those of the regular truncated form of the parent polyhedron. The necessarily more cumbersome expressions are available in full in Appendix~\ref{app_irregulars} for the more courageous of readers. While the truncated facets remain parallel to those of the parent polyhedron their edges are of arbitrary length $a$, which implies that the remnant edge, $L$, of the regular polyhedron parent is reduced from $l$ to $(l-2a)$ as illustrated in Figs,~\ref{fig_trunc_tetr} to \ref{fig_trunc_cube}, that is $L = (l-2a)$. In semi-regular polyhedra the triangular and square faces retain their regular form, that is all edges are of the length $a$, and therefore have areas of $\surd 3 /4 \, a^2$ and $a^2$, respectively. However, the hexagonal and octagonal faces are no longer regular but have alternating edge lengths of $a$ and $L$. The area of the six-sided faces is now $\surd 3/4 \, ( l^2 - 3a^2)$, which gives the area of a regular hexagon area when $l = 3a$, and the area of the eight-sided faces is now $( l^2 - a^2 )$, which corresponds to a regular octagon when $l = (\surd 2 + 1) \, a$. Although the expressions for these polyhedra cannot be reduced to simple functions of $r$, the radius of the sphere that circumscribes the particle and includes all its vertices, the particles can still be circumscribed by a sphere that encompasses all vertices. 

The mathematical expressions for the semi-regular truncated polyhedra (stT, stO amd stC) given in this section (see also Appendix~\ref{app_irregulars}) are vaild for $0 \leqslant a \leqslant l/2$,  that is the parent polyhedron edge is, at most, bisectable, which leads to the following critical conditions: \\
i) $\ \ a = 0 \ \ \ \equiv L= l \, \rightarrow$ a regular parent polyhedron, \\
ii) $\ a= l/3       \equiv L = a   \rightarrow$ a regular truncated parent polyhedron, \\
iii) $a = l/2       \equiv L = 0   \rightarrow$ a different polyhedron. \\
In the $a = l/2$ case the truncated tetrahedron solution is an octahedron, and for the truncated octahedron and truncated cube the solution is a cuboctahedron.\footnote{Hence, cO particles are not considered here because they are the stO and stC polyhedra formed by bisecting the parent polyhedron edges.} 

For each semi-regular polyhedron type, and in the same way as for the regular polyhedra of Section~\ref{sect_regulars}, we can derive the [CH]/[CH$_2$] ratios using the same expressions but substitute the new edge length expressions and the modified surface areas for the six and eight sided facets. In this case the number and type of vertices are unchanged.

%||||||||||||||||||||||||||||||||||||||||||||||||||||||||||||||||||||||||||||||||||||||||||||||||||||||||||||||||||||||||||||||||||||||||||||||||||||||||||||||||||
\subsection{Semi-regular truncated tetrahedral (stT) particles}
%||||||||||||||||||||||||||||||||||||||||||||||||||||||||||||||||||||||||||||||||||||||||||||||||||||||||||||||||||||||||||||||||||||||||||||||||||||||||||||||||||

These are similar to truncated tetrahedral particles except that the four vertices are now arbitrarily truncated into equal equilateral triangular faces of edge length $a$ (see Fig,~\ref{fig_trunc_tetr}). The total edge length is now $12a + 6(l-2a) = 6\,l$, that is truncation does not change the total edge length compared to the parent tetrahedron.\footnote{This is  because the chamfered vertices are regular tetrahedra.} The [CH]/[CH$_2$] ratio for a generalised truncated tetrahedron is then 
%\[
%\frac{\rm [CH]}{\rm [CH_2]} 
%= \frac{\rm F_{\{111\}} (\triangle\varhexagon) }{\rm E_{\rm CH_2}}
%= \frac{ \surd 3 \, a^2 \ + \ \surd 3 \, ( l^2 - 3a^2)}{6\,l}   
%\]
\begin{equation}\frac{\rm [CH]}{\rm [CH_2]} 
= \frac{\rm F_{\{111\}} (\triangle\varhexagon) }{\rm E_{\rm CH_2}}
= \frac{ \surd 3 \, a^2 + \surd 3 ( l^2 - 3a^2)}{6\,l} 
%\ \ \ \ \ \ \ \ \ \ \ \ = \frac{ \surd 3 ( l^2 \ - \ 2a^2)}{6\,l}
= \frac{ \surd 3 \left[ ( l/a )^2  - 2 \right] }{6\,(l/a^2)}.    
\end{equation}
This reduces to that for a regular tetrahedron, $(\surd 3/6) \, l$, when $a = 0$, if we ignore the change in vertex CH$_n$ composition. From this equation the behaviour of the [CH]/[CH$_2$] ratio is less obvious because of the dependency on both $l$ and $a$ but as Fig.~\ref{fig_ratios_poly} shows an increase in the truncation length leads to an increase in the ratio.

%||||||||||||||||||||||||||||||||||||||||||||||||||||||||||||||||||||||||||||||||||||||||||||||||||||||||||||||||||||||||||||||||||||||||||||||||||||||||||||||||||
\subsection{Semi-regular truncated octahedral (stO) particles}
%||||||||||||||||||||||||||||||||||||||||||||||||||||||||||||||||||||||||||||||||||||||||||||||||||||||||||||||||||||||||||||||||||||||||||||||||||||||||||||||||||

These are octahedral particles with the six vertices arbitrarily truncated into equal area square faces of edge length $a$. The total edge length is now $24a + 12(l-2a) = 12\,l$, that is truncation does not change the total edge length with respect to the parent octahedron\footnote{In this case the chamfered vertices are half of a regular octahedron.} (see Fig. \ref{fig_trunc_octa}). The surface is comprised of eight six sided \{111\} facets and six square \{100\} facets that is F$_{\{111\}}(\varhexagon) = 8 \times \surd 3 /4 \, ( l^2 - 3a^2) = 2 \surd 3 \, ( l^2 - 3a^2)$ and F$_{\{100\}}(\square) = 6 \, a^2$, and 
\begin{equation}
\frac{\rm [CH]}{\rm [CH_2]} 
= \frac{\rm F_{\{111\}} (\varhexagon) }
          {\rm F_{\{100\}} (\square) }
= \frac{2 \surd 3 \, ( l^2 - 3a^2)}{6 \, a^2}
= \frac{\surd 3}{3} \, \left[ \left( \frac{l}{a} \right)^2 - 3\right].  
\end{equation}
This expression reduces to that for a regular truncated octahedron on substituting $l = 3a$ and to a regular cuboctahedron for $l = 2a$. Fig.~\ref{fig_ratios_poly} shows that for semi-regular truncated octahedral nano-diamonds the [CH]/[CH$_2$] ratio increases with the increasing number of carbon atoms but decreases and tends to flatten with increasing truncation (increasing $a$). The latter effect is because the particle shapes are increasingly driven towards the tO and cO forms which exhibit size independent [CH]/[CH$_2$] ratios.

%||||||||||||||||||||||||||||||||||||||||||||||||||||||||||||||||||||||||||||||||||||||||||||||||||||||||||||||||||||||||||||||||||||||||||||||||||||||||||||||||||
\subsection{Semi-regular truncated cubic (stC) particles}
%||||||||||||||||||||||||||||||||||||||||||||||||||||||||||||||||||||||||||||||||||||||||||||||||||||||||||||||||||||||||||||||||||||||||||||||||||||||||||||||||||

\begin{figure}[h]
\centering
\includegraphics[width=9.5cm]{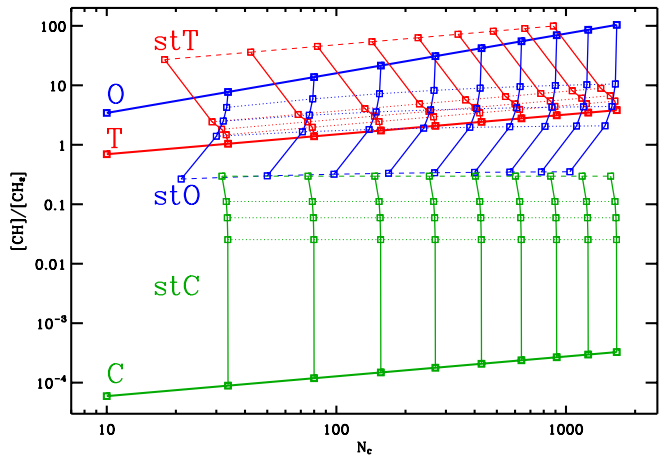}
\caption{[CH]/[CH$_2$] ratios in semi-regular truncated polyhedra, stT (red), stO (blue), and stC (green), as a function of the truncation length ($a = 0$ [thick]$, \frac{1}{6}l, \frac{1}{4}l$, $\frac{1}{3}l$ (dotted), $\frac{1}{2}l$ [dashed]),  and the number of constituent carbon atoms $N_{\rm C}$. The thick lines show the data for the un-truncated polyhedra and the thin dashed lines show the $a = l/2$ limiting cases, except for stT where $a = 0.49 \, l$. The dangling and rising thinner lines show the effects of increasing truncation. For the C polyhedra, and for illustrative purposes only, an arbitrarily low value ($10^4 \, l$) of the [CH]/[CH$_2$] ratio was assumed.} 
\label{fig_ratios_poly}
\centering
\end{figure}

These are cubes with the eight vertices arbitrarily truncated into equilateral triangular faces of edge length $a$. The total edge length is now $24a + 12 [ l-\surd 2 \, a] = [ 12 l + 12 a \,(2-\surd 2) ]$  (see Fig. \ref{fig_trunc_cube}) and therefore differs from that of the parent polyhedron in this case. The polyhedral surfaces consist of eight triangular \{111\} facets and six seight-sided \{100\} facets), F$_{\{111\}}(\triangle) = 8 \times \surd 3 /4 \, a^2 = 2 \surd 3 \, a^2$, F$_{\{100\}}({\tiny \octagon}) = 6 \times ( l^2 - a^2 )$ and 
\begin{equation}
\frac{\rm [CH]}{\rm [CH_2]} 
= \frac{\rm F_{\{111\}} (\triangle) }
          {\rm F_{\{100\}} ({\tiny \octagon}) }
= \frac{\surd 3 \, a^2}{6 ( l^2 - a^2 )} 
= \frac{\surd 3}{6 \, \left[ (l/a)^2 - 1 \right] }.  
\end{equation}
In the case of truncated cubic nano-diamonds the [CH]/[CH$_2$] ratios launch from zero because cubic particle surfaces are comprised of only CH$_2$ groups. Thereafter their  [CH]/[CH$_2$] behaviour is determined by the denominator in the above expression.Thus, with increasing truncation $(l/a)$ decreases and the [CH]/[CH$_2$] ratio increases steeply, and much more so than the increase for stT and the decrease for stO polyhedra, as can be clearly seen in  Fig.~\ref{fig_ratios_poly}. In is interesting to note the convergence of the stC and stO polyhedra in this figure, which is due to the fact that both of these forms converge to the related cO polyhedral form with increasing truncation.

% FIGURE 

%||||||||||||||||||||||||||||||||||||||||||||||||||||||||||||||||||||||||||||||||||||||||||||||||||||||||||||||||||||||||||||||||||||||||||||||||||||||||||||||||||
\subsection{The spatial properties of semi-regular polyhedra}
\label{sect_semireg_poly_props}
%||||||||||||||||||||||||||||||||||||||||||||||||||||||||||||||||||||||||||||||||||||||||||||||||||||||||||||||||||||||||||||||||||||||||||||||||||||||||||||||||||

Fig. \ref{fig_ratios_poly} shows the [CH]/[CH$_2$] ratios of semi-regular truncated polyhedra, stT (red), stO (blue), and stC (green), as a function the number of constituent carbon atoms $N_{\rm C}$. Most of what we noted in Section \ref{sect_poly_props}, pertaining to the [CH]/[CH$_2$] ratio in regular polyhedra, also holds for semi-regular polyhedra and so it will not be repeated here. 

We note that the truncation of tetrahedral (T) and octahedral (O) particles, to tT and tO, respectively, leads to an overlap in their [CH]/[CH$_2$] ratios and that truncated cubic particles (stC) always have lower ratios because of the dominance of \{100\} facets. The limiting stC particles converge with the limiting stO forms because, at maximum truncation, the stO and stC forms both converge to cO polyhedra. 

It is evident that the truncation of tetrahedral polyhedra can sequentially result in tT, O, tO, and cO forms, and with square \{100\} facet expansion, to tC and C polyhedra, and that is there is therefore a two-way transformational sequence  
\[
{\rm T\,_{0.12}^{4} \leftrightarrow tT \, _{0.40}^{8} \leftrightarrow O \, _{0.32}^{8} \leftrightarrow tO \, _{0.68}^{14} \leftrightarrow cO \, _{0.56}^{14} \leftrightarrow tC \, _{0.58}^{14} \leftrightarrow C \, _{0.37}^{6}}
\]
where the subscript is the `sphericity' (to 2 d.p.) and the superscript is the number of faces (F) in the given polyhedron. The above sequence, which was alluded to at the beginning of Section \ref{sect_irregulars}, implies that some of the `upward' tetrahedral truncation evolution in Fig. \ref{fig_ratios_poly} must lead to some exact regular polyhedral solutions. However, the counter `downward' octahedral truncation  in this figure does not represent a reversal of the above transformation, even though there is an overlap in the mapped-out parameter space. This is because there is, in general, likely to be an increase in the `sphericity' during any evolutionary sequence due to erosion, that is there will be a `blunting' of the protruding polyhedral vertices and exposed edges. This can be seen along the transformation sequence shown above as a general correlation between `sphericity' and the number of faces (F). However, cubic polyhedra, although they may appear to be a bit of an outlier to this trend, are nevertheless a part of this sequence, which is in fact a cycle. This is because a C$\leftrightarrow$T polyhedral transformation is possible via the half-truncated cube, that is a cube in which only the opposing vertices on opposing faces (indicated by the small circles in Fig.~\ref{fig_trunc_cube}) are chamfered into equilateral triangles.

Interestingly, the truncation of tetrahedra (T\,$\rightarrow$\,tT) and octahedra (O\,$\rightarrow$\,tO) initially leads to a convergence of their [CH]/[CH$_2$] ratios, that is they evolve in opposite senses, and then to an overshoot of one another to more extreme values.  Also evident in Fig. \ref{fig_ratios_poly} is a progressive flattening of this ratio as octahedral particles are progressively truncated (O\,$\rightarrow$\,tO\,$\rightarrow$\,cO), which is to be expected because, as Fig. \ref{fig_3D_geo_CHs} indicates, the [CH]/[CH$_2$] ratio in tO and cO particles is independent of size. 

The results in Fig. \ref{fig_ratios_poly} confirm the earlier conclusion that the nano-diamond [CH]/[CH$_2$] ratio likely varies over more that an order of magnitude for any given size particle. As in the previous sections, the modelling of semi-regular polyhedra is a process in which the possible particle dimensions are continuous, that is any arbitrary polyhedral dimension is valid.

% TABLE
\begin{table*}[h]
\caption{Tetrahedral particle compositions.}
\begin{center}
\begin{tabular}{ccccccccc}
     &    &    &    &    &    &    &    &   \\[-0.35cm]
\hline
\hline
     &    &    &    &    &    &    &    &    \\[-0.3cm]
  i   &  $N_{\rm C}$  &  $N_{\rm H}$  &  $N_{\rm CH}$  &   $N_{\rm CH_2}$ &  $N_{\rm 4^\circ}$  &  H/C  &  $X_{\rm H}$ &   [CH]/[CH$_2$]  \\[0.05cm]
\hline
     &    &    &    &    &    &    &    &    \\[-0.25cm]
%   TETRAHEDRAL PARTICLES  - ( 9 columns )
%  -----------------------
%    i        NC        NH       NCH       NCH2     NC4^o       H/C         X_H      [CH]/[CH2]
%   ---      ----      ----     -----     ------    -----      -----       -----     ----------      
     1   &    10   &    16   &     4   &     6   &     0   &    1.60   &    0.62   &    0.67    \\
     2   &    26   &    32   &     8   &    12   &     6   &    1.23   &    0.55   &    0.67    \\
     3   &    51   &    52   &    16   &    18   &    17   &    1.02   &    0.50   &    0.89    \\
     4   &    87   &    76   &    28   &    24   &    35   &    0.87   &    0.47   &    1.17    \\
     5   &   136   &   104   &    44   &    30   &    62   &    0.76   &    0.43   &    1.47    \\
     6   &   200   &   136   &    64   &    36   &   100   &    0.68   &    0.40   &    1.78    \\
     7   &   281   &   172   &    88   &    42   &   151   &    0.61   &    0.38   &    2.10    \\
     8   &   381   &   212   &   116   &    48   &   217   &    0.56   &    0.36   &    2.42    \\
     9   &   502   &   256   &   148   &    54   &   300   &    0.51   &    0.34   &    2.74    \\
    10   &   646   &   304   &   184   &    60   &   402   &    0.47   &    0.32   &    3.07    \\
    11   &   814   &   356   &   224   &    66   &   524   &    0.44   &    0.30   &    3.39    \\
    12   &  1011   &   412   &   268   &    72   &   671   &    0.41   &    0.29   &    3.72    \\
     &    &    &    &    &    &    &    &    \\[-0.35cm]
\hline
\end{tabular}
\tablefoot{As a function of the number of atomic layers, $i$. $N_{\rm C}$ and $N_{\rm H}$ are the total number of carbon and hydrogen atoms, $N_{\rm CH}$ and $N_{\rm CH_2}$ the number of CH$_n$ groups on the surfaces,   $N_{\rm 4^\circ}$ the number of carbon atoms in the bulk  and  $X_{\rm H}$ the atomic hydrogen fraction.}
\end{center}
\label{tetrahedral_struct}
\end{table*}

%||||||||||||||||||||||||||||||||||||||||||||||||||||||||||||||||||||||||||||||||||||||||||||||||||||||||||||||||||||||||||||||||||||||||||||||||||||||||||||||||||
\section{Diamond-bonded nano-particle networks}
\label{sect_networks}
%||||||||||||||||||||||||||||||||||||||||||||||||||||||||||||||||||||||||||||||||||||||||||||||||||||||||||||||||||||||||||||||||||||||||||||||||||||||||||||||||||

This section presents a diamond network bonding model, a complimentary, discretised approach that is more attuned to the calculation of the `molecular' structure of diamond at nano-scales, which can be directly compared with the (semi-)regular polyhedral models developed in the previous sections. The diamond lattice consists of two interwoven, face-centred cubic (fcc) lattices that are offset by one quarter of the unit cell dimension with respect to each other.\footnote{The lattice offset is one quarter of the unit cell dimension in each of the \^x, \^y, and \^z directions, which are orthogonal in the diamond lattice.} This overlapping two-fcc lattice de-construction allows for a reasonably straight-forward, cubic-grid computational description of the perfect diamond lattice. 

In this approach we consider the from-vertex `top-down' atomic layers, $i$, of regular tetrahedral and octahedral structures (where $i = 1, 2, 3, 4,$ \ldots). Construction, or rather de-construction, is such that the removal of a particular atomic layer, and of its associated  overlying atomic layers, leaves a coherent diamond(oid) particle with no `dangling' CH$_n$ groups (where $n$ can be 1,2 or 3) and with the `newly-exposed' surface passivated with only CH and CH$_2$ groups. In the following, and for simplicity, we designate the considered T, tT, O, and tO particles as diamondoids, even though they may not strictly be such. As a guide, and before we enter into the details in the following sub-sections, we present some general characteristics of this atomic layer approach. 

For tetrahedral particles the sequential number of carbon atoms per layer, $n$, is given by the series of the squares: 
\[
n = 4, 9, 16, 25, 36, 49, 64, 81, 100,  \ldots 
\]
and for the $i^{\rm th}$ atomic layer $N_{\rm C} = (i+1)^2$. Summing over this series does not yield the total number of carbon atoms in the particle because the terminating, lowest layer (largest value of $i$) has three less atoms than the series predicts.\footnote{If allowed, these C atoms would form dangling CH$_3$ groups.} Hence,
\[
N_{\rm C} =  \Bigg\{ \sum_i (i+1)^2 \Bigg\} - 3.  
\]
With this formalism the minimum tetrahedral particle, adamantane (C$_{10}$H$_{16}$), has the layer structure $4,(9-3) = 4, 6$  and a total number of carbon atoms $N_{\rm C} = (4+9)-3 = 10$). 

For octahedral particles the series for the number of carbon atoms per layer is: 
\[
n = 1, 2, 4, 6, 9, 12, 16, 20, 25, 30, 36, \ldots
\]
where this series effectively comprises alternating terms in the series with terms $j^2$ and $k(k+1)$, respectively. Reflection about a square term $\geqslant 2^2$ (i.e. 4, 9, 16, 25, 36, \ldots) yields a complete octahedral nano-diamond particle. The minimum octahedral particle is also adamantane and in this formalism has the layer structure 1, 2, 4, 2, 1 ($N_{\rm C} = 10$). For a 9-atomic layer octahedron the number of carbon atoms per layer is obtained by reflection about the $i = 5$ term ($j=3 \equiv (i-2)^2 = 3^2$) in the alternating series, i.e. 
\[
n = 1, 2, 4, 6, 9, 6, 4, 2, 1 \ \ \ \ \ \sum_n = N_{\rm C} = 35. 
\]
In the following we explore these two network structures in more detail in order to calculate their compositions and, in particular, their [CH]/[CH$_2$] group ratios.

% TABLE
\begin{table*}[h]
\caption{Truncated tetrahedral particle compositions.}
\begin{center}
\begin{tabular}{cccccccccccc}
     &    &    &    &    &    &    &    &  &    &    &    \\[-0.35cm]
\hline
\hline
     &    &    &    &    &    &    &    &    &    &    &   \\[-0.3cm]
  i   &  $N_{\rm C}$  &  $N_{\rm H}$  &  n   &  $N_{\rm C}(n)$  &  $N_{\rm H}(n)$  &  $N_{\rm CH}$  &   $N_{\rm CH_2}$ &  $N_{\rm 4^\circ}$  &  H/C  &  $X_{\rm H}$ &   [CH]/[CH$_2$]  \\[0.05cm]
\hline
     &    &    &    &    &    &    &    &    &    &    &   \\[-0.25cm]
%   TRUNCATED TETRAHEDRAL PARTICLES  - ( 12 columns )
%  ---------------------------------
%    i        NC        NH         n      NC(n)     NH(n)      NCH       NCH2     NC4^o       H/C         X_H      [CH]/[CH2]
%   ---      ----      ----      -----    -----     -----     -----     ------   -------     ------      -----     ----------          
     3   &    51   &    52   &     1   &    35   &    36   &    24   &     6   &     5   &    1.03   &    0.51   &    4.00    \\[0.2cm]
     4   &    87   &    76   &     1   &    71   &    60   &    36   &    12   &    23   &    0.85   &    0.46   &    3.00    \\[0.2cm]
     5   &   136   &   104   &     1   &   120   &    88   &    52   &    18   &    50   &    0.73   &    0.42   &    2.89    \\
         &         &         &     2   &    84   &    64   &    52   &     6   &    26   &    0.76   &    0.43   &    8.67    \\[0.2cm]
     6   &   200   &   136   &     1   &   184   &   120   &    72   &    24   &    88   &    0.65   &    0.39   &    3.00    \\
         &         &         &     2   &   148   &    96   &    72   &    12   &    64   &    0.65   &    0.39   &    6.00    \\[0.2cm]
     7   &   281   &   172   &     1   &   265   &   156   &    96   &    30   &   139   &    0.59   &    0.37   &    3.20    \\
         &         &         &     2   &   229   &   132   &    96   &    18   &   115   &    0.58   &    0.37   &    5.33    \\
         &         &         &     3   &   165   &   100   &    88   &     6   &    71   &    0.61   &    0.38   &   14.67    \\[0.2cm]
     8   &   381   &   212   &     1   &   365   &   196   &   124   &    36   &   205   &    0.54   &    0.35   &    3.44    \\
         &         &         &     2   &   329   &   172   &   124   &    24   &   181   &    0.52   &    0.34   &    5.17    \\
         &         &         &     3   &   265   &   140   &   116   &    12   &   137   &    0.53   &    0.35   &    9.67    \\[0.2cm]
     9   &   502   &   256   &     1   &   486   &   240   &   156   &    42   &   288   &    0.49   &    0.33   &    3.71    \\
         &         &         &     2   &   450   &   216   &   156   &    30   &   264   &    0.48   &    0.32   &    5.20    \\
         &         &         &     3   &   386   &   184   &   148   &    18   &   220   &    0.48   &    0.32   &    8.22    \\
         &         &         &     4   &   286   &   144   &   132   &     6   &   148   &    0.50   &    0.33   &   22.00    \\[0.2cm]
    10   &   646   &   304   &     1   &   630   &   288   &   192   &    48   &   390   &    0.46   &    0.31   &    4.00    \\
         &         &         &     2   &   594   &   264   &   192   &    36   &   366   &    0.44   &    0.31   &    5.33    \\
         &         &         &     3   &   530   &   232   &   184   &    24   &   322   &    0.44   &    0.30   &    7.67    \\
         &         &         &     4   &   430   &   192   &   168   &    12   &   250   &    0.45   &    0.31   &   14.00    \\[0.2cm]
    11   &   814   &   356   &     1   &   798   &   340   &   232   &    54   &   512   &    0.43   &    0.30   &    4.30    \\
         &         &         &     2   &   762   &   316   &   232   &    42   &   488   &    0.41   &    0.29   &    5.52    \\
         &         &         &     3   &   698   &   284   &   224   &    30   &   444   &    0.41   &    0.29   &    7.47    \\
         &         &         &     4   &   598   &   244   &   208   &    18   &   372   &    0.41   &    0.29   &   11.56    \\
         &         &         &     5   &   454   &   196   &   184   &     6   &   264   &    0.43   &    0.30   &   30.67    \\[0.2cm]
    12   &  1011   &   412   &     1   &   995   &   396   &   276   &    60   &   659   &    0.40   &    0.28   &    4.60    \\
         &         &         &     2   &   959   &   372   &   276   &    48   &   635   &    0.39   &    0.28   &    5.75    \\
         &         &         &     3   &   895   &   340   &   268   &    36   &   591   &    0.38   &    0.28   &    7.44    \\
         &         &         &     4   &   795   &   300   &   252   &    24   &   519   &    0.38   &    0.27   &   10.50    \\
         &         &         &     5   &   651   &   252   &   228   &    12   &   411   &    0.39   &    0.28   &   19.00    \\
     &    &    &    &    &    &    &    &    &    &    &   \\[-0.35cm]
\hline
\end{tabular}
\tablefoot{As a function of the number of atomic layers, $i$, and the number of atomic layer truncations, $n$.  $N_{\rm C}$ and $N_{\rm H}$ are the total number of carbon and hydrogen atoms in the parent tetrahedron and $N_{\rm C}(n)$ and $N_{\rm H}(n)$ those in the truncated tetrahedron. $N_{\rm CH}$ and $N_{\rm CH_2}$ the number of CH$_n$ groups on the surfaces,   $N_{\rm 4^\circ}$ the number of carbon atoms in the bulk  and  $X_{\rm H}$ the atomic hydrogen fraction.}
\end{center}
\label{tetrahedral_trunc}
\end{table*}

%||||||||||||||||||||||||||||||||||||||||||||||||||||||||||||||||||||||||||||||||||||||||||||||||||||||||||||||||||||||||||||||||||||||||||||||||||||||||||||||||||
\subsection{Tetrahedral and truncated tetrahedral networks}
%||||||||||||||||||||||||||||||||||||||||||||||||||||||||||||||||||||||||||||||||||||||||||||||||||||||||||||||||||||||||||||||||||||||||||||||||||||||||||||||||||

The faces, edges and vertices of tetrahedral diamondoids are comprised of CH, CH$_2$, and CH groups, respectively. In their truncated forms the four CH-terminated vertices are replaced with four \{111\} CH-passivated facets and the \{111\}/\{111\} truncation edges are alternately-directed CH bonds coherent with the adjacent \{111\} facets. 

With a bit of three-dimensional geometrical thinking it can be shown that the total number of carbon atoms, $N_{\rm C}$, in a diamondoid with $i$ atomic layers is given by 
\begin{equation}
N_{\rm C} = 6 \, (2i-1) + 4 \, (i-1)^2 + 4 + \Bigg\{ \sum_i (i-2)^2 \Bigg\}, 
\label{eq_NC_tetra}
\end{equation}
where the terms are, from left to right, the number of edge, face, vertex, and quaternary ($4^\circ$) carbon atoms. Replacing the summation with the closed formula $(i-2)^3/3 + (i-2)^2/2 +(i-2)/6$ it can be shown that Eq. (\ref{eq_NC_tetra}) reduces to
\begin{equation}
N_{\rm C} = \frac{i^3}{3} + \frac{5 \, i^2}{2} + \frac{37 \, i}{6} + 1. 
\end{equation}
The corresponding number of hydrogen atoms is 
\begin{equation}
N_{\rm H} = 12 \, i + 4 \, \frac{i \, (i-1)}{2} + 4,  
\end{equation}
where the terms are, from left to right, the number of edge CH$_2$'s, and facet and vertex CH groups, respectively, which reduces to  
\begin{equation}
N_{\rm H} = 2 \, i^2 + 10 \, i  + 4.  
\end{equation}

Truncation leads to the cumulative loss of successive layers, $i$, of carbon atoms, from each of the four vertices for equi-vertex truncation, following the atomic layers series described above, that is 16, 36, 64, 100, 144, 196, 256,  \ldots $4 \, (i+1)^2$, which highlights the increasingly rapid loss of atoms with top-down truncation in a tetrahedral pyramid. The total number of carbon atoms lost, where all four vertices are equally truncated, is given by 
\begin{equation}
N_{\rm C,loss} = 4 \sum_i (i+1)^2 = \frac{2 \, i \, ( 2 \, i^2 + 9 \, i + 13)}{3}.  
\end{equation}
The equivalent hydrogen atom loss from all four vertices is 
\begin{equation}
N_{\rm H,loss} = 4 \, \Bigg\{ \ 6 \, i + 3 \sum_i (i-1) - \sum_i (i+1) \ \Bigg\} =  i \, ( i + 3),  
\label{eq_tT_NHloss} 
\end{equation}
where the terms in brackets, from left to right, relate to the loss of CH$_2$, and the loss and gain of  CH groups, respectively, that is   
\begin{equation}
N_{\rm CH_2,loss} =   12 \, i
\label{eq_tT_CH2loss}
\end{equation}
\begin{equation}
N_{\rm CH_2,gain} =  0
\end{equation}
\begin{equation}
N_{\rm CH,loss} =   6 \, i \, (i-1) + 4
\end{equation}
\begin{equation}
N_{\rm CH,gain} =   2 \, (i+1) (i+2).
\end{equation}
The factor of two difference between the first term in Eq.~(\ref{eq_tT_NHloss}) and Eq.~(\ref{eq_tT_CH2loss}) is because the former counts the total H atom loss from CH$_2$ groups and the latter the number of CH$_2$ groups lost. 
The tetrahedral and truncated tetrahedral particle compositions are shown in Tables \ref{tetrahedral_struct} and \ref{tetrahedral_trunc}, respectively.

% TABLE
\begin{table*}[h]
\caption{Octahedral particle compositions.}
\begin{center}
\begin{tabular}{ccccccccc}
     &    &    &    &    &    &    &    &   \\[-0.35cm]
\hline
\hline
     &    &    &    &    &    &    &    &    \\[-0.3cm]
     %     &    &    &    &    &    &    &    &    \\
  i   &  $N_{\rm C}$  &  $N_{\rm H}$  &  $N_{\rm CH}$  &   $N_{\rm CH_2}$ &  $N_{\rm 4^\circ}$  &  H/C  &  $X_{\rm H}$ &   [CH]/[CH$_2$]  \\[0.05cm]
\hline
     &    &    &    &    &    &    &    &    \\[-0.25cm]
%   OCTAHEDRAL PARTICLES  - ( 9 columns )
%  ----------------------
%    i        NC        NH       NCH       NCH2     NC4^o       H/C         X_H      [CH]/[CH2]
%   ---      ----      ----     -----     ------    -----      -----       -----     ----------
     0   &    10   &    16   &     4   &     6   &     0   &    1.60   &    0.62   &    0.67    \\
     1   &    35   &    36   &    24   &     6   &     5   &    1.03   &    0.51   &    4.00    \\
     2   &    84   &    64   &    52   &     6   &    26   &    0.76   &    0.43   &    8.67    \\
     3   &   165   &   100   &    88   &     6   &    71   &    0.61   &    0.38   &   14.67    \\
     4   &   286   &   144   &   132   &     6   &   148   &    0.50   &    0.33   &   22.00    \\
     5   &   455   &   196   &   184   &     6   &   265   &    0.43   &    0.30   &   30.67    \\
     6   &   680   &   256   &   244   &     6   &   430   &    0.38   &    0.27   &   40.67    \\
     7   &   969   &   324   &   312   &     6   &   651   &    0.33   &    0.25   &   52.00    \\
     8   &  1330   &   400   &   388   &     6   &   936   &    0.30   &    0.23   &   64.67    \\ 
     &    &    &    &    &    &    &    &    \\[-0.35cm]
\hline
\end{tabular}
\tablefoot{As a function of the number of atomic layers, $i$. $N_{\rm C}$, and $N_{\rm H}$ are the total number of carbon and hydrogen atoms, $N_{\rm CH}$ and $N_{\rm CH_2}$ the number of CH$_n$ groups on the surfaces,   $N_{\rm 4^\circ}$ the number of carbon atoms in the bulk  and  $X_{\rm H}$ the atomic hydrogen fraction.}
\end{center}
\label{octahedral_struct}
\end{table*}

% TABLE
\begin{table*}[h]
\caption{Truncated octahedral particle compositions.} 
\begin{center}
\begin{tabular}{cccccccccccc}
     &    &    &    &    &    &    &    &   &    &    &   \\[-0.35cm]
\hline
\hline
     &    &    &    &    &    &    &    &    &    &    &   \\[-0.3cm]
  i   &  $N_{\rm C}$  &  $N_{\rm H}$  &  n   &  $N_{\rm C}(n)$  &  $N_{\rm H}(n)$  &  $N_{\rm CH}$  &   $N_{\rm CH_2}$ &  $N_{\rm 4^\circ}$  &  H/C  &  $X_{\rm H}$ &   [CH]/[CH$_2$]  \\[0.05cm]
\hline
     &    &    &    &    &    &    &    &    &    &    &   \\[-0.25cm]    
    %   TRUNCATED OCTAHEDRAL PARTICLES  - ( 12 columns )
%  --------------------------------
%    i        NC        NH         n      NC(n)     NH(n)      NCH       NCH2     NC4^o       H/C         X_H      [CH]/[CH2]
%   ---      ----      ----      -----    -----     -----     -----     ------   -------     ------      -----     ----------
     2   &    35   &    36   &     1   &    29   &    36   &    12   &    12   &     5   &    1.24   &    0.55   &    1.00    \\[0.2cm]
     3   &    84   &    64   &     1   &    78   &    64   &    40   &    12   &    26   &    0.82   &    0.45   &    3.33    \\
         &         &         &     2   &    66   &    64   &    16   &    24   &    26   &    0.97   &    0.49   &    0.67    \\[0.2cm]
     4   &   165   &   100   &     1   &   159   &   100   &    76   &    12   &    71   &    0.63   &    0.39   &    6.33    \\
         &         &         &     2   &   147   &   100   &    52   &    24   &    71   &    0.68   &    0.40   &    2.17    \\
         &         &         &     3   &   123   &   100   &    28   &    36   &    59   &    0.81   &    0.45   &    0.78    \\[0.2cm]
     5   &   286   &   144   &     1   &   280   &   144   &   120   &    12   &   148   &    0.51   &    0.34   &   10.00    \\
         &         &         &     2   &   268   &   144   &    96   &    24   &   148   &    0.54   &    0.35   &    4.00    \\
         &         &         &     3   &   244   &   144   &    72   &    36   &   136   &    0.59   &    0.37   &    2.00    \\
         &         &         &     4   &   208   &   144   &    36   &    54   &   118   &    0.69   &    0.41   &    0.67    \\[0.2cm]
     6   &   455   &   196   &     1   &   449   &   196   &   172   &    12   &   265   &    0.44   &    0.30   &   14.33    \\
         &         &         &     2   &   437   &   196   &   148   &    24   &   265   &    0.45   &    0.31   &    6.17    \\
         &         &         &     3   &   413   &   196   &   124   &    36   &   253   &    0.47   &    0.32   &    3.44    \\
         &         &         &     4   &   377   &   196   &    88   &    54   &   235   &    0.52   &    0.34   &    1.63    \\
         &         &         &     5   &   323   &   196   &    52   &    72   &   199   &    0.61   &    0.38   &    0.72    \\
         &         &         &     6   &   251   &   196   &     4   &    96   &   151   &    0.78   &    0.44   &    0.04    \\[0.2cm]
     7   &   680   &   256   &     1   &   674   &   256   &   232   &    12   &   430   &    0.38   &    0.28   &   19.33    \\
         &         &         &     2   &   662   &   256   &   208   &    24   &   430   &    0.39   &    0.28   &    8.67    \\
         &         &         &     3   &   638   &   256   &   184   &    36   &   418   &    0.40   &    0.29   &    5.11    \\
         &         &         &     4   &   602   &   256   &   148   &    54   &   400   &    0.43   &    0.30   &    2.74    \\
         &         &         &     5   &   548   &   256   &   112   &    72   &   364   &    0.47   &    0.32   &    1.56    \\
         &         &         &     6   &   476   &   256   &    64   &    96   &   316   &    0.54   &    0.35   &    0.67    \\
         &         &         &     7   &   380   &   256   &    16   &   120   &   244   &    0.67   &    0.40   &    0.13    \\[0.2cm]
     8   &   969   &   324   &     1   &   963   &   324   &   300   &    12   &   651   &    0.34   &    0.25   &   25.00    \\
         &         &         &     2   &   951   &   324   &   276   &    24   &   651   &    0.34   &    0.25   &   11.50    \\
         &         &         &     3   &   927   &   324   &   252   &    36   &   639   &    0.35   &    0.26   &    7.00    \\
         &         &         &     4   &   891   &   324   &   216   &    54   &   621   &    0.36   &    0.27   &    4.00    \\
         &         &         &     5   &   837   &   324   &   180   &    72   &   585   &    0.39   &    0.28   &    2.50    \\
         &         &         &     6   &   765   &   324   &   132   &    96   &   537   &    0.42   &    0.30   &    1.38    \\
         &         &         &     7   &   669   &   324   &    84   &   120   &   465   &    0.48   &    0.33   &    0.70    \\
         &         &         &     8   &   549   &   324   &    24   &   150   &   375   &    0.59   &    0.37   &    0.16    \\
     &    &    &    &    &    &    &    &   &    &    &    \\[-0.35cm]
\hline
\end{tabular}
\tablefoot{As a function of the number of atomic layers, $i$, and the number of atomic layer truncations, $n$.  $N_{\rm C}$ and $N_{\rm H}$ are the total number of carbon and hydrogen atoms in the parent octahedron and $N_{\rm C}(n)$ and $N_{\rm H}(n)$ those in the truncated octahedron. $N_{\rm CH}$ and $N_{\rm CH_2}$ the number of CH$_n$ groups on the surfaces,   $N_{\rm 4^\circ}$ the number of carbon atoms in the bulk  and  $X_{\rm H}$ the atomic hydrogen fraction.}
\end{center}
\label{octahedral_trunc}
\end{table*}

%||||||||||||||||||||||||||||||||||||||||||||||||||||||||||||||||||||||||||||||||||||||||||||||||||||||||||||||||||||||||||||||||||||||||||||||||||||||||||||||||||
\subsection{Octahedral and truncated octahedral networks}
%||||||||||||||||||||||||||||||||||||||||||||||||||||||||||||||||||||||||||||||||||||||||||||||||||||||||||||||||||||||||||||||||||||||||||||||||||||||||||||||||||

The faces, edges and vertices of octahedral diamondoids are comprised of CH, alternately-directed CH and CH$_2$ groups, respectively. In the truncated form the \{100\} truncation facets are CH$_2$ covered and the \{100\}/\{111\} truncation edges and vertices comprise CH bonds coherent with the adjacent \{111\} facets. 

With yet another bout of three dimensional gymnastics it is possible to show that the total number of carbon atoms in a diamondoid with $(2i+1)$ atomic layers is given by 
\begin{equation}
N_{\rm C} = 2 \, \sum_i i^2 + 2 \, \sum_i i \, (i+1)^2 + (i+1)^2, 
\end{equation}
where the first two terms on the left give the number of mid-plane to vertex atoms and the right hand term is the number of atoms in the square mid-plane. Replacing the summations with their closed forms this reduces to
\begin{equation}
N_{\rm C} = \frac{4 \, i^3}{3} + 4 \, i^2 + \frac{11 \, i}{3} + 1. 
\end{equation}
The corresponding total number of hydrogen atoms is 
\begin{equation}
N_{\rm H} = (i+3)^2.  
\end{equation}
The number of vertex CH$_2$ groups, $N_{\rm CH_2}$, is constant at 6 (i.e. $N_{\rm H,vertex} = 12$) and the number of CH groups in \{111\} facets and along their edges is then $N_{\rm CH} = N_{\rm H} - 12$. 

Truncation leads to the cumulative loss of successive layers, $i$, of carbon atoms, from each of the six vertices for equi-vertex truncation, following the atomic layers series described above, that is 12, 24, 36, 54, 72, 96, 120,  \ldots, which again shows a  rapid loss of atoms with increasing truncation. The total number of carbon atoms lost, where all vertices are equally truncated, is given by 
\begin{equation}
N_{\rm C,loss} = 6 \Bigg\{  \sum_{k {\rm \ even}}^{\rm limit} \frac{(i-k)[(i-k)+1]}{2}+ \frac{i \, (i+1)}{2}  \Bigg\},  
\end{equation}
where the upper `limit' $= 2[(i+2)/2-1]$. All of the above assume integer calculations adopting the lowest integer result in each case. In the octahedral diamondoid case truncation results in no net hydrogen atom loss if the truncated facets are hydrogen-passivated with CH$_2$ and CH groups. The CH$_n$ group losses and gains, per vertex, are in this case
\begin{equation}
N_{\rm CH_2,loss} =   1
\end{equation}
\begin{equation}
N_{\rm CH_2,gain} =   \left( \frac{i}{2} + 1\right) \Bigg\{ \frac{(i+1)}{2} + 1 \Bigg\}
\end{equation}
\begin{equation}
N_{\rm CH,loss} =  2 \, N_{\rm CH_2,gain} - 2 
\end{equation}
\begin{equation}
N_{\rm CH,gain} =   0.
\end{equation}
The octahedral and truncated octahedral particle compositions are shown in Tables \ref{octahedral_struct} and \ref{octahedral_trunc}, respectively.

% FIGURE 
\begin{figure}[h]
\centering
\includegraphics[width=9.5cm]{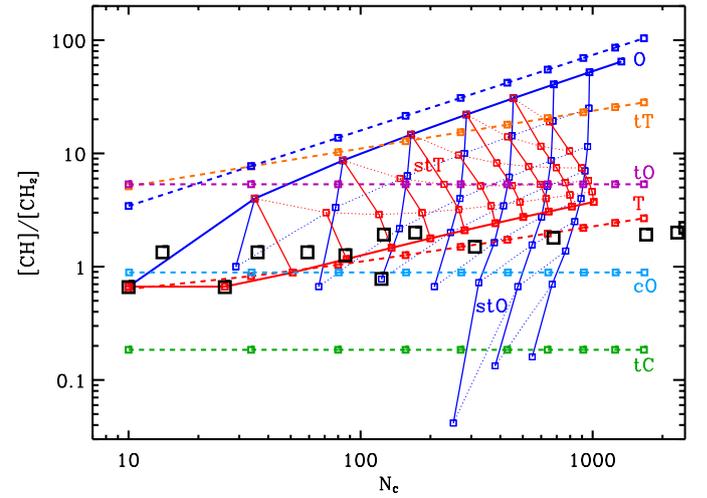}
\caption{CH]/[CH$_2$] ratios in diamond-bonded networks of polyhedral form as a function of the truncation length (thin solid lines) increasing with distance from the thick solid red and blue lines, as a function of the number of constituent carbon atoms $N_{\rm C}$: the thick solid lines are for the regular tetrahedral (red) and octahedral (blue) forms and the dangling and rising thinner lines show the effects of increasing truncation. The thick dashed lines show the [CH/[CH$_2$] ratios for the polyhedral models:  T (red), tT (orange), O (blue), tO (purple), cO (cobalt), and tC (green). For cubic particles (C) this ratio is zero. The data points show the only possible particle solutions with this truncation scheme. The black squares show the ratios for `spherical' nano-diamonds for comparison.}
\label{fig_ratios_netwk}
\centering
\end{figure}

%||||||||||||||||||||||||||||||||||||||||||||||||||||||||||||||||||||||||||||||||||||||||||||||||||||||||||||||||||||||||||||||||||||||||||||||||||||||||||||||||||
\subsection{The spatial properties of diamond-bonded networks}
%||||||||||||||||||||||||||||||||||||||||||||||||||||||||||||||||||||||||||||||||||||||||||||||||||||||||||||||||||||||||||||||||||||||||||||||||||||||||||||||||||

In Fig. \ref{fig_ratios_netwk} we show the [CH]/[CH$_2$] ratios in diamond-bonded networks, of overall polyhedral form, as a function the number of constituent carbon atoms $N_{\rm C}$. Perhaps the first thing of note is the wide spread in the ratios and, secondly, how the truncation of tetrahedral (T) networks can lead to exact and terminal octahedral particle (O) solutions (as noted in Section \ref{sect_semireg_poly_props}). The latter are reflected in the [CH]/[CH$_2$] ratios, that is the coincident red and blue squares on the thick blue line in Fig. \ref{fig_ratios_netwk}. However, the reverse is not true, in that octahedral particles cannot be truncated to exact tetrahedral forms and so there are no coincident solutions in this truncation direction and neither are there any truncated forms close to tetrahedral. In the latter case the truncated forms are truncated octahedron and cuboctahedron, as evidenced by the tendency of the truncated forms to flatter [CH]/[CH$_2$] ratios and to stray into the regions occupied by these polyhedra in Fig. \ref{fig_ratios_netwk}. 

For the network-modelled tetrahedral and octahedral nano-diamond particles, and their truncated forms, the [CH]/[CH$_2$] ratios are shown in Tables \ref{tetrahedral_struct} to \ref{octahedral_trunc}. Unlike the regular and semi-regular polyhedral modelling the diamond-bonded network is a discrete modelling process, in that only certain lattice-allowed dimensions are valid, as shown in Fig. \ref{fig_ratios_netwk} and the tables. 

In comparing the T and O polyhedral forms with the tT and tO network forms it can clearly be seen that increasing T network truncation tends towards O polyhedra and O network truncation tends towards, and indeed underpasses, tT polyhedra, tending towards tO and even underpassing tC polyhedra in some cases.  Yet again, as can be seen in Fig. \ref{fig_ratios_netwk} this modelling clearly indicates that the nano-diamond [CH]/[CH$_2$] ratio likely varies over more that an order of magnitude for any given size particle. 

Using the diamond network approach we constructed spherical nano-diamond particles of radius $a_{\rm nd}$ by filling the specified volume with a diamond lattice of a given number of C atoms, $N_{\rm C}$, and then passivating the exposed surface atom dangling bonds with CH and CH$_2$ groups as required.\footnote{This approach ensures that there are no dangling bonds or pendant $-$CH$_3$ methyl groups.} The particle properties and  [CH]/[CH$_2$] ratios for these `spherica'l nano-diamonds are shown in Table \ref{table_sphl_nds} and Fig. \ref{fig_ratios_netwk} (black squares). We again point out that this is a discrete modelling process in that only certain particle dimensions are allowed, which is reflected in the values in the table because these data represent exact solutions to the filling of a given spherical volume with a regular diamond lattice. 
%In Fig. \ref{fig_ratios_netwk} the black squares indicate the [CH]/[CH$_2$] ratios for `spherical' nano-diamonds,  
The black squares in Fig. \ref{fig_ratios_netwk} indicate that spherical nano-diamonds would appear to have a rather narrow range in [CH]/[CH$_2$], that is from a value of 0.67 for adamantane (C$_{10}$H$_{16}$) up to $\simeq 2.2$ for particles with $N_{\rm C} \leq 2500$. A close look at these data shows that spherical nano-diamonds  exhibit [CH]/[CH$_2$] ratios approximating those of the tetrahedral polyhedral and network particles. This was indeed speculated by \cite{2007ApJ...661..919P} who found that the [CH]/[CH$_2$] ratios of large diamondoid molecules or nano-diamonds ought to resemble those of the $T_d$ point group, the largest subgroup of the $O_h$ point group. We therefore concur with their speculation. Interestingly, the low [CH]/[CH$_2$] ratios for spherical nano-diamonds, would seem to imply that any (erosional) process that leads to a rounding of euhedral/polyhedral nano-diamond facets would tend to suppress the [CH]/[CH$_2$] ratio, with respect to most non-cuboid polyhedral particle forms, if surface passivation with hydrogen is maintained.

%||||||||||||||||||||||||||||||||||||||||||||||||||||||||||||||||||||||||||||||||||||||||||||||||||||||||||||||||||||||||||||||||||||||||||||||||||||||||||||||||||
\section{The CH/CH$_2$ ratio in nano-diamonds as a ruler} 
\label{sect_CH-CH2}
%||||||||||||||||||||||||||||||||||||||||||||||||||||||||||||||||||||||||||||||||||||||||||||||||||||||||||||||||||||||||||||||||||||||||||||||||||||||||||||||||||

The aim of this work is to explore whether it is possible to use the observed $3.53\,\mu$m/$3.43\,\mu$m (nano-)diamond IR band ratio as a proxy for the ratio of the surface concentrations or abundances of CH and CH$_2$ groups, [CH]/[CH$_2$], on nano-diamonds, and hence as a ruler to measure their sizes in circumstellar (and interstellar) environments. Firstly, we need to consider whether the measured or observed $3.53\,\mu$m and $3.43\,\mu$m nano-diamond IR bands directly map the surface concentrations of CH and CH$_2$. This appears to be so but must be qualified because, although a direct mapping appears to be true for tetrahedral diamondoids, was shown by \cite{2007ApJ...661..919P}, this has not yet be demonstrated for larger species or for other (polyhedral) forms. Secondly, assuming that the direct mapping issue is a given, we need to determine if it is possible to use the ratio as a ruler. To this there can remain, as yet, no definitive answer within the astronomical context. This is because there is an inherent degeneracy in [CH]/[CH$_2$] between particle form and size. To break this degeneracy, or at least to reduce the uncertainties to a manageable degree,  therefore requires some direct knowledge of the particle form and/or size. Currently, and if the nano-diamond form is unknown, a particular [CH]/[CH$_2$] ratio can spread over orders of magnitude in size (and the cube of this in mass!). 

%The first question is: Do the measured or observed $3.53\,\mu$m and $3.43\,\mu$m nano-diamond IR bands directly map the surface concentrations of CH and CH$_2$? The answer to this appears to be an unequivocal but qualified yes, qualified because a direct mapping appears to be true for tetrahedral diamondoids, as demonstarted by \cite{2007ApJ...661..919P}. However, whether this remains true for larger species and other (polyhedral) forms has seemingly yet to be demonstrated. 
%Assuming that the direct mapping issue is a given, the second question is can we use the ratio as a ruler? To this there can remain, as yet, no definitive answer within the astronomical context. This is because there is an inherent degeneracy in [CH]/[CH$_2$] between particle form and size. To break this degeneracy, or at least to reduce the uncertainties to a manageable degree,  therefore requires some direct knowledge of the particle form and/or size. Currently, and if the nano-diamond form is unknown, a particular [CH]/[CH$_2$] ratio can spread over orders of magnitude in size (and the cube of this in mass!). 

Clearly, it would help to have some idea of the form or the likely range of forms. As \cite{2007ApJ...661..919P} have demonstrated, it appears that small diamondoids could tend towards tetrahedral forms for $N_{\rm C} \leq 140$. While at the other size extreme, that is at micronic scales, images of synthetic nano-diamonds and CVD diamond coatings indicate that a wide range of particle shapes is possible, including: truncated octahedral (tO), cuboctahedral (cO), tetrahedral (T), and cubic (C) particles. Looking at these likely particle shapes, and the data presented in the figures here, it would therefore seem to be a lost cause to try and break the form/size degeneracy. For example, for the favoured tO and cO nano-particle forms the [CH]/[CH$_2$] ratio is, unfortunately, independent of size (see for example Figs. \ref{fig_3D_geo_CHs} and \ref{fig_ratios_netwk}). Thus, it appears that a restriction to the tO and cO forms still does not sufficiently reduce the degeneracy to any useful degree because their [CH]/[CH$_2$] ratios are fixed in each case, independent of size, and are separated by about an order of magnitude (three orders of magnitude in mass) 

In optical constant modelling it would seem that we therefore have little choice but to focus on a range of polyhedral forms, that is tetrahedral, truncated tetrahedral, octahedral, truncated octahedral, and cuboctahedron (T, tT, O, tO, and cO).  Ideally we could use the semi-regular forms stT and stO, to replace all of these but this approach would entail the introduction of the difficult to constrain truncation lengths, $a$, as free parameters into the mix. We would therefore like to be able to adopt a more restricted range of fixed forms, which can then be matched against the observed [CH]/[CH$_2$] values. Thus, it appears that the usual astronomical practice of assuming spherical particles is probably the most viable solution for nano-diamonds.

% TABLE
\begin{table}
\caption{Spherical nano-diamond particle compositions.}
\begin{center}
\begin{tabular}{cccccc}
     &    &    &    &    &    \\[-0.35cm]
\hline
\hline
     &    &    &    &    &    \\[-0.3cm]
   radius [nm]  &  $N_{\rm C}$  &  $N_{\rm H}$  &  $N_{\rm CH}$  &   $N_{\rm CH_2}$ &   $\frac{\rm [CH]}{\rm [CH_2]}$  \\[0.1cm]
\hline
     &    &    &    &    &    \\[-0.25cm]
0.27 & 10 & 16 & 4 & 6 & 0.67 \\
0.31 & 14 & 20 & 8 & 6 & 1.34 \\
0.36 & 26 & 32 & 8 & 12 & 0.67 \\
0.40 & 36 & 40 & 16 & 12 & 1.34 \\
0.44 & 59 & 60 & 24 & 18 & 1.34 \\
0.53 & 123 & 100 & 28 & 36 & 0.78 \\ 
0.76 & 311 & 178 & 76 & 151 & 1.50 \\
0.98 & 678 & 296 & 140 & 78 & 1.80 \\
1.33 & 1700 & 552 & 276 & 138 & 2.00 \\
1.46 & 2316 & 694 & 340 & 177 & 1.92 \\
1.51 & 2509 & 732 & 384 & 174 & 2.20 \\
2.31 & 9177 & 1756 & 868 & 444 & 1.96 \\
4.98 & 91820 & 8160 & 4320 & 1920 & 2.25 \\
8.99 & 539011 & 26556 & 14400 & 6078 & 2.39 \\
     &    &    &    &    &    \\[-0.35cm]
\hline
\end{tabular}
\tablefoot{Theoretically-constructable nano-diamond structures with only surface CH and CH$_2$ groups, i.e. there are no pendant $-$CH$_3$ groups nor dangling bonds.}
\end{center}
\label{table_sphl_nds}
\end{table}

%||||||||||||||||||||||||||||||||||||||||||||||||||||||||||||||||||||||||||||||||||||||||||||||||||||||||||||||||||||||||||||||||||||||||||||||||||||||||||||||||||
\section{Stability and dehydrogenation considerations}
\label{sect_dehydrogenation}
%||||||||||||||||||||||||||||||||||||||||||||||||||||||||||||||||||||||||||||||||||||||||||||||||||||||||||||||||||||||||||||||||||||||||||||||||||||||||||||||||||

Perhaps the first, and most fundamental, issue to be resolved is whether nano-diamonds actually are the most stable form of carbon at nano-scales. This was addressed by \cite{1990Natur.343..244B} who compared the heats of formation of small diamond and graphitic  clusters. Following \cite{1987Ap&SS.139..103N,1987Natur.329..589N} these authors concluded that surface stabilisation plays a critical role and, consequently, that diamonds with hydrogen-terminated surfaces and radii smaller than $\sim 1.5$\,nm are energetically favoured over polycyclic aromatics. However, since this early work the fullerene allotrope of carbon was discovered, which has added to the possible carbon nanoparticle forms that need to be considered. \cite{2003JChPh.118.5094B} re-considered this issue, in the light of these more recent developments, and found that at the nanoscale diamond is not necessarily the most stable phase but that there is a `window' of stability for nano-diamonds with radii $\sim 0.9-2.6$\,nm ($1,127 < N_{\rm C} < 24,398$). Further, \cite{2003JChPh.118.5094B} found that fullerenes are the more stable form for smaller carbon clusters  ($a < 0.9$\,nm, $20 < N_{\rm C} < 1,127$), while graphite is the more stable form for larger clusters ($a > 2.6$\,nm, $N_{\rm C} > 24,398$). 

Nevertheless, the story does not end here because nano-diamonds are known to take several polyhedral forms, which have differing stability in their (de-)hydrogenated states \citep[e.g.][]{2004JChPh.121.4276B}.  This question was investigated in detail by \cite{2004JChPh.121.4276B} who found that, for hydrogenated nano-diamonds with $N_{\rm C} > 10^4$ ($a \gtrsim 2.4$\,nm) the cubic form is the most stable form, followed by the sphere, cuboctahedron or octahedron, and truncated octahedron.  For dehydrogenated nano-diamonds the equivalent stability order is: truncated octahedron, cuboctahedron or sphere, octahedron, and cube but for smaller sizes the spherical form is the most stable with the cuboctahedron and truncated octahedron becoming more stable as size increases. For particles with up to $10^6$ atoms ($a \lesssim 10$\,nm), the dehydrogenated cubic and octahedral forms are higher in energy making them unlikely forms in the larger size range \citep{2004JChPh.121.4276B}. These shape- and size-dependent behaviours are principally due to the differences in the particle surface energies. 

Given that in excited regions the nano-diamond surface hydrogenation may be less than complete, that is  close to hot stars where they may undergo extreme heating and/or direct surface CH bond photo-dissociation, in our follow-up work we introduce a fractional surface H atom coverage factor, $f_{\rm H}$, where $0 \leq f_{\rm H} \leq 1$. A critical issue  is then whether the different surface facets or, more generally, the different CH and CH$_2$ surface groups lose hydrogen atoms through the same processes and at the same rates. 

A definitive answer to this is probably not yet possible but we can perhaps garner clues from some of the work published on the related issues of CH$_n$ groups, surface re-structuration/relaxation, deprotonation potentials, and proton affinities \cite[e.g.][]{1997PhRvB..55.1838Z,1997SurSc.374..333Z,2003DRM....12.1867B,2014Nanot..25R5702B}. The findings of these works are consistent with other theoretical and experimental studies in this area. For instance, the theoretical work of \cite{2003DRM....12.1867B} shows, for dehydrogenated particles with $a \leq 0.5$\,nm, that is less than a few hundred carbon atoms, that octahedral and cuboctahedral particle \{111\} facets preferentially transform from sp$^3$ to sp$^2$ bonding, that is they exfoliate, resulting in onion-like structures with carbon clusters at their core (`bucky-diamonds'). They also found that \{100\} (cubic) nano-diamonds are stable against exfoliation re-structuring.   

Surface re-structuring must obviously have an effect on the CH$_n$ surface group IR band strengths and positions. \cite{1997PhRvB..55.1838Z,1997SurSc.374..333Z} used classical molecular dynamics simulations to study diamond surface structural transformations of \{111\} surfaces, designated C\{111\}(1$\times$1)H structures in their terminology, which exhibit a single peak at $4.28\,\mu$m in their simulations. They explored the transformation of dehydrogenated bulk \{111\} surfaces into C\{111\}(2$\times$1) structures, that consist of $\pi$-bonded chains containing five- and seven-membered rings. With hydrogen adsorption, to form C\{111\}(2$\times$1)H surfaces, an additional peak appears at $3.49\,\mu$m due to a metastable structure. As the hydrogen coverage in the simulations increases this peak increases in intensity, up to half monolayer coverage, and then disappears in favour of the C\{111\}(1$\times$1)H $4.28\,\mu$m peak. They also found that C\{111\}(2$\times$1) bulk surfaces graphitise upon heating to 2300\,K. 

In their later theoretical work \citep{2014Nanot..25R5702B} studied the deprotonation potentials and proton affinities\footnote{The deprotonation energy is the enthalpy change of the reaction \newline \hspace*{1.0cm} nano-diamond-H $\rightleftharpoons$ nano-diamond$^-$ + H$^+$. \\ The proton affinity is the negative of the enthalpy change of the reaction \newline \hspace*{1.0cm} nano-diamond + H$^+$ $\rightleftharpoons$ nano-diamond-H$^+$.} of $\sim 0.9-1.4$\,nm radius nano-diamonds and found that they generally decrease with particle size but exhibit strong shape- and facet-dependencies. Differences of the order of 1\,eV were found for 1\,nm size variations and proton loss was found to be inhomogeneous over the nano-diamond surfaces. As \cite{2014Nanot..25R5702B} point out, tertiary ions ($>$\hspace*{-0.25cm}$-$C$^-$) being more stable that secondary ions ($>$C$<^-_{\rm H}$) results in tertiary (3$^\circ$) CH bonds ($>$\hspace*{-0.25cm}$-$C$-$H) being more easily deprotonated than secondary CH bonds ($>$C$<^{\rm H}_{\rm H}$) on facets, edges, and vertices. Thus, deprotonation, that is hydrogen abstraction or dehydrogenation by H$^+$ loss, occurs preferentially from facets. The opposing process, proton affinity, was found to be strongly facet-dependent while the deprotonation potential was edge/vertex-dependent. The proton affinity was found to be higher, that is protonation is preferred, on \{110\} facets, followed by \{111\} facets and then \{100\} facets. From this work we can probably conclude that deprotonation  (and by inference dehydrogenation) will tends to preferentially occur from tertiary CH on \{111\} facets, while  protons preferentially attach to, and therefore preferentially rehydrogenate, edges and vertices.  

The theoretical and experimental observations described in the preceding paragraphs may be summarised into the following broad scenarios for the evolution of nano-diamond properties: \\
%\begin{itemize}
%\item 
{\bf stability:} they are most stable for radii from 0.9 to 2.6\,nm (for smaller [larger] sizes fullerenes [graphite] are more stable), \\
%\item 
{\bf dehydrogenation:} occurs preferentially via the dissociation of 3$^\circ$ CH bonds on \{111\} facets (before edges and vertices) leading to carbon atom sp$^3$ to sp$^2$ transformation (i.e. aromatisation), \\
%\item 
{\bf re-structuring:} aromatisation triggers the formation of $\pi$-bonded chains with 5- and 7-membered rings and ultimately to outer layer exfoliation as edge-anchored, aromatic sheets that will shift absorption to longer (visible) wavelengths,  \\
%\item 
{\bf shape:} the preference for hydrogenated particles is: cube $>$ sphere $>$ cuboctahedon/octahedon $>$ truncated octahedon. For dehydrogenated particles the order is almost reversed but as their surfaces transform sp$^3 \rightarrow $\, sp$^2$ its relevance is moot, and \\
%\item 
{\bf rehydrogenation:} occurs preferentially at edges and vertices, leading to a hysteresis with dehydrogenation because H atoms do not necessarily (re)attach to the sites they were removed from. \\
%\end{itemize}
The above summary indicates that modelling the thermal- and photo-processing of nano-diamonds in circumstellar regions must ideally try and  include these complex surface re-structurations and compositional changes. 

%In a follow-up paper we derived the complex indices of refraction ($n$ and $k$) of `spherical' nano-diamonds, as a function of their size and degree of hydrogenation, using the optEC$_{\rm (s)}$(a)  methodology \citep{2012A&A...540A...1J,2012A&A...540A...2J,2012A&A...542A..98J}.
% developed for the aliphatic-rich, wide band gap, hydrogenated amorphous carbons, materials that most closely resemble nano-diamonds in composition and structure. 

%||||||||||||||||||||||||||||||||||||||||||||||||||||||||||||||||||||||||||||||||||||||||||||||||||||||||||||||||||||||||||||||||||||||||||||||||||||||||||||||||||
\section{Discussion and speculations}
\label{sect_discussion}
%||||||||||||||||||||||||||||||||||||||||||||||||||||||||||||||||||||||||||||||||||||||||||||||||||||||||||||||||||||||||||||||||||||||||||||||||||||||||||||||||||

The key parameters in determining the ratio of the CH and CH$_2$ IR band strengths are the surface structure, the euhedral (polyhedral) form and the edge to surface ratio where faces and/or edges exhibit different crystal facet properties. For example, regular tetrahedral (T), truncated tetrahedral (tT) and octahedral (O) particles (see Fig.~\ref{fig_shapes}) exhibit only \{111\} facets and \{111\}/\{111\} edges\footnote{However, it should be noted that not all \{111\}/\{111\} edges are equivalent. For example,  these edges are lined with CH$_2$ groups on tetrahedral particles but with CH groups on octahedral particles, the CH groups in the latter case actually being an integral part of the adjacent facets.} and their [CH]/[CH$_2$] ratios are size-dependent and differ between the three polyhedral forms (see Fig.\ref{fig_3D_geo_CHs}). In contrast,  regular truncated octahedral (tO), cuboctahedral (cO), and truncated cubic (tC) particles exhibit both \{111\} and \{100\} facets and \{111\}/\{100\} edges, and also \{111\}/\{111\} edges in the case of tO (Fig.~\ref{fig_shapes}). However, the [CH]/[CH$_2$] ratios for these particles do not depend on size and are spread over more than an order of magnitude (Fig.\ref{fig_3D_geo_CHs}). For cubic (C) particles the [CH]/[CH$_2$] ratio is zero because they exhibit no tertiary CH bonds on their exclusively \{100\} surfaces.

For fully surface-hydrogenated nano-diamonds larger than typical diamondoids (e.g. $N_C > 100$) the most stable, and therefore the most likely, forms are the cube (C), followed by the sphere, cuboctahedron(cO)/octahedron(O), and truncated octahedron (tO)\citep{2004JChPh.121.4276B}. For the cube [CH]/[CH$_2$]$ = 0$ and so is of no help to us here, even though it may be the most stable form. Of the  remaining four only two exhibit size-dependent [CH]/[CH$_2$] ratios, the sphere because it mimics tetrahedral particles (see Section \ref{sect_semireg_poly_props}) and the octahedron (O). The [CH]/[CH$_2$] ratios for the cuboctahedron (cO) and truncated octahedron (tO) are fixed at $\sim 0.9$ and $\sim 5.3$, respectively. 

\begin{figure*}[h]
\centering
\includegraphics[width=18cm]{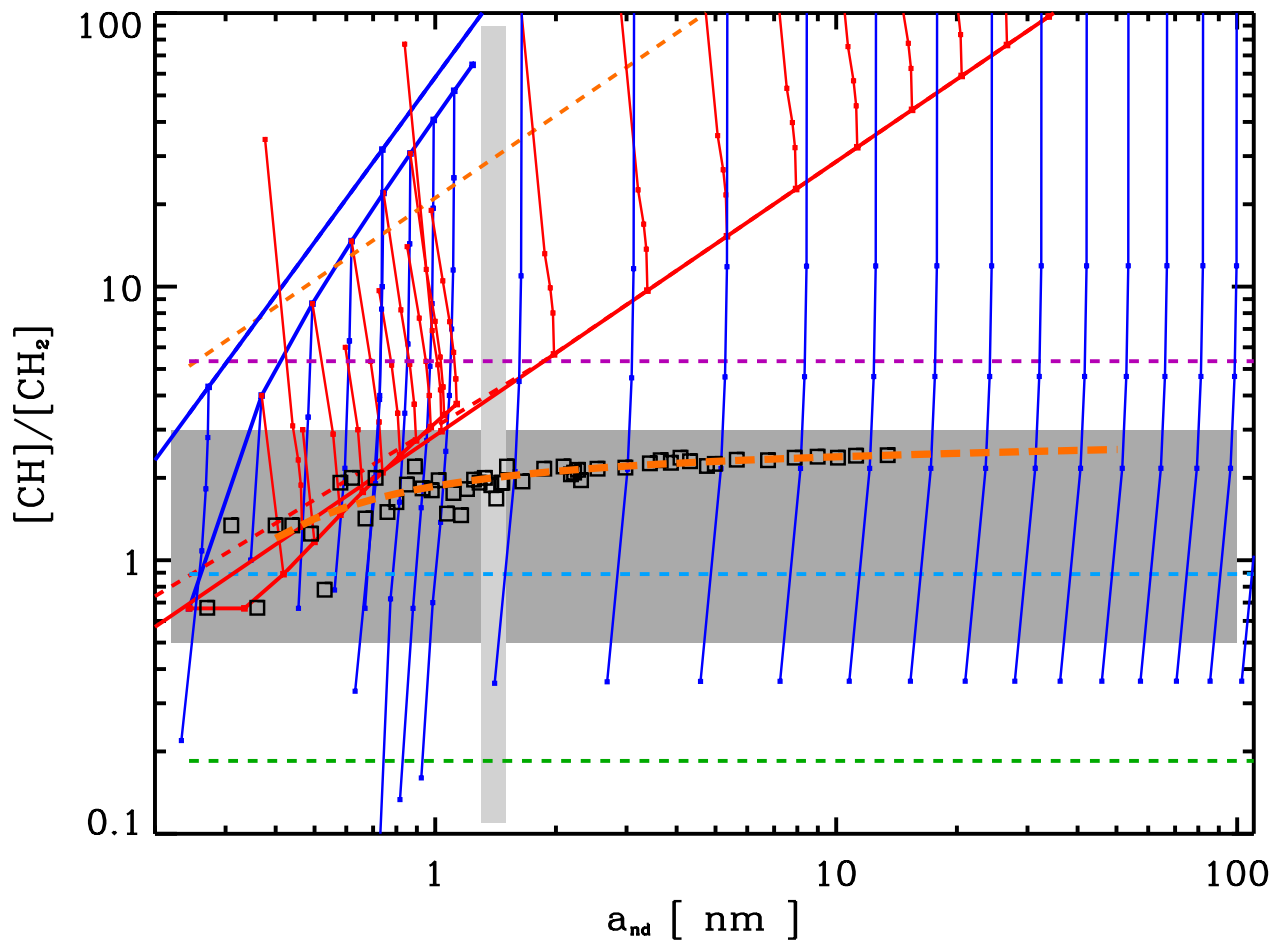}
\caption{Ranges of the [CH]/[CH$_2$] ratios for all of the considered models: tetrahedra and octahedra including their truncated forms, solid red and blue lines, respectively.  The dangling and rising thinner lines show the effects of increasing truncation. The dashed lines show the [CH/[CH$_2$] ratios for the regular polyhedra:  T (red), tT (orange), O (blue), tO (purple), cO (cobalt), and tC (green). For cubic particles (C) this ratio is zero. The black squares show the ratios for `spherical' nano-diamonds and the orange dashed line shows an analytical approximation to these data (see text for details). The grey shaded gives an indication of  the observed and experimental values ($0.5-3$) and the lighter grey vertical band indicates the typical radii of the most abundant pre-solar nano-diamonds ($1.3-1.5$\,nm).}
\label{fig_ratios_poly_disc}
\centering
\end{figure*}

Fig. \ref{fig_ratios_poly_disc} summarises the [CH]/[CH$_2$] ratios for all of the modelled polyhedra and network structures, the data are the same as in the previous figures but are here plotted as a function of the equivalent sphere radius, $a_{\rm n d}$, rather than against the number of carbon atoms, $N_{\rm C}$. These are compared with the typical range ($0.5-3$) for the experimental and observational data (grey shaded area). From this figure we conclude that, for particles with radii larger than $\sim 1$\,nm, only spherical nano-diamonds (black squares) and truncated octahedral/cuboctahedral nano-diamonds in their semi-regular forms (`vertical' blue lines) appear to be consistent with the measured data. However, for nano-diamonds with radii smaller than $\sim 1$\,nm it appears that almost any shape could be consistent with the data. Thus, from a consideration of the [CH]/[CH$_2$] ratio it seems that spherical and truncated octahedral/cuboctahedral  nano-diamonds are the most probable forms. It is interesting that \cite{2004JChPh.121.4276B} basically come to the same conclusion but from a completely different and energetic point of view. Thus, and given that regular tetrahedral (T) and octahedral (O) forms are apparently not favoured, nano-diamonds in astrophysical environments must therefore include spherical, semi-regular and truncated octahedral family (cO, tO an stO) particle forms in order to allow for the observed variations in the CH to CH$_2$ band ratio both in the laboratory and in space (see Fig. \ref{fig_ratios_poly_disc} and also Sections \ref{sect_poly_props} and \ref{sect_semireg_poly_props} for a discussion of these effects).

% FIGURE 

% FIGURE 
\begin{figure}[h]
\centering
\includegraphics[width=9.5cm]{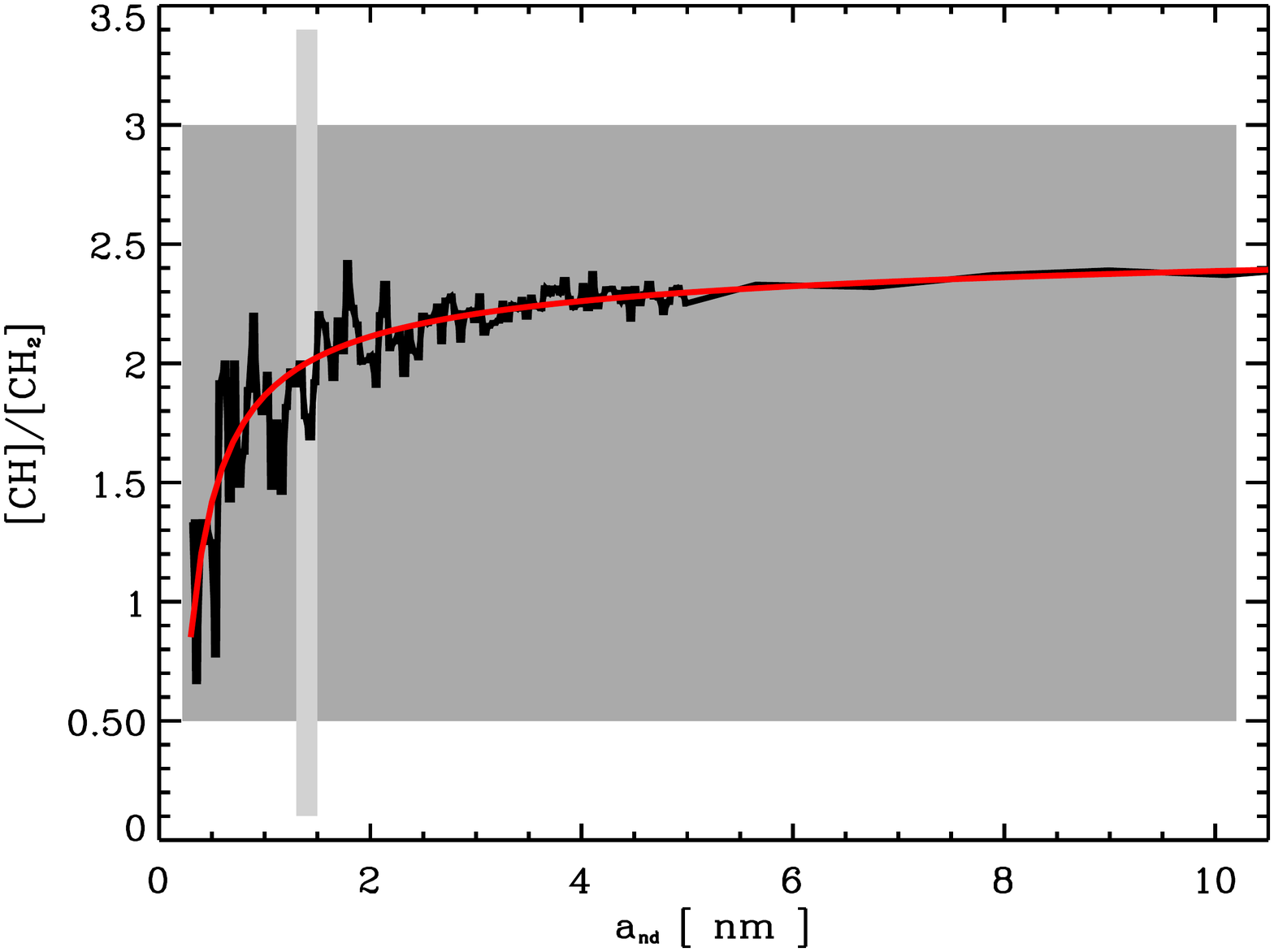}
\caption{[CH]/[CH$_2$] ratios for `spherical' nano-diamonds with actual radii $\leq 10$\,nm (black line), showing the inherent dispersion in the data. The red line shows the analytical approximation to these data (see text for details). The grey shaded gives an indication of  the observed and experimental values ($0.5-3$) and the lighter grey vertical band indicates the typical radii of the most abundant pre-solar nano-diamonds ($1.3-1.5$\,nm).}
\label{fig_ratios_tester}
\centering
\end{figure}

Determining nano-diamond sizes, both in space and in the laboratory, from only their $3-4\,\mu$m spectra will be difficult unless the particle shapes are well-determined. Therefore, an increase in useful information, such as further laboratory data and the modelling of any specific nano-diamond bands at longer wavelengths, is critical for interpreting their spectra with future instruments, such as the James Webb Space Telescope (JWST), that will give us a more complete spectral coverage than currently available. However, it is likely that most longer wavelength (i.e. mid-IR) bands are due to (nitrogen) hetero-atom impurity and structural defects in diamond \cite[e.g.][]{1998A&A...336L..41H,2000A&A...355.1191J,2000M&PS...35...75B} and so may not be particularly size-diagnostic. 

Nevertheless, the spherical nano-diamond data do indicate an apparent convergence in the [CH]/[CH$_2$] ratio (to $\simeq 2.4$) for the larger sizes (see Fig. \ref{fig_ratios_poly_disc}). However, at the smaller (actual) sizes ($a_{\rm nd} \leq 10$\,nm), as shown in Fig.~\ref{fig_ratios_tester}), there is an inherent and unavoidable dispersion in the CH$_n$ group abundance data, which is due to trying to shoehorn a discretised, 3D diamond network lattice into a volume imposed by a given radius.\footnote{In the diamond network calculations here the actual particle radius is determined by the structure discretisation and not by the imposed radius.} The red line in Fig.~\ref{fig_ratios_tester} (and the long-dashed orange line in Fig.~\ref{fig_ratios_poly_disc}) is a by-eye fit to these data with the following function
\begin{equation}
\frac{{\rm [CH]}}{{\rm [CH_2]}} = 2.265 \, a_{\rm nd}^{0.03} - \left( \frac{ 1 }{ 2.5 \, a_{\rm nd}} \right), 
\label{eq_ratio_fit}
\end{equation}
where the nano-diamond radius $a_{\rm nd}$ is in nm.  This equation may thus provide a means to avoid the `ups and downs' in determining the [CH]/[CH$_2$] ratio by exact diamond network calculations (see Fig.~\ref{fig_ratios_tester}), which introduce unpredictable `noise' into the derivation. In essence, Eq.~(\ref{eq_ratio_fit}) therefore provides a sort of statistical averaging of the [CH]/[CH$_2$] ratios and may yield a means of estimating nano-diamond sizes from their IR spectra. Even though Eq.~(\ref{eq_ratio_fit}) is only indicative it does show that, if circumstellar nano-diamonds are quasi-spherical, there may be some hope in estimating the sizes of the larger nano-diamonds ($a_{\rm nd} \gtrsim 2$\,nm) from their [CH]/[CH$_2$] ratios,  albeit with some uncertainty because of the rather flat dependence of [CH]/[CH$_2$] on radius. The utility of such a simple expression clearly rests upon adopting the critically fundamental assumption of spherical nano-diamonds. Unfortunately, given the wide dispersion in their properties, the sizes and shapes of circumstellar (and interstellar?) nano-diamonds with radii $\lesssim 2$\,nm are probably always going to be somewhat poorly constrained by their [CH]/[CH$_2$] ratios. 

The survivability of nano-diamonds in intense radiation fields close to bright stars is determined by their absorption of energetic (UV to EUV) photons versus their ability to shed this absorbed energy via thermal emission, be it through stochastic or thermal equilibrium emission. The nano-diamond surface-to-volume ratio and also the nature of the particle surfaces, in particular the degree of surface hydrogenation, are critical in determining the absolute balance between absorption and emission. In this sense, smaller nano-diamonds have both high surface-to-volume ratios and high surface hydrogen atom fractions and are, perhaps paradoxically, more resilient than their larger relations. While larger, fully hydrogenated nano-diamonds will be more stable than in their de-hydrogenated forms, if they were to lose their surface hydrogen through extreme heating they would undergo surface aromatisation and likely runaway heating leading to their rapid erosion and destruction. 

At large distances from bright stars fully-hydrogenated nano-diamonds of all sizes should be stable against extreme thermal processing. Nevertheless, given that the analysed pre-solar nano-diamonds are the most abundant of all pre-solar grains and have approximately log-normal size distributions peaking at diameters \o\ $\simeq 3$\,nm and extending out to \o\ $\sim 10$\,nm \citep{Lewis_etal_1987,1989Natur.339..117L,1996GeCoA..60.4853D}, it would seem that nano-diamonds with radii as large as 100\,nm are not the norm. Indeed, it appears that the pre-solar nano-diamond sizes fall well within the nano-diamond stability window \citep[\o\ $ = 1.9-5.2$\,nm,][]{2003JChPh.118.5094B}. If large nano-diamonds were common we would surely see signs of them in the pre-solar grains but we do not. A possible explanation for this is that the nano-diamonds are actually formed in the inner regions of circumstellar regions, by some as yet unspecified process,\footnote{As noted by \cite{1996GeCoA..60.4853D}, the pre-solar nano-diamonds were most probably formed via some vapour phase condensation process.} and that they are there size-sorted as a result of thermal processing, which leads to the observed size distribution biased towards smaller nano-diamonds. Indeed we know that the analysed pre-solar nano-diamonds, or at least some fraction of them, must be of extra-solar origin based upon their anomalous Xe isotopic component (Xe-HL), which is considered characteristic of the nucleosynthetic processes in supernovae \citep{Lewis_etal_1987}. This extra-solar origin is also supported by the fact that they exhibit  $^{15}$N depletions and low C/N ratios that are consistent with carbon-rich stellar environments \citep{1997AIPC..402..567A}. 

There has always been something of a question mark over the origin of the anomalous Xe-HL in pre-solar nano-diamonds. It has been proposed that it arises from implantation \citep[e.g.][]{2000LPI....31.1804V}. However, it is hard to understand how this process could lead to the trapping of Xe atoms in such small particles, when it would be expected that the incident heavy, Xe ions would likely traverse the particle rather than be implanted. If small nano-diamonds are, however, formed by the erosion of much larger particles in intense radiation field environments, it is possible that the heavier Xe atoms could be retained in the grain during down-sizing as a result of progressive sublimation. Such an effect would, as required, explain Xe atom trapping in nano-diamonds and also result in a concentration effect that would increase the number of Xe atoms per unit nano-diamond mass. 

Taking the Xe-HL to carbon ratio in nano-diamonds, $f(\frac{\rm Xe}{\rm C})_{\rm nd}^\odot$, to be $1.7 \times 10^{-3}$ of the solar ratio \citep{1993Metic..28..490A}, the solar abundances of xenon, [Xe]$_\odot\ = 1.9 \times 10^{-10}$,  and of carbon, [C]$_\odot\ = 3.2 \times 10^{-4}$ \citep{2014pacs.book...15P},  we can estimate the quantity of cosmic carbon likely tied up in nano-diamonds. Given that the most abundant pre-solar nano-diamonds have radii $\simeq 1.5$\,nm and $\sim 2,500$ carbon atoms per grain ($N_{\rm C,nd} $, see Table \ref {table_sphl_nds}) then the fraction of nano-diamonds that contain a Xe-HL atom, $f_{\rm Xe}$, is
\[
f_{\rm Xe} = \frac{ {\rm [Xe] } }{ {\rm [C] } } f\left(\frac{\rm Xe}{\rm C}\right)_{\rm nd}^\odot \ N_{\rm C,nd}  =  
\frac{  1.9 \times 10^{-10} }{ 3.2 \times 10^{-4} } \times 1.7 \times 10^{-3} \times 2,500  
\]
\begin{equation}
\ \ \ \ \ \ \, =  2.5 \times 10^{-6}, 
\end{equation}
which implies that only one in $400,000$ nano-diamonds actually contains a xenon atom. From this we can conclude that there must, inescapably, be a large reservoir of pre-solar nano-diamonds that are Xe-free but must be associated with the nano-diamonds that do contain  Xe atoms, that is all of these other nano-diamonds are `guilty by association'  and have to come from the same source or sources. 

Expressing this in another way the fraction of solar carbon that must be associated with the Xe-containing pre-solar nano-diamonds, $f_{\rm C,nd}$, can be estimated from the Xe/C ratio in nano-diamonds with respect to the solar Xe/C ratio $f(\frac{\rm Xe}{\rm C})_{\rm nd}^\odot\  (= 0.0017)$ \citep{1993Metic..28..490A}, which is equivalent to $\approx 0.2$\% of the solar carbon abundance. Given that the heavy Xe atom trapping efficiency into nanoparticles, $f_{\rm trap}$, by whatever mechanism, cannot be a very efficient process ($f_{\rm trap} < 1$) the actual fraction of carbon tied up in nano-diamonds in  circumstellar and interstellar media must be significantly larger, that is $f_{\rm C,nd} > 0.2$\%. In their experiments \cite{2001Natur.412..615K} found that the efficiency for He, Ne, Kr, and Xe atom trapping into nano-diamonds is of the order of 10\%, which implies that $f_{\rm C,nd}$ must be at least of the order of 2\%. If the origin of the Xe-HL atoms is by implantation at keV energies, then this inefficiency is compounded by the experimental observation that nano-diamonds with radii less than $\sim 4$\,nm can be completely destroyed during the Xe implantation process \citep{Shiryaev:2018aa}. Consequently the Xe record would be preferentially preserved in the larger size fraction of the pre-solar nano-diamonds. This experimental result is a further indication of the inefficient  trapping of Xe atoms in nano-diamonds. The initial nano-diamond reservoir must therefore have been considerably larger than that indicated by the measured Xe content, implying that $f_{\rm C,nd} > 2$\% or, given that $\simeq 50$\% of carbon is in dust, that $>4$\% of the cosmic carbonaceous dust ought to be in the form of nano-diamonds. Thus, if a significant fraction of them are indeed interstellar, as well as circumstellar, as is an unavoidable conclusion given that a significant fraction were associated with distant supernov\ae\ and have therefore traversed the interstellar medium, then it would seem that they have to be  `multi-talented'\footnote{Nano-diamonds in the ISM would have to express their presence in multiple, but indirect, ways in order to most efficiently use the rather limited supply of carbon available \cite[e.g.][]{2000A&A...355.1191J}, i.e. they would have to contribute to the FUV extinction, the $3-15\,\mu$m IR emission bands, the mid-IR emission, \ldots} and well hidden with(in) other (carbon) dust components in order to have avoided widespread detection in interstellar and circumstellar media.

%||||||||||||||||||||||||||||||||||||||||||||||||||||||||||||||||||||||||||||||||||||||||||||||||||||||||||||||||||||||||||||||||||||||||||||||||||||||||||||||||||
\section{Summary and conclusions}
\label{sect_conclusions}
%||||||||||||||||||||||||||||||||||||||||||||||||||||||||||||||||||||||||||||||||||||||||||||||||||||||||||||||||||||||||||||||||||||||||||||||||||||||||||||||||||

We developed several different approaches to the calculation of the CH and CH$_2$ group abundances on nano-diamonds: regular and semi-regular polyhedral shapes, and diamond bonding networks. As a function of the particle size and shape, and for the different calculation methods, we derived the relative abundance ratio [CH]/[CH$_2$], which can then be weighted by the laboratory-measured IR band intensity ratio in order to interpret and/or predict the observed $3.53\,\mu$m/$3.43\,\mu$m (nano-)diamond IR band ratio. We found that the various methods give good qualitative agreement, within the likely uncertainties.

Overall we found that the [CH]/[CH$_2$] ratio, and therefore the observed IR band intensity ratios, ought to strongly depend upon both the particle size and shape. For a given particle size or shape the [CH]/[CH$_2$] ratio varies over more than an order of magnitude. Thus, it appears that it will be somewhat difficult to constrain the sizes of the observed nano-diamonds solely on the basis of their observed infrared spectra in the $3-4\,\mu$m region. This conclusion remains valid even if we restrict ourselves to the most probable and most stable nano-diamonds forms, that is spherical and the family of cO, tO, and stO (cuboctahedral and truncated octahedral) particles.  

If we, justifiably, make the strong assumption that circumstellar (and interstellar) nano-diamonds are (quasi-)spherical, then there may be some hope in estimating the sizes of nano-diamonds larger than $\simeq 2$\,nm from the observed ratios of the CH and CH$_2$ IR bands at $3.53\,\mu$m and $3.43\,\mu$m, respectively. Although, the uncertainties are still likely to be rather large because of the very flat dependence of their [CH]/[CH$_2$] ratio on radius ($\propto a_{\rm nd}^{0.03}$). If this same statistically-averaged, size-dependent behaviour is assumed to hold for the smallest nano-diamonds ($a_{\rm nd} < 10$\,nm) then we may have some hope estimating their mean sizes, with the critical caveat that the surfaces are fully hydrogenated and/or that de-hydrogenation processes do not affect the surface CH and CH$_2$ groups abundance ratios.   

Further laboratory and modelling data of the longer wavelength (mid-IR) bands specific to nano-diamonds is therefore essential for the coming JWST era if we are to understand and extract the maximum amount of data from circumstellar nano-diamond spectra.  With these new spectroscopic data, we will perhaps discover nano-diamonds in new astronomical objects, other than circumstellar discs regions, and that they even exist in the interstellar medium.  Although, and as a caveat, it is probable that as interesting as these bands will be they may not be particularly size-specific because they are predominantly, and perhaps exclusively, due to impurities such as nitrogen and to structural defects in the diamond lattice.

As something of an aside, based upon their hetero-atom Xe content, it is here speculated that nano-diamonds may actually be quite abundant in the ISM; of the order of several percent of cosmic carbon could be in the form of nano-diamonds. However, given that they have not yet been detected there they must be well-hidden with(in) other dust components, either that or we do not yet recognise their spectroscopic signatures in the ISM because they are confused with other carbonaceous dust features.

Interestingly, this modelling indicates that spherical nano-diamonds exhibit a narrower range in [CH]/[CH$_2$] than shown by the regular and semi-regular polyhedral forms and also by the polyhedral nano-diamond network models. Further, spherical nano-diamonds  exhibit [CH]/[CH$_2$] ratios approximating those of the tetrahedral polyhedral and network particles, as has already been speculated, that is the ratios for large diamondoid molecules or nano-diamonds resemble those of the $T_d$ point group, the largest subgroup of the $O_h$ point group.

%||||||||||||||||||||||||||||||||||||||||||||||||||||||||||||||||||||||||||||||||||||||||||||||||||||||||||||||||||||||||||||||||||||||||||||||||||||||||||||||||||
\begin{acknowledgements}
The author particularly wishes to thank Nathalie Ysard for a very careful reading of the manuscript and an extremely thorough and painstaking verification of the equations. \\
The author is especially grateful to the anonymous referee for the critical suggestions that led to a fundamental clarification of the presentation.\\ \\ 
This work is dedicated to Keith, my long-suffering brother of more than 63 years. Taken from us too soon he will remain forever in our hearts and minds. \\ Keith Edward Jones  ( 25$^{th}$ February 1957 $-$  29$^{th}$ October 2020 ) 
\end{acknowledgements}
%||||||||||||||||||||||||||||||||||||||||||||||||||||||||||||||||||||||||||||||||||||||||||||||||||||||||||||||||||||||||||||||||||||||||||||||||||||||||||||||||||

%||||||||||||||||||||||||||||||||||||||||||||||||||||||||||||||||||||||||||||||||||||||||||||||||||||||||||||||||||||||||||||||||||||||||||||||||||||||||||||||||||
\bibliographystyle{bibtex/aa} % style aa.bst
\bibliography{../Ant_bibliography} 
%||||||||||||||||||||||||||||||||||||||||||||||||||||||||||||||||||||||||||||||||||||||||||||||||||||||||||||||||||||||||||||||||||||||||||||||||||||||||||||||||||

\appendix

%||||||||||||||||||||||||||||||||||||||||||||||||||||||||||||||||||||||||||||||||||||||||||||||||||||||||||||||||||||||||||||||||||||||||||||||||||||||||||||||||||
\section{Regular polyhedral particles}
\label{app_regulars}
%||||||||||||||||||||||||||||||||||||||||||||||||||||||||||||||||||||||||||||||||||||||||||||||||||||||||||||||||||||||||||||||||||||||||||||||||||||||||||||||||||

Here we consider all of the relevant equations pertaining to regular and regular truncated polyhedral particles where all particle edges are of length, $l$. Following the modelling of regular and semi-regular polyhedra and their truncated forms presented in this work we propose the following rules, which generalise the observations presented in Section~\ref{sect_shape}, for determining the CH$_n$ group terminations of nano-diamond polyhedral vertices, edges, and facets: 
\begin{itemize}
\item Vertices (V) 
   \begin{itemize}
      \item with an even number of intersecting facets are CH$_2$, 
      \item with an odd number of intersecting \{111\} facets are CH, 
      \item with one or more intersecting \{100\} facets are CH$_2$, 
   \end{itemize}
\item Edges (E)
   \begin{itemize}
      \item between facets of the same type are CH$_2$, 
      \item of triangular facets are CH$_2$, 
      \item between a square and a hexagonal facet are CH, 
   \end{itemize}
\item Facets (F)
   \begin{itemize}
      \item with three and six sides are \{111\}, 
      \item with four and eight sides are \{100\},    
      \item \{111\} are covered in coherently-directed CH bonds and 
      \item \{100\} are CH$_2$ covered. 
   \end{itemize}
\end{itemize}
These rules can be used to determine the CH$_n$ group structures comprising the surfaces of polyhedral nano-diamonds.

%||||||||||||||||||||||||||||||||||||||||||||||||||||||||||||||||||||||||||||||||||||||||||||||||||||||||||||||||||||||||||||||||||||||||||||||||||||||||||||||||||
\subsection{ Regular tetrahedral (T) particles}
%||||||||||||||||||||||||||||||||||||||||||||||||||||||||||||||||||||||||||||||||||||||||||||||||||||||||||||||||||||||||||||||||||||||||||||||||||||||||||||||||||

We first describe the properties of regular tetrahedral particles, the simplest regular polyhedra, where the total edge length is 6\,$l$ (see Fig. \ref{fig_shapes} and Tables \ref{table_struct_summary} and \ref{table_struct_heirarchy}). An encompassing sphere,  that is one that is equally arrayed around and intersects all four of the particle vertices, is of radius $r = \surd 6/4 \, l = 0.612 \, l$. The surface area of a tetrahedron, with four equivalent, triangular \{111\} facets, each of area $\surd 3/4 \, l^2$, is:  
\begin{equation}
  A_{\rm T}  =  \surd 3  \, l^2  = \surd 3 \, \left( \frac{4}{\surd 6} \right)^2  r^2 =  \frac{8\surd 3}{3}  r^2 = 4.619 \, r^2. 
\end{equation}
Given that the entire surface area is in \{111\} facets, $f_{\rm s \{111\}} = 1$, which does not include the four CH-terminated vertices. All edges are distinct in structure from the faces, are CH$_2$-terminated and can be considered to be of \{100\}-type with a \{100\} edge length $E_{\{100\}} = 6 \, l = 4 \surd 6 \, r = 9.798 \, r$.  The ratio of the total edge length, $E_{\rm T}$, to the total surface area of such a particle is
\begin{equation} 
 \frac{ E_{\rm T} }{ A_{\rm T} } =  \frac{6 \, l}{ \surd 3 \, l^2} =  \frac{ 2 \surd 3 }{ l } =  \frac{ 3 \surd 2 }{ 2 \, r } =  \frac{ 2.121 }{ r } 
\end{equation} 
and the ratio of the surface area of a tetrahedron to that of its encompassing sphere is 
\begin{equation} 
  \frac{ A_{\rm T} }{ A_{\rm sphere} } =  \frac{ 8 \surd 3 / 3 \, r^2 }{ \, 4 \pi r^2 } = \frac{ 2 }{ \pi \surd 3} 
  =  0.368. 
  \end{equation}
 The volume of a tetrahedron is given by:
\begin{equation}
  V_{\rm T}  =   \frac{\surd 2}{12}  \, l^3  =  \frac{\surd 2}{12} \left( \frac{4}{\surd 6} \right)^3 \, r^3  =  \frac{ 8 }{ 9 \surd 3 } \, r^3 =  0.513 \, r^3  
\end{equation}
and the ratio of the volume of a tetrahedron to that of its encompassing sphere is: 
\begin{equation}
  \frac{ V_{\rm T} }{ V_{\rm sphere} }  =   =  \frac{[8 \surd 3 / 9] \, r^3 }{ (4/3) \pi r^3  } = \frac{ 2 }{ \pi \, 3 \surd 3} =  0.123. 
\end{equation}
Hence, and because for convenience we usually define the particle mass in terms of a radius, the volume of a tetrahedron, with circumscribed sphere of effective radius $r_{\rm eff}$, must be normalised to that of a spherical particle of the same volume $ V_{\rm T}(r_{\rm eff})$, that is the radius, $a_{\rm nd}$, of a "spherical nano-diamond" is defined by
\begin{equation}
\frac{4}{3} \pi a_{\rm nd}^3 =   \frac{ 8 }{ 9 \surd 3 } \, r_{\rm eff}^3  
\end{equation}
and hence 
\begin{equation}
  r_{\rm eff}  =  \left( \frac{  \pi \, 3 \surd 3 }{ 2 } \right)^\frac{1}{3}  a_{\rm nd}  =  2.103  \, a_{\rm nd}. 
\end{equation}
We note that $r_{\rm eff}$ is greater than $a_{\rm nd}$ because any polyhedron is, individually, less efficient at space filling than a sphere.

%||||||||||||||||||||||||||||||||||||||||||||||||||||||||||||||||||||||||||||||||||||||||||||||||||||||||||||||||||||||||||||||||||||||||||||||||||||||||||||||||||
\subsection{Regular truncated tetrahedral (tT) particles}
%||||||||||||||||||||||||||||||||||||||||||||||||||||||||||||||||||||||||||||||||||||||||||||||||||||||||||||||||||||||||||||||||||||||||||||||||||||||||||||||||||

In the same manner as for tetrahedral particles we now look to the equivalent properties of regular truncated tetrahedral particles, that is tetrahedral particles with the four vertices truncated into equilateral triangular faces. In this case the total edge length is 18\,$l$ (see Fig. \ref{fig_shapes} and Tables \ref{table_struct_summary} and \ref{table_struct_heirarchy}) and the  encompassing sphere radius $r = \surd 22/4 \, l = 1.173 \, l$. The surface area of a truncated tetrahedron, with four equivalent, triangular \{111\} facets (each of area $\surd 3/4 \, l^2$), and four hexagonal \{111\} facets each of area $3\surd 3/2 \, l^2$, is:  
\begin{equation}
  A_{\rm tT}  = 7 \surd 3 \, l^2   = 7 \surd 3 \, \left( \frac{4}{\surd 22} \right)^2  r^2 =  \frac{ 56 \surd 3 }{ 11 }  r^2 = 8.818 \, r^2. 
\end{equation}
Given that the entire surface area is in triangular and hexagonal \{111\} facets, $f_{\rm s \{111\}} = 1$. 
However, unlike tetrahedra, not all edges are \{100\}-type, twelve are CH-terminated and therefore an integral part of the adjacent \{111\} facets. Thus, only six of the truncated form edges are of \{100\} CH$_2$-type and $E_{\{100\}} = 6 \, l = ( 24 / \surd 22 ) \, r = 5.117\, r$.  The ratio of the total edge length, $E_{\rm tT}$, to the total surface area of such a particle is
\begin{equation} 
 \frac{ E_{\rm tT} }{ A_{\rm tT} } =  \frac{18 \, l}{ 7 \surd 3 \, l^2} =  \frac{ 6 \surd 3 }{ 7 l } =  \frac{ 3 \surd 3 \surd 22 }{ 14 \, r } =  \frac{ 1.741 }{ r } 
\end{equation} 
and the ratio of the surface area to that of the encompassing sphere is 
\begin{equation} 
  \frac{ A_{\rm tT} }{ A_{\rm sphere} } =  \frac{ 56 \surd 3 / 11 \, r^2 }{ 4 \pi r^2 } = \frac{ 14 \surd 3}{ 11 \pi}   =  0.702. 
  \end{equation}
 The volume of a truncated tetrahedron is given by:  
\begin{equation}
 V_{\rm tT.}  =   \frac{ 23 \surd 2}{12}  \, l^3  =  \frac{ 23 \surd 2}{12} \left( \frac{4}{\surd 22} \right)^3 \, r^3   
  =  \frac{ 184 }{ 33 \surd 11 } \, r^3 =  1.681 \, r^3  
\end{equation}
 and the ratio of the volume of a truncated tetrahedron to that of its encompassing sphere is: 
\begin{equation}
  \frac{ V_{\rm tT} }{ V_{\rm sphere} }  =   \frac{[ 184 / (33 \surd 11) ] \, r^3 }{ (4/3) \pi r^3  } = \frac{ 46 }{ \pi 11 \surd 11 } =  0.401. 
\end{equation}
Normalising to a spherical particle of the same volume $ V_{\rm tetra.}(r_{\rm eff})$, the radius, $a_{\rm nd}$, of the equivalent "spherical nano-diamond" is  from 
\begin{equation}
\frac{4}{3} \pi a_{\rm nd}^3 =   \frac{ 184  }{ 33 \surd 11 } \, r_{\rm eff}^3  
\end{equation}
whence  
\begin{equation}
  r_{\rm eff}  =  \left( \frac{ \pi 11 \surd 11 }{ 46 } \right)^\frac{1}{3}  a_{\rm nd}  =  1.356  \, a_{\rm nd}. 
\end{equation}

%||||||||||||||||||||||||||||||||||||||||||||||||||||||||||||||||||||||||||||||||||||||||||||||||||||||||||||||||||||||||||||||||||||||||||||||||||||||||||||||||||
\subsection{Regular octahedral (O) particles}
%||||||||||||||||||||||||||||||||||||||||||||||||||||||||||||||||||||||||||||||||||||||||||||||||||||||||||||||||||||||||||||||||||||||||||||||||||||||||||||||||||

We now consider the properties of regular octahedral particles, with a total edge length of 12\,$l$ (see Fig. \ref{fig_shapes} and Tables \ref{table_struct_summary} and \ref{table_struct_heirarchy}) and an encompassing sphere of radius $r = \surd 2 /2 \, l = 0.707 \, l$. The surface area of an octahedron, which exhibits eight equivalent triangular \{111\} facets is:  
\begin{equation}
  A_{\rm O}  =  8 \frac{\surd 3}{4}  \, l^2  = 2 \surd 3 \, \left( \frac{2}{\surd 2} \right)^2  r^2 =  4 \surd 3 \, r^2 = 6.928 \, r^2. 
\end{equation}
As per tetrahedral particles, the surface of octahedral particles is entirely in triangular \{111\} facets, that is $f_{\rm s \{111\}} = 1$. However, in this case, all the \{111\}/\{111\} edges are comprised of alternately-facing CH bonds forming an integral part of those adjacent facets.\footnote{For regular octahedral particles there are only six CH$_2$ groups present, one to be found on each of the particle vertices.} Hence, $E_{\{100\}} = 0$ and the total edge length $E_{\rm O} = 12 \, l = 12 \surd 2 \, r = 16.971 \, r$. The ratio of the total edge length, $E_{\rm O}$, to the total surface area of such a particle is
\begin{equation} 
 \frac{ E_{\rm O} }{ A_{\rm O} } =  \frac{12 \, l}{ 2 \surd 3 \, l^2} =  \frac{ 2 \surd 3 }{ l } =  \frac{ 2 \surd 3 \surd 2 }{ 2 r }=  \frac{ \surd 6 }{ r } =  \frac{ 2.450 }{ r }. 
\end{equation} 
The ratio of the surface area to that of its encompassing sphere is  
\begin{equation} 
  \frac{ A_{\rm O} }{ A_{\rm sphere} } =  \frac{4 \surd 3 \, r^2}{4 \pi r^2} =  \frac{\surd 3 }{\pi} = 0.551. 
  \end{equation}
 The volume of an octahedron is 
\begin{equation}
  V_{\rm O}  =   \frac{ \surd 2}{3} \, l^3 =  \frac{ \surd 2}{3} \left( \frac{2}{\surd 2} \right)^3 \, r^3 = \frac{4}{3}  \, r^3 = 1.333  \, r^3 
\end{equation}
and the ratio of its volume to that of the encompassing sphere is: 
\begin{equation}
  \frac{ V_{\rm O} }{ V_{\rm sphere} }  =   \frac{(4/3) r^3}{(4/3) \pi r^3} = \frac{1}{\pi} = 0.318.
\end{equation}
Normalising the volume to that of a sphere of radius, $a_{\rm nd}$ we have
\begin{equation}
\frac{4}{3} \pi a_{\rm nd}^3 =   \frac{4}{3}  \, r_{\rm eff}^3 
\end{equation}
and  
\begin{equation}
  r_{\rm eff}  =  \pi^\frac{1}{3} \, a_{\rm nd} =  1.465 \, a_{\rm nd}.  
\end{equation}

%||||||||||||||||||||||||||||||||||||||||||||||||||||||||||||||||||||||||||||||||||||||||||||||||||||||||||||||||||||||||||||||||||||||||||||||||||||||||||||||||||
\subsection{Regular truncated octahedral (tO) particles}
%||||||||||||||||||||||||||||||||||||||||||||||||||||||||||||||||||||||||||||||||||||||||||||||||||||||||||||||||||||||||||||||||||||||||||||||||||||||||||||||||||

Our attention now turns to a common nano-diamond particle shape, a regular truncated octahedron \cite[e.g.][]{Barnard:2005dt}, an octahedron with its six vertices truncated into square \{100\} facets, with a total edge length of 36\,$l$ (see Fig. \ref{fig_shapes} and Tables \ref{table_struct_summary} and \ref{table_struct_heirarchy}) and an encompassing sphere of radius $r = \surd(5/2) \, l = 1.581 \, l$. The surface of a truncated octahedron is comprised of eight hexagonal \{111\} facets (each of area $3 \surd 3 / 2 \, l^2$) and six square \{100\} facets (each of area $l^2$) is:  
\[
A_{\rm tO}  =  8 \left( \frac{3 \surd 3 }{2} \right) \, l^2 + 6 \, l^2   =  6 \, (2 \surd 3  + 1 ) \, l^2  
\]
\begin{equation}
\ \ \ \ \ \ \ \ \ \ \ \ \  =  \frac{ 12 \, (2 \surd 3  + 1 ) }{ 5 } \, r^2
=  10.714 \, r^2. 
\end{equation}
Here the surface is in hexagonal \{111\} facets, with $f_{\rm s \{111\}} = 0.776$,  and square \{100\} facets, with $f_{\rm s \{100\}} = 0.224$.  
In truncated octahedral particles both \{111\}/\{111\} and \{111\}/\{100\} edges exist (see Fig. \ref{fig_shapes}).  \{111\}/\{111\} edges are alternating CH-terminated and an integral part of \{111\} facets, while \{111\}/\{100\} edges are CH$_2$-terminated and form part of the \{100\} facets.  
The ratio of the total edge length, $E_{\rm tO} = 36 \, l = 36 \surd (2/5) \, r = 22.768 \, r$, to the total surface area of the particle is
\begin{equation} 
 \frac{ E_{\rm tO} }{ A_{\rm tO} } =  \frac{36 \, l}{ 6 \, (2 \surd 3  + 1 ) \, l^2} = \frac{6}{(2 \surd 3  + 1 ) l}  =  \frac{ 3 \surd 2 \surd 5 }{(2 \surd 3  + 1 ) r} = \frac{ 2.125 }{ r }. 
\end{equation} 
The ratio of the surface area to that of an encompassing sphere is 
\begin{equation} 
  \frac{ A_{\rm tO} }{ A_{\rm sphere} } =  \frac{ (12/5) ( 2 \surd 3 + 1 ) \, r^2 }{ \, 4 \pi r^2 } 
  =  \frac{ 3 ( 2 \surd 3 + 1 ) }{ \, 5 \pi } =  0.853. 
\end{equation}
 The volume of a truncated octahedron is given by:
\begin{equation}
  V_{\rm tO}  =  8 \surd 2 \, l^3  =  8 \surd 2 \, [ \surd( 2 / 5 ) \, r ]^3  =  \frac{ 32 }{ 5 \surd5} \, r^3  = 2.862 \, r^3
\end{equation}
and its volume with respect to that of an encompassing sphere of the same radius $r$ is: 
\begin{equation}
  \frac{ V_{\rm tO} }{ V_{\rm sphere} }  =  \frac{[32/(5 \surd 5 )] \, r^3 }{ (4/3) \pi r^3  } = \frac{24}{\pi 5 \surd 5} =  0.683. 
\end{equation}
Normalising to the volume of a sphere with radius $a_{\rm nd}$ we have
\begin{equation}
\frac{4}{3} \pi a_{\rm nd}^3 =  \frac{ 32 }{ 5 \surd5} \ r_{\rm eff}^3  
\end{equation}
and
\begin{equation}
  r_{\rm eff}  =  \left( \frac{ \pi 5 \surd 5 }{ 24 } \right)^\frac{1}{3}  a_{\rm nd}  =  1.135  \, a_{\rm nd}. 
\end{equation}

%||||||||||||||||||||||||||||||||||||||||||||||||||||||||||||||||||||||||||||||||||||||||||||||||||||||||||||||||||||||||||||||||||||||||||||||||||||||||||||||||||
\subsection{Regular cuboctahedral (cO) particles}
%||||||||||||||||||||||||||||||||||||||||||||||||||||||||||||||||||||||||||||||||||||||||||||||||||||||||||||||||||||||||||||||||||||||||||||||||||||||||||||||||||

Another common nano-diamond shape is the cuboctahedron with a total edge length of 24\,$l$ (see Fig. \ref{fig_shapes} and Tables \ref{table_struct_summary} and \ref{table_struct_heirarchy}) and an encompassing sphere of radius $r = l$. The surface area of a cuboctahedron, which exhibits eight triangular \{111\} facets  (each of area $ [\surd 3/4] l^2$) and six square facets (each of area $l^2$) is:  
\begin{equation}
  A_{\rm cO}  =  8 \frac{\surd 3}{4}  \, l^2  + 6 l^2 = 2 (\surd 3 + 3 ) \, l^2 = 2 ( \surd 3 + 3 ) \, r^2 =  9.464 \, r^2. 
\end{equation}
In cuboctahedral particles the surface is in triangular \{111\} facets, with $f_{\rm s \{111\}} = 0.366$,  and square \{100\} facets, with $f_{\rm s \{100\}} = 0.634$.  
All edges are \{111\}/\{100\} and equivalent (see Fig. \ref{fig_shapes}) and their CH$_n$ groups form an integral parts of their adjacent \{111\} and \{100\} facets. 
The ratio of the total edge length, $E_{\rm cO} = 24 \, l = 24 \, r$, to the total surface area of such a particle is
\begin{equation} 
 \frac{ E_{\rm cO} }{ A_{\rm cO} } =  \frac{24 \, l}{ 2 ( \surd 3  + 3 ) \, l^2} = \frac{12}{( \surd 3  + 3 ) l }  =  \frac{12}{( \surd 3  + 3 ) r } = \frac{ 2.536 }{ r }. 
\end{equation} 
The ratio of the surface area of a cuboctahedron to that of its encompassing sphere is 
\begin{equation} 
  \frac{ A_{\rm cO} }{ A_{\rm sphere} } =  \frac{ 2 ( \surd 3 + 3 ) \, r^2 }{ \, 4 \pi r^2 } =  \frac{ ( \surd 3 + 3 ) }{ \, 2 \pi } =  0.753. 
 \end{equation}
 The volume of a cuboctahedron is given by:
\begin{equation}
  V_{\rm cO}  =   \frac{ 5 \surd 2 }{ 3 } \, l^3 =  \frac{ 5 \surd 2 }{ 3 } \, r^3 = 2.357 \, r^3
\end{equation}
and the ratio of its volume to that of the encompasing sphere of the same radius $r$ is: 
\begin{equation}
  \frac{ V_{\rm cO} }{ V_{\rm sphere} }  =   \frac{ 5 \surd 2 / 3  \, r^3 }{ (4/3) \pi r^3 } = \frac{ 5 \surd 2 }{ 4  \pi } = 0.563
\end{equation}
and normalising to a sphere of radius $a_{\rm nd}$ 
\begin{equation}
\frac{4}{3} \pi a_{\rm nd}^3 = \frac{ 5 \surd 2 }{ 3 } \, r_{\rm eff}^3 
\end{equation}
the effective radius ($r_{\rm eff}$) of a cuboctahedron with the same volume as a sphere of radius $a_{\rm nd}$ is then 
\begin{equation}
  r_{\rm eff}  =  \left( \frac{ 4 \pi }{ 5 \surd 2 } \right)^\frac{1}{3}  a_{\rm nd}  =  1.211  \, a_{\rm nd}. 
\end{equation}

%||||||||||||||||||||||||||||||||||||||||||||||||||||||||||||||||||||||||||||||||||||||||||||||||||||||||||||||||||||||||||||||||||||||||||||||||||||||||||||||||||
\subsection{Regular cube (C) particles}
%||||||||||||||||||||||||||||||||||||||||||||||||||||||||||||||||||||||||||||||||||||||||||||||||||||||||||||||||||||||||||||||||||||||||||||||||||||||||||||||||||

For geometrical completeness we here describe the properties of a regular cube, with a total edge length of 12\,$l$ (see Fig. \ref{fig_shapes} and Tables \ref{table_struct_summary} and \ref{table_struct_heirarchy}) and an encompassing sphere of radius $r = \surd 3 / 2 \, l = 0.866 \, l$. The surface area of a cube, comprised of six square \{100\} facets of area $l^2$, is:  
\begin{equation}
A_{\rm C}  =  6 \, l^2 = 8 \, r^2.  
\end{equation}
The ratio of the total edge length, $E_{\rm C} = 12 \, l = 24 \surd 3 \, r$, to the total surface area is
\begin{equation} 
 \frac{ E_{\rm C} }{ A_{\rm C} } =  \frac{12 \, l}{  6 \, l^2} = \frac{2}{l} = \frac{ \surd 3 }{ r } = \frac{1.732}{r}. 
 \end{equation} 
The ratio of the surface area of a cube to that of its encompassing sphere is 
\begin{equation}
  \frac{ A_{\rm C} }{ A_{\rm sphere} } =  \frac{ 8 \, r^2 }{ \, 4 \pi r^2 } = \frac{2}{\pi} \, r = 0.637 \, r. 
\end{equation}
 The volume of the cube ($l^3 = 1.540 \, r^3$) with respect to that of an encompassing sphere of the same radius $r$ is: 
\begin{equation}
\frac{ V_{\rm C} }{ V_{\rm sphere} }  =  \frac{ l^3 }{ (4/3) \pi r^3  } = \frac{ 8 / ( 3 \surd 3 ) \, r^3 }{ (4/3) \pi r^3  } = \frac{ 2 }{ \pi \surd 3  } = 0.368. 
\end{equation}
Normalising to the volume of a sphere with radius $a_{\rm nd}$ we have
\begin{equation}
\frac{4}{3} \pi a_{\rm nd}^3 =  \frac{ 8 }{ 3 \surd 3 } \ r_{\rm eff}^3  
\end{equation}
and
\begin{equation}
  r_{\rm eff}  =  \left( \frac{ \pi \surd 3 }{ 2 } \right)^\frac{1}{3} a_{\rm nd}  =  1.396  \, a_{\rm nd}. 
\end{equation}

%||||||||||||||||||||||||||||||||||||||||||||||||||||||||||||||||||||||||||||||||||||||||||||||||||||||||||||||||||||||||||||||||||||||||||||||||||||||||||||||||||
\subsection{Regular truncated cube (tC) particles}
%||||||||||||||||||||||||||||||||||||||||||||||||||||||||||||||||||||||||||||||||||||||||||||||||||||||||||||||||||||||||||||||||||||||||||||||||||||||||||||||||||

In order to be fully complete we now describe a regular truncated cube, a cube with its eight vertices truncated into triangular facets, with a total edge length of 36\,$l$ (see Fig. \ref{fig_shapes} and Tables \ref{table_struct_summary} and \ref{table_struct_heirarchy}) and an encompassing sphere of radius $r = \frac{1}{2} \surd( 7 + 4 \surd 2 ) \, l = 1.779 \, l$. The surface of a truncated cube is thus comprised of eight triangular \{111\} facets (each of area $\surd 3 / 4 \, l^2$) and six octagonal \{100\} facets (each of area $ 2[ \surd 2 + 1] \, l^2$) and its surface area is:  
\[
A_{\rm tC}  =  8  \frac{\surd 3 }{4} \, l^2 + 6 [ 2 ( \surd 2 + 1 ) ] \, l^2   = 2 [ 6 \, ( \surd 2  + 1 ) + \surd 3 ] \, l^2  
\]
\begin{equation}
\ \ \ \ \ \ \ \ \ \ \ \ \ \ \  =  \frac{ 8 [ 6 \, ( \surd 2  + 1 ) + \surd 3 ] }{  7 + 4 \surd 2 } \, r^2 =  10.251 \, r^2. 
\end{equation}
Here the surface is in triangular \{111\} facets, with $f_{\rm s \{111\}} = 0.107$,  and octagonal \{100\} facets, with $f_{\rm s \{100\}} = 0.893$.  
The edges of truncated cube particles are \{111\}/\{100\} and \{100\}/\{100\} and their CH$_n$ groupings an integral part of their adjacent facets. The ratio of the total edge length, $E_{\rm tC} = 36 \, l = 72 / (\surd [ 7 + 4 \surd 2]) \, r$, to the total surface area of such a particle is
\[
 \frac{ E_{\rm tC} }{ A_{\rm tC} } =  \frac{36 \, l}{ 2 [ 6 \, ( \surd 2  + 1 ) + \surd 3 ]  \, l^2} = \frac{3}{[( \surd 2  + 1 ) + \surd 3/6] \, l }  
 \]
\begin{equation} 
\ \ \ \ \ \ \ \ \ \ \ \ \ \ \  =  \frac{ 3 \surd ( 7 + 4 \surd 2 ) }{ 2 [ ( \surd 2  + 1 ) + \surd 3/6 ]\, r} = \frac{ 2.126 }{ r }. 
\end{equation} 
The ratio of the surface area of a truncated cube to that of its encompassing sphere is 
\[
  \frac{ A_{\rm tC} }{ A_{\rm sphere} } =  \frac{ 8 [ 6 \, ( \surd 2  + 1 ) + \surd 3 ] / ( 7 + 4 \surd 2 ) \, r^2 }{ \, 4 \pi r^2 } 
  \]
\begin{equation}
\ \ \ \ \ \ \ \ \ \ \ \ \ \ \  =  \frac{ 2 [ 6 \, ( \surd 2  + 1 ) + \surd 3 ] }{ \pi ( 7 + 4 \surd 2 ) } 
= 0.816. 
\end{equation}
 The volume of the truncated cube is given by:
\begin{equation}
  V_{\rm tC}  =  \frac{ 7 ( 3 + 2 \surd 2 ) }{ 3 } \, l^3  =  \frac{ 56 ( 3 + 2 \surd 2 ) }{ 3 [\surd( 7 + 4 \surd 2 )]^3 } r ^3  = 2.416 \, r^3   
\end{equation}
and its volume with respect to that of an encompassing sphere of the same radius $r$ is: 
\[
\frac{ V_{\rm tC} }{ V_{\rm sphere} }  =  \frac{ 56 ( 3 + 2 \surd 2 ) / \{ 3 [\surd( 7 + 4 \surd 2 )]^3 \} \, r^3 }{ (4/3) \pi r^3  } 
\]
\begin{equation}
\ \ \ \ \ \ \ \ \ \ \ \ \ \ \  = \frac{ 14 ( 3 + 2 \surd 2 ) }{ \pi [\surd( 7 + 4 \surd 2 )]^3 } =  0.577. 
\end{equation}
Normalising to the volume of a sphere with radius $a_{\rm nd}$ we have
\begin{equation}
\frac{4}{3} \pi a_{\rm nd}^3 =  \frac{ 56 ( 3 + 2 \surd 2 ) }{ 3 [\surd( 7 + 4 \surd 2 )]^3 } \ r_{\rm eff}^3  
\end{equation}
and
\begin{equation}
  r_{\rm eff}  =  \frac{ \surd (7 + 4 \surd 2 ) \, \pi^{\frac{1}{3}} }{ [ 14 ( 3 + \surd 2 ) ]^\frac{1}{3} } a_{\rm nd}  =  1.201  \, a_{\rm nd}. 
\end{equation}

%||||||||||||||||||||||||||||||||||||||||||||||||||||||||||||||||||||||||||||||||||||||||||||||||||||||||||||||||||||||||||||||||||||||||||||||||||||||||||||||||||
\subsection{The spatial properties of regular polyhedra}
%||||||||||||||||||||||||||||||||||||||||||||||||||||||||||||||||||||||||||||||||||||||||||||||||||||||||||||||||||||||||||||||||||||||||||||||||||||||||||||||||||

% FIGURE 
\begin{figure}[h]
\centering
\includegraphics[width=9.5cm]{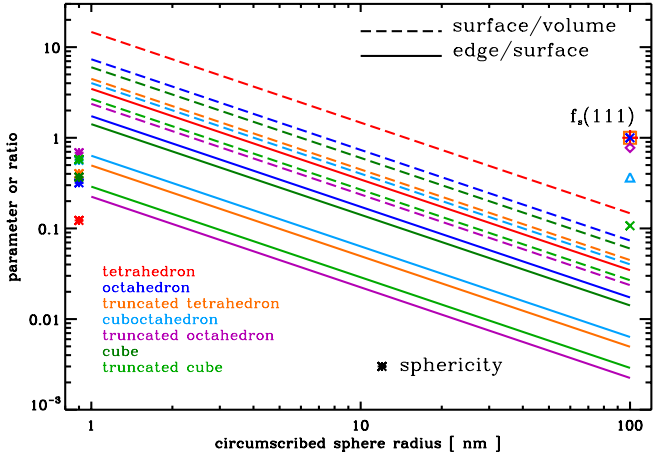}
\caption{Spatial properties of regular polyhedra. The coloured dashed lines show the surface area to volume ratios of T (red), tT (orange), O (blue), tO (purple), cO (cobalt), C (dark green), and tC (green) particle forms, and the solid lines show the edge/surface area ratio. The asterisk symbols on the far left show their `sphericities' and the different coloured symbols on the right indicate the fraction of the particle surface in \{111\} facets, which is zero for a cube (C).}
\label{fig_3D_geo}
\centering
\end{figure}

In the preceding sub-sections we described the surface and volume properties of known and other possible nano-diamond polyhedral forms. Fig. \ref{fig_3D_geo} shows the surface area to volume ratios of T (red), tT (orange), O (blue), tO (purple), and cO (cobalt) particle forms (dashed lines ), and their edge/surface area ratios (solid lines). Also shown are their `sphericities' (asterisks on the left), the ratio of the volume of the polyhedron to that of its circumscribed sphere, and the surface fraction in \{111\} facets, $f_{s\{111\}}$ (different shaped and coloured symbols on the right, these data also appear in Fig. \ref{fig_shapes}).

Given that the primary motivation for this study is to better understand the [CH]/[CH$_2$] abundance ratio on nano-diamond surfaces and the associated $3.53\,\mu$m(CH)/$3.43\,\mu$m(CH$_2$) IR band intensity ratio, perhaps the most interesting result in Fig. \ref{fig_3D_geo} is the wide variation in the CH-covered, \{111\} facet surface fraction, $f_{s\{111\}}$. For each of the seven regular polyhedra considered in this study $f_{s\{111\}}$ is independent of size and spreads over almost an order of magnitude ($0.107-1.000$). Nevertheless, $f_{s\{111\}}$ is not a good indication of the surface CH$_n$ composition because it does not take account disproportionate edge effects. 

Other things to note in Fig.~\ref{fig_3D_geo} are that: the particle `sphericity' roughly increases with the number of polyhedral faces, that is particle complexity, and the surface-to-volume and edge-to-surface ratios decrease linearly in a log-log plot with increasing size. 
%and the [CH]/[CH$_2$] ratio generally seems to follow a power law dependency for any given form. 

%||||||||||||||||||||||||||||||||||||||||||||||||||||||||||||||||||||||||||||||||||||||||||||||||||||||||||||||||||||||||||||||||||||||||||||||||||||||||||||||||||
\section{Semi-regular polyhedral particles}
\label{app_irregulars}
%||||||||||||||||||||||||||||||||||||||||||||||||||||||||||||||||||||||||||||||||||||||||||||||||||||||||||||||||||||||||||||||||||||||||||||||||||||||||||||||||||

We now consider the more complex case of semi-regularly truncated polyhedra with truncated facets that are of arbitrary size, which remain parallel to those of the regularly truncated parent polyhedron. In this case the expressions are necessarily more cumbersome and cannot be reduced to relatively straight-forward expressions of, $r$, the radius of the sphere that circumscribes the particle and includes all its vertices.\footnote{However, all particles can still be circumscribed by an all-vertex encompassing sphere of radius $r$.} The truncated facet edges are assumed to be of arbitrary length $a$, which implies that the remnant edge, $L$, of the regular polyhedron parent is reduced from $l$ to $(l-2a)$ as illustrated in Figs,~\ref{fig_trunc_tetr} to \ref{fig_trunc_cube}, that is $L = (l-2a)$. 

As will become clear in the following sections, for the expressions to hold in the case of each of the three truncated polyhedra that we consider, the following condition must hold 
\begin{equation} 
\frac{a}{l} \leqslant \frac{1}{2},
\end{equation}
that is the parent polyhedron edge length is maximally-truncatable at its centre and we note that when 
\[
a = 0 \ \ \ \ (\equiv  L= l) \ \ \, \rightarrow \ {\rm a\ regular\ parent\ polyhedron},
\]
\[
a= l/3 \ \ (\equiv L = a ) \  \rightarrow  \ {\rm a\ regular\ truncated\ parent\ polyhedron,} 
\]
\begin{equation} 
a = l/2 \ \ (\equiv  L = 0) \ \rightarrow \ {\rm a\ different\ polyhedron}.
\end{equation}
In the $L=0$ case the truncated tetrahedron solution is an octahedron, and for a truncated octahedron or cube the solution is a cuboctahedron.

%||||||||||||||||||||||||||||||||||||||||||||||||||||||||||||||||||||||||||||||||||||||||||||||||||||||||||||||||||||||||||||||||||||||||||||||||||||||||||||||||||
\subsection{Semi-regular truncated tetrahedral (stT) particles}
%||||||||||||||||||||||||||||||||||||||||||||||||||||||||||||||||||||||||||||||||||||||||||||||||||||||||||||||||||||||||||||||||||||||||||||||||||||||||||||||||||

These are similar to truncated tetrahedral particles except that the four vertices are now arbitrarily truncated into equal equilateral triangular faces of edge length $a$ (see Fig,~\ref{fig_trunc_tetr}). The total edge length is now $12a + 6(l-2a) = 6\,l$, that is truncation does not change the total edge length compared to the parent tetrahedron. In this case the encompassing sphere radius is 
\begin{equation}
r =  \left( \frac{3}{8}L^2 + \frac{1}{2}aL + \frac{1}{2}a^2  \right)^\frac{1}{2}, 
\end{equation}
where $L=(l-2a)$, and which gives the values for the regular tetrahedron, truncated tetrahedron, and octahedron for $a=0$, $a=L$, and $L=0$ , respectively. The above equation can, for later convenience, be re-arranged as a function of the remnant edge length $L$ and the edge length ratio $(a/L)$, to 
\begin{equation}
r =  L \, \Bigg\{ \frac{3}{8} + \frac{1}{2} \left( \frac{a}{L} \right) +  \frac{1}{2} \left( \frac{a}{L} \right)^2  \Bigg\}^\frac{1}{2}.  
\label{eq_solver1}
\end{equation}
The surface area of an semi-regular truncated tetrahedron, with four equivalent, triangular \{111\} facets, each of area $\surd 3/4 \, a^2$, and four six-sided facets, each of area $\surd 3/4 \, ( l^2 - 3a^2)$, is:  
\begin{equation}
  A_{\rm stT}  = \surd 3 \, ( l^2 - 2 a^2).   
\end{equation}
In this case $a = l/2$ leads to a singularity and so we instead set $a = 0.49 \, l$ in order to provide indicative data points for this limiting case in the figures. Given that the entire surface area is in triangular and six-sided \{111\} facets, $f_{\rm s \{111\}} = 1$. However, as noted above for tT polyhedra, not all edges are of \{100\} CH$_2$-type, only 6 are and hence $E_{\{100\}} = 6 \, L$.  The ratio of the total edge length, $E_{\rm stT}$, to the total surface area is
\begin{equation} 
 \frac{ E_{\rm stT} }{ A_{\rm stT} } =  \frac{6 \, l}{ \surd 3 ( l^2 - 2 a^2) } 
 =  \frac{ 2 \surd 3 \, l}{ ( l^2 - 2 a^2) } 
\end{equation} 
and the ratio of the surface area to that of the encompassing sphere is 
\begin{equation} 
\frac{ A_{\rm stT} }{ A_{\rm sphere} } =  \frac{ \surd 3 ( l^2 - 2 a^2) }{ 4 \pi r^2 } = \frac{ \surd 3 ( l^2 - 2 a^2) }{ 4 \pi \left( \frac{3}{8}L^2 + \frac{1}{2}aL + \frac{1}{2}a^2  \right) } . 
\end{equation}
The volume of a semi-regular truncated tetrahedron is given by:  
 \begin{equation}
V_{\rm stT}  =   \frac{ \surd 2}{12} ( l^3 - 4 a^3 )    
\end{equation}
 and the ratio of the volume of a semi-regular truncated tetrahedron (radius $r$) to that of its encompassing sphere of the same radius $r$ is: 
\begin{equation}
  \frac{ V_{\rm stT} }{ V_{\rm sphere} }  =  \frac{ \frac{ \surd 2}{12} ( l^3 - 4 a^3 ) }{ \frac{4}{3} \pi r^3} =  \frac{ \surd 2 ( l^3 - 4 a^3 ) }{ 16 \, \pi \left( \frac{3}{8}L^2 + \frac{1}{2}aL + \frac{1}{2}a^2  \right)^\frac{3}{2}  }.  
\end{equation}
Equating $V_{\rm stT}$ to a spherical particle of the same volume to determine the radius, $a_{\rm nd}$, of the equivalent "spherical nano-diamond" we have 
\begin{equation}
\frac{4}{3} \pi a_{\rm nd}^3 =   \frac{ \surd 2}{12} ( l^3 - 4 a^3 ). 
\label{eq_a_nd_solver-tT}  
\end{equation}
This expression is not directly solvable for $r_{\rm eff}$ as for regular polyhedra and so we need to adopt a different approach. Substituting $l = ( L + 2 a)$ and re-arranging the above equation to the following form we have 
\begin{equation}
  L_{\rm eff}  =  \Bigg\{ \frac{ \pi \, 8 \surd 2 }{ 1 + 6 (a/L) + 12 (a/L)^2 + 4 (a/L)^3 } \Bigg\}^\frac{1}{3}  a_{\rm nd},  
\end{equation}
where the ratio $(a/L)$ is defined for the particular truncated particle shape under consideration.\footnote{N.B., $(a/L)$ can take any positive value: for $a = 0$, $a=L$, and $L=0$ we have $(a/L) = 0, 1$ and $\infty$, respectively.} We can determine the remnant polyhedron effective edge length, $L_{\rm eff}$, and substitute this and $(a/L)$ into Eq.~(\ref{eq_solver1}) and thus obtain the  value of $r_{\rm eff}$ that corresponds to the required nano-diamond radius $a_{\rm nd}$. Alternatively we can bypass $r_{\rm eff}$ and calculate $a_{\rm nd}$ directly by solving Eq. (\ref{eq_a_nd_solver-tT}), that is 
\begin{equation}
a_{\rm nd} =   \Big\{ \frac{ \surd 2}{16 \, \pi} ( l^3 - 4 a^3 ) \Big\}^{\frac{1}{3}}.   
\end{equation}

%||||||||||||||||||||||||||||||||||||||||||||||||||||||||||||||||||||||||||||||||||||||||||||||||||||||||||||||||||||||||||||||||||||||||||||||||||||||||||||||||||
\subsection{Semi-regular truncated octahedral (stO) particles}
%||||||||||||||||||||||||||||||||||||||||||||||||||||||||||||||||||||||||||||||||||||||||||||||||||||||||||||||||||||||||||||||||||||||||||||||||||||||||||||||||||

These are octahedral particles with the six vertices arbitrarily truncated into equal square faces of edge length $a$. The total edge length is now $24a + 12(l-2a) = 12\,l$, that is truncation does not change the total edge length with respect to the parent octahedron (see Fig. \ref{fig_trunc_octa}). The encompassing sphere radius is 
\begin{equation}
r =  \left( \frac{1}{2}L^2 + aL + a^2  \right)^\frac{1}{2}, 
\end{equation}
where $L=(l-2a)$, and which gives the values for the regular octahedron, truncated octahedron and cuboctahedron for $a=0$, $a=L$, and $L=0$ , respectively. As above, and for later convenience, the above equation can be re-arranged as a function of the remnant edge length $L$ and the edge length ratio $(a/L)$, to 
\begin{equation}
r =  L \, \Bigg\{  \frac{1}{2} + \frac{a}{L} + \left( \frac{a}{L}  \right)^2   \Bigg\}^\frac{1}{2}.  
\label{eq_solver2}
\end{equation}
The surface area of a semi-regular truncated octahedron, with six equivalent, square \{100\} facets, each of area $a^2$, and eight six-sided \{111\} facets, each of area $\surd 3/4 \, ( l^2 - 3a^2 )$, is:  
\begin{equation}
  A_{\rm stO}  = 8 \frac{ \surd 3}{4} \surd 3 (l^2 - 3a^2 ) + 6a^2 =  2 \, \Big\{ \surd 3 \, l^2 - 3 ( \surd 3 -1 ) \, a^2 \Big\}.   
\end{equation}
The fraction of the surface in \{111\} facets, $f_{\rm s \{111\}}$, is 
\begin{equation}
f_{\rm s \{111\}}  = \frac{ 2 \surd 3 (l^2 - 3a^2 ) }{ 2 \surd 3 (l^2 - 3a^2 ) + 6a^2 } = \Bigg\{ \frac{ 1 - 3 (a/l)^2 }{ 1 - ( 3 - \surd 3 ) (a/l)^2 } \Bigg\}
\end{equation}
and, trivially, $f_{\rm s \{100\}} = ( 1 - f_{\rm s \{111\}})$. The square truncated facet edges, of total length $ 24 \, a $, are of \{100\}-type and are therefore included in the \{100\} facet CH$_2$ groups. However, the twelve edges of length $L = (l-2a)$ of the six-sided faces are \{111\}/\{111\}  edges, hence $E_{\{111\}} = 12 \, L = 12 (l-2a)$, and their CH groups form part of the adjacent \{111\} facets. The ratio of the total edge length, $E_{\rm stO}$, to the total surface area is
 \begin{equation}  
 \frac{ E_{\rm stO} }{ A_{\rm stO} } =  \frac{12 \, l}{ 2 \, \Big\{ \surd 3 \, l^2 - 3 ( \surd 3 -1 ) \, a^2 \Big\} } 
= \Big\{ \frac{ \surd 3 }{ 6 } \, l - \frac{1}{2} ( \surd 3 -1 ) \frac{a^2}{l} \Big\}^{-1}
 \end{equation} 
and the ratio of the surface area to that of the encompassing sphere is 
\begin{equation}
%\[
\frac{ A_{\rm stO} }{ A_{\rm sphere} } =  \frac{ 2 \, \Big\{ \surd 3 \, l^2 - 3 ( \surd 3 -1 ) \, a^2 \Big\} }{ 4 \pi r^2 } 
%\]
%\begin{equation}
%  \ \ \ \ \ \ \ \ \,  
  = \frac{  \surd 3 \, l^2 - 3 ( \surd 3 -1 ) \, a^2 }{ 2 \pi \left( \frac{1}{2}L^2 + aL + a^2  \right) }. 
\end{equation}
The volume of a semi-regular truncated octahedron is given by:  
 \begin{equation}
V_{\rm stO}  =   \frac{ \surd 2}{3} ( l^3 - 3 a^3 )    
\end{equation}
and the ratio of its volume (radius $r$) to that of its encompassing sphere of the same radius $r$ is: 
\begin{equation}
  \frac{ V_{\rm stO} }{ V_{\rm sphere} }  =  \frac{ \frac{ \surd 2}{3} ( l^3 - 3 a^3 ) }{ \frac{4}{3} \pi r^3} 
 =  \frac{ \surd 2 ( l^3 - 3 a^3 ) }{ 4 \, \pi \left( \frac{1}{2}L^2 + aL + a^2  \right)^\frac{3}{2}  }.  
\end{equation}
Equating $V_{\rm stO}$ to a spherical particle of the same volume to determine the radius, $a_{\rm nd}$, of the equivalent `spherical nano-diamond' we have 
\begin{equation}
\frac{4}{3} \pi a_{\rm nd}^3 =   \frac{ \surd 2}{3} ( l^3 - 3 a^3 ). 
\label{eq_a_nd_solver_tO}  
\end{equation}
This expression is not directly solvable for $r_{\rm eff}$ as for regular polyhedra and so as in the preceding case we substitute $l = ( L + 2 a)$ and re-arrange to the following form
\begin{equation}
  L_{\rm eff}  =  \Bigg\{ \frac{ \pi \, 2 \surd 2 }{ 1 + 6 (a/L) + 12 (a/L)^2 + 5 (a/L)^3 } \Bigg\}^\frac{1}{3}  a_{\rm nd},  
\end{equation}
where the ratio $(a/L)$ is defined for the particular truncated particle shape under consideration. We can determine the remnant polyhedron effective edge length, $L_{\rm eff}$, and substitute this and $(a/L)$ into Eq.~(\ref{eq_solver2}) and thus obtain the  value of $r_{\rm eff}$ that corresponds to the required nano-diamond radius $a_{\rm nd}$. Again, if we do not need $r_{\rm eff}$, we can calculate $a_{\rm nd}$ by solving Eq. (\ref{eq_a_nd_solver_tO}), that is 
\begin{equation}
a_{\rm nd} =   \Big\{ \frac{ \surd 2}{4 \, \pi} ( l^3 - 3 a^3 ) \Big\}^{\frac{1}{3}}.   
\end{equation}

%||||||||||||||||||||||||||||||||||||||||||||||||||||||||||||||||||||||||||||||||||||||||||||||||||||||||||||||||||||||||||||||||||||||||||||||||||||||||||||||||||
\subsection{Semi-regular truncated cubic (stC) particles}
%||||||||||||||||||||||||||||||||||||||||||||||||||||||||||||||||||||||||||||||||||||||||||||||||||||||||||||||||||||||||||||||||||||||||||||||||||||||||||||||||||

These are cubes with the eight vertices arbitrarily truncated into equilateral triangular faces of edge length $a$. The total edge length is now $24a + 12 [ l-\surd 2 \, a] = [ 12 l + 12 a \,(2-\surd 2) ]$ (see Fig. \ref{fig_trunc_cube}). Note that in this case truncation does change the total edge length with respect to the parent cube. The encompassing sphere radius is 
\begin{equation}
r =  \left( \frac{3}{4}L^2 + \surd 2 \,aL + a^2  \right)^\frac{1}{2}, 
\end{equation}
where $L=(l-2a)$, and which gives the values for the regular cube, truncated cube and cuboctahedron for $a=0$, $a=L$,  and $L=0$ , respectively. Once again for later convenience we re-arrange the above equation as a function of the remnant edge length $L$ and the edge length ratio $(a/L)$, to 
\begin{equation}
r =  L \, \Bigg\{  \frac{3}{4} + \surd 2 \frac{a}{L} + \left( \frac{a}{L}  \right)^2   \Bigg\}^\frac{1}{2}.  
\label{eq_solver3}
\end{equation}
The surface area of an semi-regular truncated cube, with eight equivalent, triangular \{111\} facets, each of area $(\surd 3 / 4) \, a^2$, and six eight-sided \{100\} facets, each of area $( l^2 - a^2 )$, is:  
\begin{equation}
%\[
  A_{\rm stC}  = 8 \, (\surd 3 / 4) \, a^2 + 6 \, ( l^2 - a^2 )
%\]
%\begin{equation}
%\ \ \ \ \ \ \ \, 
= 6 \, l^2 + 2 ( \surd 3 - 3) \, a^2.   
\end{equation}
The fraction of the surface in \{111\} facets, $f_{\rm s \{111\}}$, is 
\begin{equation}
%\[
f_{\rm s \{111\}}  = \frac{ 8 (\surd 3 / 4) \, a^2 }{ 6 \, l^2 + 2 ( \surd 3 - 3) \, a^2 } 
%\]
%\begin{equation}
% \ \ \ \ \ \ \ \ \ \ 
= \surd 3 \, \Bigg\{ 3 \left( \frac{l}{a} \right)^2 +  \surd 3 -6 \Bigg\}^{-1}
\end{equation}
and, trivially, $f_{\rm s \{100\}} = ( 1 - f_{\rm s \{111\}})$. The triangular truncated \{111\} facet edges, of total length $ 24 \, a $, all border \{100\} facets are of  \{100\}-type and their CH$_2$ groups are therefore included in the \{100\} facets. All other edges are \{100\}/\{100\} edges, hence $E_{\{111\}} = 0$. The ratio of the total edge length, $E_{\rm stC}$, to the total surface area is
 \begin{equation} 
%\[
 \frac{ E_{\rm stC} }{ A_{\rm stC} } =  \frac{ [ 12\,l + 12\,a (2 - \surd 2) ] }{ 6 \, l^2 + 2 ( \surd 3 - 3) \, a^2 } 
% \]
% \begin{equation} 
% \ \ \ \ \ \ \ 
= \frac{ 2 [ \left( l/a \right) + 2 - \surd 2 ]}{ (l^2 / a) + (\surd 3 / 3 -1 ) \, a }
 \end{equation} 
and the ratio of the surface area to that of the encompassing sphere is 
\[
\frac{ A_{\rm stC} }{ A_{\rm sphere} } =  \frac{ 6 \, l^2 + 2 ( \surd 3 -3) \, a^2 }{ 4 \pi r^2 } 
%\]
%\[
%  \ \ \ \ \ \ \ \ \,  
= \frac{  6 \, l^2 + 2 ( \surd 3 -3)) \, a^2 }{ 4 \pi \left( \frac{3}{4}L^2 + \surd 2 \,aL + a^2 \right) }  
\]
\begin{equation}
  \ \ \ \ \ \ \ \ \ \ \ \,  = \frac{  (l/a)^2 + \surd 3 / 3 -1 }{ \frac{2}{3} \pi \left[ \frac{3}{4} (L/a)^2 + \surd 2 \, (L/a) + 1 \right] }. 
\end{equation}
The volume of a semi-regular truncated cube is given by:  
 \begin{equation}
V_{\rm stC}  =   \frac{1}{3} ( 3 l^3 - \surd 2 a^3 )    
\end{equation}
and the ratio of its volume (radius $r$) to that of its encompassing sphere of the same radius $r$ is: 
\begin{equation}
  \frac{ V_{\rm stC} }{ V_{\rm sphere} }  =  \frac{ \frac{1}{3} ( 3 l^3 - \surd 2 a^3 ) }{ \frac{4}{3} \pi r^3} 
 =  \frac{ ( 3 l^3 - \surd 2 a^3 ) }{ 4 \, \pi \left( \frac{3}{4}L^2 + \surd 2 \,aL + a^2  \right)^\frac{3}{2}  }.  
\end{equation}
Equating $V_{\rm stC}$ to a spherical particle of the same volume to determine the radius, $a_{\rm nd}$, of the equivalent "spherical nano-diamond" we have 
\begin{equation}
\frac{4}{3} \pi a_{\rm nd}^3 =   \frac{1}{3} ( 3 l^3 - \surd 2 a^3 ). 
\label{eq_a_nd_solver_tC}  
\end{equation}
Once again we have an equation that is not directly solvable for $r_{\rm eff}$ and so we substitute $l = ( L + 2 a)$ and re-arrange to the following 
\begin{equation}
  L_{\rm eff}  =  \Bigg\{ \frac{ 4 \pi }{ 3 [1 + 6 (a/L) + 12 (a/L)^2 + (8 + \surd 2 / 3 ) (a/L)^3 ] } \Bigg\}^\frac{1}{3}  a_{\rm nd},  
\end{equation}
where the ratio $(a/L)$ is defined for the particular truncated particle shape under consideration. This again allows us to  determine the remnant polyhedron effective edge length, $L_{\rm eff}$, and substitute this and $(a/L)$ into Eq.~(\ref{eq_solver3}) and thus obtain the  value of $r_{\rm eff}$ that corresponds to the required nano-diamond radius $a_{\rm nd}$. Again we can bypass $r_{\rm eff}$ and calculate $a_{\rm nd}$ by solving Eq. (\ref{eq_a_nd_solver_tC}), that is 
\begin{equation}
a_{\rm nd} =   \Big\{ \frac{ 1 }{4 \, \pi} ( 3 l^3 - \surd 2 a^3 ) \Big\}^{\frac{1}{3}}.   
\end{equation}

\end{document}